\documentclass[a4paper,10pt]{article}
   \ifdefined\XeTeXversion
     \ifdefined\pdfoutput
       \pdfoutput=0
     \fi
     \PassOptionsToPackage{xetex}{hyperref}
   \else
     \ifdefined\pdfoutput
       \pdfoutput=1
     \fi
   \fi
\pdfoutput=1 

\usepackage{jhepmod}
\usepackage[T1]{fontenc}

\usepackage{setspace}

\usepackage{amsthm}
\usepackage{mathtools}
\usepackage{subcaption}
\usepackage{physics2}
\usepackage{dsfont}
\usepackage{tensor}
\usepackage[normalem]{ulem}
\usepackage{xcolor}
\usepackage{graphicx}
\usepackage[export]{adjustbox}
\usepackage{eulervm}
\usepackage{comment}
\usepackage{enumitem}
\usepackage{microtype}
\usepackage{todonotes}
\usepackage{standalone}
\usepackage{comment}

\usepackage{tikz}
\usetikzlibrary{shapes.geometric}
\usetikzlibrary{decorations.markings}

\colorlet{darkblue}{blue!70!black}
\colorlet{darkgreen}{green!50!black}
\colorlet{midgreen}{green!60!black}
\colorlet{chillred}{red!60!white}

\usepackage{pgfplots}
\usepackage{svg}
\usepackage{tikz-cd}

\usetikzlibrary{pgfplots.fillbetween}
\usetikzlibrary{intersections}
\usetikzlibrary{shapes.geometric}
\usetikzlibrary{calc}
\usetikzlibrary{cd}

\usepackage[status=draft]{fixme}
\fxsetup{layout=margin}

\let\oldfxnote\fxnote

\renewcommand{\fxnote}[2][]{%
  {\reversemarginpar\oldfxnote[#1]{#2}}%
}

\newif\iffocusmode
\focusmodefalse
\iffocusmode
  \excludecomment{lotusfocusdraft}
\else
  
\fi
\newcommand{\VF}[6]{\mathds{F}_{#1#2} \left[ \begin{matrix} #3 & #4 \\ #5 & #6 \\ \end{matrix} \right]}

\title{\boldmath A Holographic Map from $\text{AdS}_{3}$ to $\text{CFT}_{2}$}

\author[a,b]{Manish Ramchander}
\author[c]{ and Ronak M Soni}
\affiliation[a]{The Institute of Mathematical Sciences, IV Cross Road, C.I.T. Campus, Taramani, Chennai, India 600113}
\affiliation[b]{Homi Bhabha National Institute, Training School Complex, Anushakti Nagar, Mumbai, India 400094}
\affiliation[c]{Chennai Mathematical Institute, H1, SIPCOT IT Park, Siruseri, Kelambakkam 603103, India}
\emailAdd{physicophillic@gmail.com}
\emailAdd{ronakmsoni@gmail.com}

\abstract{
	We propose a holographic map from the semiclassical Hilbert space of pure general relativity in $\text{AdS}_{3}$ to that of $\text{CFT}_{2}$.
	We define the bulk Hilbert space by semiclassically quantising the phase space in the basis of a fixed-area network on a Cauchy slice $\Sigma$.
	A fixed-area network is a maximal non-intersecting set of geodesics on $\Sigma$ whose lengths and angular momenta have been fixed.
	Our holographic map differentiates between `external' geodesics that are homotopic to a connected component of $\partial \Sigma$, and `internal' geodesics which are not.
	The lengths and angular momenta of external geodesics become conformal weights of primaries in the Hilbert space of the CFT living on the corresponding component of $\partial \Sigma$.
	The fixed-area network determines the wavefunction, which is given by a network of OPE coefficients of primaries whose weights are determined by the corresponding lengths.
	For sufficiently semiclassical states, there is an agreement between bulk and boundary inner products.

	We apply this proposal to various physics questions.
	The boundary dual of a bulk gauge turns out to be an emergent basis for sufficiently semiclassical states.
	We also define boundary operators that measure the lengths of geodesics behind the horizon, again in semiclassical states.
	Finally, we apply our map to closed universes and find a failure of semiclassicality in simple cases, which can be partially alleviated by the addition of a massive probe.
}

\begin{document}
\maketitle
\flushbottom

\section{Introduction} \label{sec:intro}

One of the striking developments in AdS/CFT over the last ten years or so has been an enhanced appreciation of the unreasonable effectiveness of semiclassical Euclidean gravity.
While the first example of this unreasonable effectiveness was probably the Gibbons-Hawking calculation of black hole entropy \cite{Gibbons:1976ue}, these recent developments show that it is far more effective than even that.
Not only does Euclidean gravity know about the total entropy, wormhole contributions also encode non-perturbative effects that allow us to find the Page curve \cite{Almheiri:2019qdq,Penington:2019npb}, calculate spectral statistics of black hole microstates \cite{Saad:2019lba}, sum up exponentially small inner products to solve the bag-of-gold paradox \cite{Penington:2019kki,Hsin:2020mfa,Balasubramanian:2022gmo,Balasubramanian:2022lnw}, and precisely count microstates of BPS black holes \cite{Iliesiu:2022kny}, among other things.

One of the by-products of this unreasonable effectiveness is a refinement of our understanding of the AdS/CFT dictionary.
The wormhole contributions to the gravitational path integral (GPI) foil any attempts to interpret the semiclassical GPI as the exact partition function of a CFT.\footnote{
	In fact, these issues were already apparent due to the duality between the thermofield double and the two-sided eternal black hole, as explained in \cite{Mathur:2014dia}.
}
This is completely explicit in 2d, where JT gravity is exactly dual to an ensemble average over theories \cite{Saad:2019lba}.
Pure 3d gravity is also believed to be dual to an ensemble, see e.g. \cite{Chandra:2022bqq,Belin:2023efa,Hartman:2025cyj,Hartman:2025ula,Jafferis:2026gzn}.

This does not contradict the exact AdS/CFT conjecture, which states that a holographic CFT is dual to \emph{full} quantum gravity.
Semiclassical gravity is of course just a coarse-graining of quantum gravity and so we should expect it to be dual to a coarse-graining of the CFT.
From this perspective, ensemble averaging can be thought of as merely a tractable toy model for the required coarse-graining.
This is important, since there is a massive amount of arbitrariness in coarse-graining and it is a very hard problem to make explicit the coarse-graining that leads to a given EFT description of the bulk.
This problem is mostly unsolved, though see \cite{Liu:2025ikq} for an axiomatic approach and \cite{Wu:2026zop} for an application to $\text{AdS}_{3}/\text{CFT}_{2}$.\footnote{
	There have also been attempts to make the coarse-graining explicit, like \cite{Chandra:2023dgq,Chandra:2023rhx,Soni:2025qau,Lin:2026wvf}.
}

\paragraph{An Introduction to Holographic Maps}

Amazingly, however, \cite{Iliesiu:2024cnh} showed that the problem is much more tractable at the level of states, if one goes the other way.
They defined a \emph{fine-graining} ``holographic map,'' which maps states in a given EFT to a state, in a single boundary theory, which is extremely well-approximated by the bulk state (up to exponentially small corrections).
It can be defined as the map that makes the following diagram commute up to small corrections:
\begin{equation}
	\begin{tikzcd}[sep=large,cells={inner ysep=1.75ex, inner xsep=1.5ex}]
		\text{\small Semiclassical CFT states} \ar[r,hook] \ar[d, "\text{\tiny Coarse-Graining}"] & \mathcal{H}_{\mathrm{CFT}} \ar[d, leftarrow, "V"] \\
		\text{\small Semiclassical bulk states} \ar[r,hook] & \mathcal{H}_{\mathrm{bulk}}
	\end{tikzcd}
\end{equation}
Semiclassical bulk states are coarse-grainings of semiclassical CFT states.
We define a bulk ``Hilbert space'' which contains these semiclassical bulk states; as we will see, this is not a true Hilbert space.\footnote{
	It will be a vector space where the inner product function maps pairs of states not to complex numbers but to transseries.
}
This bulk Hilbert space is an EFT Hilbert space, and not every state in this Hilbert space should be expected to have meaning in the physical Hilbert space.
The simple reason is that the set of states where any semiclassical EFT is valid does not form a linearly closed set, whereas $\mathcal{H}_{\mathrm{bulk}}$ by definition does.
The holographic map $V$ is a linear map from $\mathcal{H}_{\mathrm{bulk}}$ to the CFT Hilbert space such that semiclassical bulk states map to the corresponding semiclassical CFT state.
Where non-semiclassical states in $\mathcal{H}_{\mathrm{bulk}}$ map to in the CFT Hilbert space is restricted only by the linearity conditions above.

The reason this map is useful is that it separates out a `universal' part of the semiclassical AdS/CFT dictionary from all the messy decisions that need to be made.
In any theory, there are many sets of semiclassical states (for example, the squeezed coherent states with different squeezing parameters are all semiclassical states for the harmonic oscillator).
Each such set in the EFT will be a different coarse-graining; the construction of the auxiliary $\mathcal{H}_{\mathrm{bulk}}$ that contains all of them allows us to have one map that embeds them all in the fundamental theory.
Further, if one starts from the fundamental theory, we have to study for any given problem what EFT is valid and what is not.
Here, we pick an EFT and live with the fact that the result of the map will be nonsense if the EFT is not valid for the problem in question.

In this work, we apply this idea to pure 3d gravity with asymptotically AdS boundary conditions and end-of-the-world (ETW) branes, leading to the holographic map described in section \ref{sec:V}.
Note that, in keeping with the philosophy just outlined, we will map pure 3d GR to a \emph{true} holographic CFT.
Any calculations done with our map are only valid when pure GR is a good effective theory for the problem under consideration.

The classical states in pure 3d gravity are multi-boundary wormholes \cite{Brill:1995jv,Brill:1995yc,Aminneborg:1997pz,Aminneborg:1998si,Aminneborg:2008sa}, which are $n$ black holes connected behind the horizon.
Adding ETW branes, the classical states are similar, except that some of the asymptotic boundaries may be intervals instead of circles and the interior can have ETW-brane boundaries.
The construction of these classical states from Euclidean path integral states (EPI) in the CFT has been explored in \cite{Krasnov:2001va,Skenderis:2009ju,VanRaamsdonk:2010pw,Balasubramanian:2014hda,VanRaamsdonk:2018zws}, among others.
We go beyond these works in that we provide a holographic map also for semiclassical states, which may not be dual to CFT states prepared by path integrals of fixed moduli.
For this, we use recent developments in the bootstrap in holographic 2d CFTs; an incomplete set of references is \cite{Collier:2019weq,Belin:2020hea,Belin:2021ryy,Anous:2021caj,Kusuki:2021gpt,Numasawa:2022cni,Kusuki:2022ozk,Belin:2025nqd} and a pedagogical introduction is \cite{Kusuki:2024gtq}.
In particular, it has been shown that the statistics of OPE coefficients have detailed agreement with bulk actions \cite{Kusuki:2022ozk,Chandra:2022bqq,Belin:2023efa,Jafferis:2025yxt,Jafferis:2025jle,Wang:2025bcx,Hartman:2025cyj,Hartman:2025ula,Belin:2026pko}, and this will be the main evidence for our proposal.
In fact, our proposal is essentially implicit in these works, and our job is mainly to collect some of the results and make them explicit.

\paragraph{Summary of our Results}

Let us now describe the setup a little more carefully.
Firstly, note that since gravity is not renormalisable, we do not have a true Hilbert space in the bulk.
However, we can still define a semiclassical bulk Hilbert space using the Wheeler-De Witt (WdW) formalism.
There are two prefatory comments about this object.
This semiclassical Hilbert space is not a true Hilbert space: the inner products we define will not be numbers but \emph{transseries} (sums over asymptotic expansions around different saddles) in $G_{N}$.
We will abuse terminology and call it a Hilbert space anyway.
Secondly, the reason this is a semiclassical Hilbert space is that it follows from a procedure of semiclassical quantisation; this does not mean that all the elements of the Hilbert space are semiclassical.
The existence of both semiclassical and non-semiclassical states in this Hilbert space will play an important role in our discussion.

The WdW formalism is as follows.
We define a set of basis states on a slice $\Sigma$ by specifying a Poisson-commuting set of phase space coordinates; let us denote such a basis element by $\pbra{\Sigma}$ for now.
This is the natural generalisation of the $\bra{x}$ states in $L^{2} (\mathds{R})$.
These bases states are not elements of the Hilbert space but of its algebraic dual, again similar to $\bra{x}$ states; Hilbert space states $\pket{\Psi}$ are given by smooth wavefunctions $\pip{\Sigma}{\Psi}$.
Then, we interpret these basis states $\pbra{\Sigma}$ as boundary conditions for the GPI, which we evaluate as a sum over saddle-point expansions.
Crucially, the inner product is taken to be, see e.g. \cite{Teitelboim:1981ua} for a pedagogical explanation,
\begin{equation}
  \pmel{\Sigma'}{\eta} {\Sigma} = \int_{\partial \mathcal{M} = \Sigma \cup \Sigma'}^{*} Dg \, e^{- I[\mathcal{M},g]},
  \label{eqn:WdW-inner-product}
\end{equation}
where the $*$ denotes that we evaluate it in a semiclassical expansion.
$I$ is the Euclidean action, but $\mathcal{M}$ is allowed to be Lorentzian; the Euclidean action just becomes complex in this case.

\begin{figure}[h!]
  \centering
  \includegraphics[width=.4\textwidth]{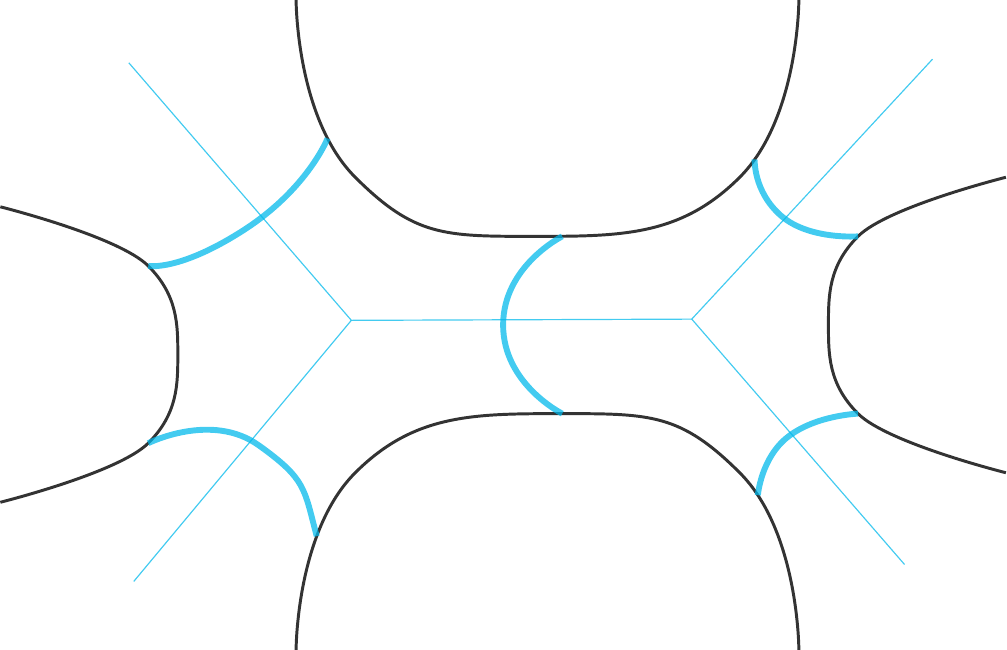}
  \caption{An example of a four-boundary wormhole. Thick cyan lines are the geodesics whose lengths and angular velocities we are fixing. The thin cyan lines are an auxiliary `framing graph' whose purpose will be explained in the main text. The open ends are asymptotic boundaries; the four geodesics homotopic to the asymptotic boundaries are external and the fifth one in the middle is internal.}
  \label{fig:four-bd-eg}
\end{figure}

The basis we take is what we misleadingly call a maximally fixed-area basis.
For $\Sigma$ of any given topology, we take a maximal set of non-intersecting geodesics on $\Sigma$ and fix both their lengths as well as their angular velocities.
For example, one can consider $\Sigma$ to be a slice in a four-boundary wormhole, as shown in figure \ref{fig:four-bd-eg}, in which case there are five non-intersecting geodesics.
Four of them are `external' geodesics homotopic to the boundaries and the fifth one is an `internal' geodesic which we have taken to be in the $s$ homotopy class (surrounding boundaries $1$ and $2$) for illustration.
In figure \ref{fig:four-bd-eg}, we have also drawn a `framing' graph; while its main purpose is to keep track of the quantity conjugate to the angular velocity, we also use this graph to denote the boundary condition itself; our basis state is
\begin{equation}
		\pbra{\input{figs/fa-4-bd-s.tex}}.
  \label{eqn:fa-s-state-intro}
\end{equation}
We have set the angular velocities to $0$ for the time being, so the lines are only labelled by the lengths.

Our holographic map takes such a state to a state in the primary Hilbert space (the part of the Hilbert space of the CFT that only keeps track of the primary sector)
\begin{align}
	V \pket{ 
		\begin{tikzpicture}[baseline]
			\draw[cyan] (0,0) -- node[below] {\tiny $\ell_{s}$} ++(  0:1);
			\draw[cyan] (0,0) -- node[pos=1,above] {\tiny $\ell_{1}$} ++(120:.5);
			\draw[cyan] (0,0) -- node[pos=1,above] {\tiny $\ell_{2}$} ++( 60:.5);
			\draw[cyan] (1,0) -- node[pos=1,above] {\tiny $\ell_{3}$} ++(120:.5);
			\draw[cyan] (1,0) -- node[pos=1,above] {\tiny $\ell_{4}$} ++( 60:.5);
		\end{tikzpicture}
	} 
		&= \sum_{(P_{I}, P_{I}) \in \mathrm{prim}} N(P)  \left[ \prod_{I = 1\dots 4,s} \delta(\ell_{I} - 4 \pi b P_{I}) \right]
		  C_{12s} C_{s34} \otimes_{i=1}^{4} \ket{(P_{i}, P_{i})}.
		\label{eqn:hol-map-intro}
\end{align}
Here, $P$ is related to the conformal weight of the primary (non-rotating geodesics are dual to scalars where $\bar{P} = P$), $b$ is $\sqrt{4 G_{N}}$ and the $C_{abc}$s are OPE coefficients of the theory.
$N(P)$ are important normalisation factors containing primary densities of states and averages of three-point functions.
This map
\begin{enumerate}[label=\alph*.]
  \item checks whether all the lengths (divided by $4\pi b$) are weights of primaries in the primary spectrum; if not, the state is annihilated,

  \item associates to each pair of pants in $\Sigma$ (i.e. a vertex in our fixed-area network) an OPE coefficient,
  
  \item and, finally, picks the states in the primary Hilbert space with weights related to the lengths of the external geodesics.
  
\end{enumerate}
There are various things to note.
The first is that the lengths of the internal geodesics only appear in the wavefunction, so that the images of basis elements with different values of $\ell_{s}$ are proportional to each other.
Secondly, this map is highly non-isometric, annihilating all but a measure-zero set of basis elements.
It becomes approximately isometric only for states where all the lengths are smeared over $\mathcal{O} (G_{N})$ `microcanonical' (or larger!) windows with slowly varying phases.
We call such states semiclassical.
This is a concrete realisation of the fact discussed above that most of $\mathcal{H}_{\mathrm{bulk}}$ is not obligated to have much meaning.
\eqref{eqn:hol-map-intro} and its generalisation to arbitrary $\Sigma$ is the central proposal of our paper.

We discuss the construction of the bulk Hilbert space in section \ref{sec:H-bulk} and known facts about holographic 2d CFTs in section \ref{sec:H-CFT}.
The holographic map is then introduced in section \ref{sec:V}.
Armed with this new tool, we then explore various applications in the subsequent sections before concluding in section \ref{sec:disc}.
These applications should be thought of as initial forays rather than complete analyses.

The most interesting application of our map is that it provides us with a clearer understanding of how bulk gauge symmetry embeds into the boundary Hilbert space, as discussed in section \ref{sec:gauge}.
Different choices of non-intersecting geodesics turn out to be related by Hamiltonian gauge transformation in the bulk and a transformation between \emph{emergent} bases in the boundary.
This corroborates and refines the observations of \cite{Parrikar:2025xmz}.

We also discuss bulk length operators in section \ref{sec:rel-ops}.
These bulk length operators are non-perturbatively relational, in a sense that we explain in section \ref{sec:rel-ops}.
We are able to find boundary reconstructions of these operators: linear operators whose matrix elements agree with the bulk operators for semiclassical states.

Finally, we study closed universes in section \ref{sec:cus}.
We find a failure of semiclassicality, in that semiclassical pure states are exponentially close to parallel, in agreement with \cite{Belin:2025ako,Abdalla:2026mxn}.
No bulk phase space can emerge from approximate orthogonality of coherent states if this is the case.
This can be relieved either by considering mixed states as in \cite{Belin:2025ako} or by including massive probes.

\section{The Bulk Hilbert Space} \label{sec:H-bulk} 

In this section, we formulate the Hilbert Space of 3d GR in the bulk using a basis that we call a maximally fixed-area basis.
The basis is labelled by ``fixed-area networks,'' a choice of a maximal set of non-intersecting geodesics whose lengths and angular velocities have been fixed.
Inner products between these basis elements can be calculated using a few known elementary pieces.
Superpositions of these basis elements with microcanonical smearing functions give us the fixed-area states discussed in the literature \cite{Dong:2018seb,Akers:2018fow,Dong:2019piw,Marolf:2020vsi,Dong:2023xxe,Dong:2022ilf}.

\subsection{3d GR with ETW Branes} \label{ssec:gr-defn}

\begin{figure}
  \centering
  \includegraphics[width=0.45\linewidth]{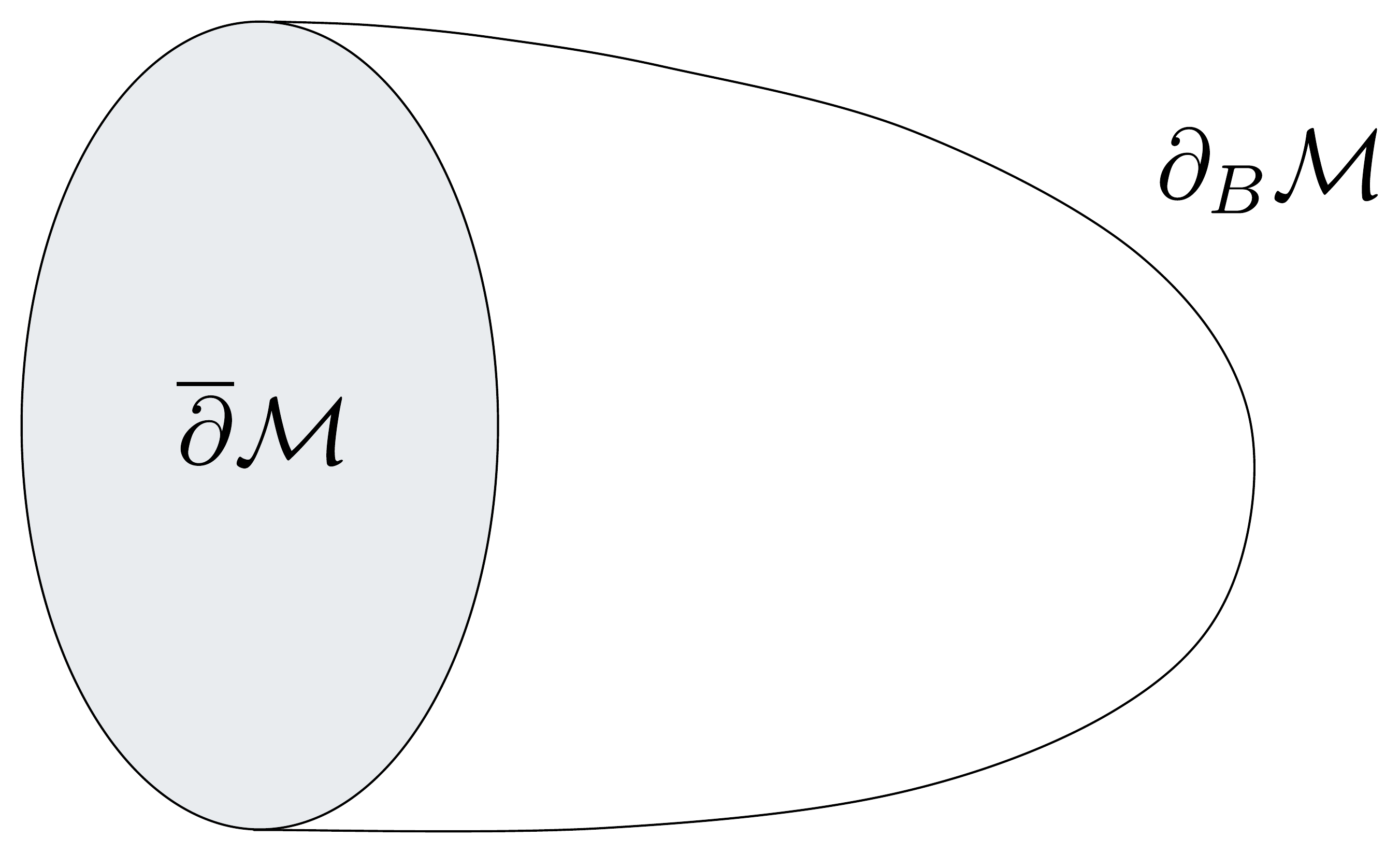}
  \caption{The manifold $\mathcal{M}$'s depiction, with $\partial \mathcal{M}$ splitting into two components. }
	\label{fig:bd-types}
\end{figure}

In this work, we will take the bulk theory to be 3d GR with end-of-the-world (ETW) branes.
Consider a 3d manifold $\mathcal{M}$, which we take to be Euclidean for concreteness.
Its boundary $\partial \mathcal{M}$ can be split into an asymptotic boundary $\bar{\partial} \mathcal{M}$ and a brane boundary $\partial_{B} \mathcal{M}$, as shown in figure \ref{fig:bd-types}.
The Euclidean action is \cite{Takayanagi:2011zk,Fujita:2011fp}
\begin{equation}
  I_{0} = - \frac{1}{16 \pi G_{N}} \int_{\mathcal{M}} \sqrt{g} (R+2) - \frac{1}{8 \pi G_{N}} \int_{ \bar{\partial} \mathcal{M}} \sqrt{h} (K-1) - \frac{1}{8 \pi G_{N}} \int_{\partial_{B} \mathcal{M}} \sqrt{h} (K-T).
  \label{eqn:euclidean-action}
\end{equation}
Here, $g$ is the metric on $\mathcal{M}$ and $h$ is the induced metric on the corresponding boundary.
We have taken the AdS radius $L_{\mathrm{AdS}} = 1$, and will continue to do so throughout this work.
The parameter $T$ is known as the tension.
We assume $T > 0$ for simplicity, since there are certain additional subtleties in the case of negative tension which will distract from the main flow of this work \cite{Kusuki:2022ozk}.
Different parts of $\partial_{B} \mathcal{M}$ can have different tensions but we have taken them all to be equal for simplicity.
Our results will generalise to the case of different tension in a straightforward fashion, using the results of \cite{Wang:2025bcx}.

The boundary conditions are standard asymptotically AdS boundary conditions on $\bar{\partial} \mathcal{M}$; on the ETW branes, they are
\begin{equation}
  K_{ab} = (K - T) h_{ab} \quad \implies \quad K = 2T, K_{ab} = T h_{ab}.
  \label{eqn:ETW-bd-cond}
\end{equation}
The boundaries $\partial \bar{\partial} \mathcal{M} = \partial \partial_{B} \mathcal{M}$ are also boundaries of the dual CFT where we impose Cardy boundary conditions \cite{Fujita:2011fp}.
The tension is related to the Affleck-Ludwig boundary entropy of the Cardy boundary condition by
\begin{equation}
  S_{\mathrm{bdy}} = \log g = \frac{c}{6} \tanh^{-1} T.
  \label{eqn:g-T}
\end{equation}
$g$ is also the inner product between the vacuum of the CFT and the boundary state $\oket{\sigma}$.

We will construct the semiclassical bulk Hilbert space of this theory and study its properties for the rest of this section.
We begin with a discussion of the phase space of the theory in section \ref{ssec:phase-space}.
We use a set of Poisson-commuting coordinates on this phase space to define a basis set for the Hilbert space in section \ref{ssec:H-bulk} and look at inner products between these basis elements in section \ref{ssec:inner-products}.

\subsection{The Phase Space} \label{ssec:phase-space}

We begin with a review of the phase space of this theory.
The phase space consists of two almost decoupled parts: the boundary gravitons whose phase space was studied in \cite{Brown:1986nw}, and the `bulk' phase space which is the space of multi-boundary wormholes.
Here, we review this bulk phase space.
Though we will use a slightly different gauge in our quantisation, various tools from this discussion will be useful.
This discussion is mainly a repackaging of existing results; the works we consulted most closely are \cite{mess2007lorentz,Scarinci:2011np,Krasnov:2005dm}.

We work in the initial value formulation of phase space, where we specify data on a codimension-one Cauchy slice $\Sigma$.
Indices on $\Sigma$ will be denoted by $i,j$ and those in the full 3d manifold by $\mu,\nu$.

In GR, the extended phase space is characterised by a metric $h_{ij}$ on a Cauchy slice $\Sigma$ and its canonical conjugate
\begin{equation}
	\pi^{ij} = \frac{ \sqrt{h}}{16 \pi G_{N}} (K^{ij} - g^{ij} K),
\end{equation}
where $K_{ij}$ is the extrinsic curvature, $K \equiv K^{i}_{i}$ and $h \equiv \det h_{ij}$.
This is not the physical phase space due to gauge-invariance: an arbitrary set $(h_{ij}, \pi^{ij})$ are not valid initial data, and different valid sets $(h_{ij}, \pi^{ij})$ can represent the same solution.
The physical phase space is given by a quotient of a constraint submanifold.
The constraint submanifold, which consists of valid data $(h_{ij}, \pi^{ij})$, is defined by the two constraints
\begin{align}
	 \qquad \frac{16 \pi G_{N}}{ \sqrt{h}} (\pi^{ij} \pi_{ij} - \pi^{2}) - \frac{ \sqrt{h}}{16 \pi G_{N}} (R[h] + 2) &=  0 \quad\quad \text{(Hamiltonian constraint)}\nonumber\\
	 \qquad \grad_{i}^{(h)} \pi^{ij} &= 0 \quad\quad\text{(Momentum constraint)}.
  \label{eqn:constraints}
\end{align}
While the two constraints vanish on the constraint surface, they still generate non-trivial flows on the surface via Poisson brackets.
These flows are known as gauge transformations, and the orbits of gauge transformations are known as gauge orbits.
Different points on a gauge orbit represent the same physical solution.
The physical phase space is the space of gauge orbits; in other words, the physical phase space is the quotient of the constraint surface by gauge transformations.

To perform this quotient, and therefore obtain the physical phase space, we pick a gauge: a slice of the constraint submanifold that, ideally, intersects each orbit only once.
The most convenient gauge to study this phase space is the maximal slice gauge
\begin{equation}
  \pi \propto K = 0.
  \label{eqn:max-slice-gauge}
\end{equation}
Every asymptotically AdS solution has a unique maximal slice intersecting a given boundary Cauchy slice, see e.g. \cite{Scarinci:2011np}, and so this completely fixes the gauge freedom generated by the Hamiltonian constraint.
The diffeomorphism constraints generate diffeomorphisms on $\Sigma$, and will be handled below.

The following approach will allow us to describe each point in the physical phase space.
For a start, take complex coordinates on $\Sigma$, such that
\begin{equation}
  \dd{s}_{\Sigma}^{2} = e^{2 \Omega} \dd{z} \dd{ \bar{z}}.
  \label{eqn:Sigma-h}
\end{equation}
There will typically not be a single complex coordinate chart that covers all of $\Sigma$; what we really mean is a system of charts with holomorphic transition functions on their intersections.
We may replace the notion of valid complex coordinates by a complex structure $J^{i}_{\ j} \equiv \sqrt{h} \epsilon^{ik} h_{kj}$; in any chart $J^{z}_{\ z} = i, J^{ \bar{z}}_{\ \bar{z}} = -i$.
Denoting $K_{zz}, K_{ \bar{z} \bar{z}}$ by $K, \bar{K}$ respectively, the constraints reduce to
\begin{align}
  4 \partial \bar{\partial} \Omega + 4 K \bar{K} &=  e^{2 \Omega} 
	\label{eqn:ham-max-gauge}
	\\
  \bar{\partial} K = \partial \bar{K} &= 0.
  \label{eqn:diff-const-max-gauge}
\end{align}
Given a complex structure and a choice of $K (z), \bar{K} ( \bar{z})$, we can use the first equation to solve for the Weyl factor $\Omega (z, \bar{z})$.
So a point in the phase space is given by (a) a topological manifold $\Sigma$, (b) a complex structure and (c) a holomorphic and an antiholomorphic quadratic differential $K, \bar{K}$.\footnote{
	A holomorphic quadratic differential is defined as something that is $K(z) \dd{z}^{2}$ in any local complex coordinate chart.
}
This description of the phase space can be further decomposed into a bulk part and a boundary graviton part.

We first deal with the case without ETW branes.
Consider a homotopy class of closed curves $\gamma$ in $\Sigma$.
This homotopy class contains a curve that is geodesic in $\Sigma$ (it does not need to be a spacetime geodesic).
Here as well as below, we will denote both a geodesic as well as its homotopy class by the same symbol.
We can choose coordinates in a neighbourhood of the geodesic $\gamma$ so that the metric takes the form
\begin{equation}
  \dd{s}^{2} = \dd{\rho}^{2} + r(\rho)^{2} \dd{\phi}^{2}, \qquad \phi \sim \phi + 2\pi,
  \label{eqn:gamma-nbhd-coords}
\end{equation}
where the geodesic is at $\rho = 0$.
We denote $r(0)$ by $r_{+,\gamma}$, so that the length of this geodesic is
\begin{equation}
  \ell_{\gamma} = 2\pi r_{+,\gamma}.
  \label{eqn:l-gamma}
\end{equation}
Each geodesic also has a twist $\vartheta_{\gamma}$.
We can cut the surface at $\gamma$ into $\Sigma_{l,r}$; denote the $\phi$ coordinate on $\partial_{\gamma} \Sigma_{l,r}$ as $\phi_{l,r}$.
Now paste $\Sigma_{l}$ to $\Sigma_{r}$ by the identification $\phi_{l} \sim \phi_{r} + \vartheta_{\gamma}$.
This results in a new hyperbolic surface that is not diffeomorphic to $\Sigma$; the parameter $\vartheta_{\gamma}$ is known as the twist.

The conjugate to the length is an appropriately integrated contribution to $K^{\rho\rho}$ in these coordinates.
The conjugate to the twist is the angular momentum of the geodesic, which is encoded in $K^{\rho\phi}$ as \cite{Soni:2024aop}\footnote{
	We are assuming that the normal is future-pointing; for a past-pointing normal the angular momentum and angular velocity get a sign.
}
\begin{equation}
	J_{\gamma} = \frac{r_{+,\gamma} K_{\rho\phi}}{4 G_{N}} = \frac{\ell_{\gamma} \Omega_{\gamma}}{2 G_{N}}.
  \label{eqn:ang-vel-relns}
\end{equation}
$\Omega_{\gamma}$ is the angular velocity.
We also define
\begin{equation}
  r_{-,\gamma} = \frac{2 G_{N} J_{\gamma}}{r_{+,\gamma}}.
  \label{eqn:rm-gamma}
\end{equation}
If the spacetime were a BTZ black hole and $\gamma$ was a spacetime geodesic, $r_{+,\gamma}$ and $\abs{r_{-,\gamma}}$ would be the outer and inner horizon radii respectively.
Here, we allow $r_{-,\gamma}$ to also be negative to take into account the direction of rotation.

Now, consider a slice $\Sigma$ such that $\partial \Sigma$ consists of $n$ disconnected circles $B_{1\dots n}$ at asymptotic infinity.
Choose a maximal set of non-intersecting homotopy classes $\left\{ \gamma \right\}$ of closed curves.
By non-intersecting, we mean that the homotopy classes admit representatives that don't intersect; in particular, the corresponding geodesics will not intersect.
If $\Sigma$ is a connected surface of genus $g$, there are $3g - 3 + 2n$ such classes assuming that $\chi_{\Sigma} = 3 - 3 g - n < 0$; we will assume this for now and deal with the remaining cases later.
$n$ of the classes are `external,' in that they contain the $n$ asymptotic boundaries, and the rest are `internal.'
If $\Sigma $ has multiple connected components, then we can apply the analysis below to each connected component separately.
The bulk part of the phase space will be parametrised by the lengths and twists of each geodesic along with their conjugates.

These quantities are neatly repackaged by the Mess map \cite{mess2007lorentz}.
We begin by defining the two auxiliary Mess metrics
\begin{equation}
	h^{\pm}_{ij} = h_{kl}\cdot (\delta^{k}_{i} \pm J^{k}_{\ m} K^{m}_{i}) \cdot (\delta^{l}_{j} \pm J^{l}_{\ n} K^{n}_{j}).
  \label{eqn:mess-map}
\end{equation}
Remarkably, $h^{\pm}_{ij}$ are hyperbolic metrics, i.e. $R[h^{\pm}] = -2$ everywhere iff the constraints \eqref{eqn:ham-max-gauge}, \eqref{eqn:diff-const-max-gauge} are satisfied \cite{mess2007lorentz,Krasnov:2005dm}.
These metrics also connect to the description of 3d GR as $PSL(2, \mathds{R}) \times PSL (2, \mathds{R})$ Chern-Simons theory.
The spatial components of the $PSL(2, \mathds{R})$ gauge fields $\mathds{A}_{i}, \bar{\mathds{A}}_{i}$ are functions of $h^{\pm}$ respectively, up to a gauge transformation on $\Sigma$ \cite{Scarinci:2011np}.

An elementary calculation shows that if $\gamma$ is a geodesic in the metric $h$ and if the constraints are satisfied, then $\gamma$ is also a geodesic in the two metrics $h^{\pm}$ introduced in \eqref{eqn:mess-map}.
In each of these, $\gamma$ has a length and a twist.
The two lengths take the values\footnote{
	There is an unfortunate clash of notation here, but hopefully common sense and the placement of the $\pm$ will be enough to keep the reader oriented.
}
\begin{equation}
  \ell_{\gamma}^{\pm} = 2\pi (r_{+,\gamma} \pm r_{-,\gamma}).
  \label{eqn:lpm-vals}
\end{equation}
The symplectic form is
\begin{equation}
  \Omega_{g,n} = \sum_{\gamma = 1}^{3g - 3 + 2n} \sum_{s = \pm} \delta \ell_{\gamma}^{s} \wedge \delta \vartheta_{\gamma}^{s} + \sum_{i=1}^{n} \Omega_{\mathrm{bdgrav},i} + \bar{\Omega}_{\mathrm{bdgrav},i}.
  \label{eqn:symplectic-form}
\end{equation}
$\Omega_{\mathrm{bdgrav}} + \bar{\Omega}_{\mathrm{bdgrav}}$ is the symplectic form of the boundary gravitons, which we will not need explicitly.
Notice that the bulk part of the phase space has dimension $6 g - 6 + 4n$.
This is the dimension of moduli space of hyperbolic spaces with boundaries, where the boundaries have marked points.\footnote{
	Without the marked points, the dimension is $6 g - 6 + 3n$; this cannot be a phase space for obvious reasons.
}
These marked points are important to keep track of the twists of the external geodesics, as we will elaborate upon soon.

Notice that the description of the phase space in terms of lengths and twists in these auxiliary hyperbolic metrics is invariant under small diffeomorphisms of $\Sigma$, since the notion of a geodesic in a homotopy class is a diffeomorphism-invariant one.
The choice of $\left\{ \gamma \right\}$ gauge-fixes the large diffeomorphisms of $\Sigma$ (known as the mapping class group), which exchange internal geodesics with each other.
Thus, we have found a true parametrisation of the phase space, where we have gauge-fixed the Hamiltonian constraint and the mapping class group (MCG) of $\Sigma$ and the rest of the parameters label small diffeomorphism gauge orbits.

We also have to deal with the case that $\chi_{\Sigma} \ge 0$.
In the case of no boundaries $n = 0$, we can have $g = 0,1$; there are no classical solutions of either sort.
In the case of one boundary and $g = 0$, there is the global AdS solution; the symplectic form here is
\begin{equation}
  \Omega_{0,1} = \Omega'_{\mathrm{bdgrav}} + \bar{\Omega}'_{\mathrm{bdgrav}}.
  \label{eqn:symplectic-form-gads}
\end{equation}
The space of boundary gravitons here is slightly different, because of the $SL(2, \mathds{R}) \times SL(2, \mathds{R})$ invariance of global AdS.
In the case of two boundaries and $g=0$, the solution is the two-sided wormhole and there is only one geodesic, giving
\begin{equation}
  \Omega_{0,2} = \sum_{s = \pm} \delta \ell^{s} \wedge \delta \tau^{s} + \sum_{i=1}^{2} \Omega_{\mathrm{bdgrav},i} + \bar{\Omega}_{\mathrm{bdgrav},i}.
  \label{eqn:symplectic-form-er}
\end{equation}

If there are ETW branes, then there are additionally geodesics stretching between ETW branes.
These geodesics have the topology of the interval $I$.
Since the twist deformation of the surface depended on the $U(1)$ symmetry of the closed geodesics, we cannot twist along these open geodesics and so they are labelled only by their lengths.
In terms of our auxiliary metrics $\ell^{+} = \ell^{-}$ is just the length, and $\tau^{+} = \tau^{-}$ depends on $K^{\rho\rho}$ in a neighbourhood of this geodesic.
For any component of $\partial\Sigma$ which has topology $I$, the boundary graviton symplectic form is only $\Omega_{\mathrm{bdgrav}}$.
Putting it all together, we have the total symplectic form for a fixed $\Sigma$
\begin{equation}
  \Omega = \sum_{\gamma \text{ closed}} \sum_{s = \pm} \delta \ell_{\gamma}^{s} \wedge \delta \vartheta_{\gamma}^{s} + \sum_{\gamma \text{ open}} \delta \ell_{\gamma}^{+} \wedge \delta \tau^{+} + \Omega_{\partial \Sigma},
  \label{eqn:symplectic-form-total}
\end{equation}
where all boundary graviton contributions have been repackaged into $\Omega_{\partial \Sigma}$.
Below, we will quantise using a basis of fixed $\ell^{\pm}$ (though in a slightly different gauge).

Physically, we are interested in the space of solutions with $n_{c}$ closed boundaries and $n_{o}$ open boundaries.
This phase space splits into different connected components, labelled by the topology of $\Sigma$,
\begin{equation}
  \mathcal{P}_{n_{\mathrm{c}}, n_{\mathrm{o}}} = \coprod_{\text{Different topologies for } \Sigma} \mathcal{P}_{n_{\mathrm{c}}, n_{\mathrm{o}}, \Sigma}.
  \label{eqn:phase-space-disjoint-union}
\end{equation}
For example, with two closed boundaries, one component is given by two copies of global AdS, another by a wormhole, a third by a copy of global AdS and a higher-genus one-boundary black hole, etc.
With open boundaries, there can be topologically distinct ways for the ETW branes to connect the various open boundaries.
Finally, we can also have an arbitrary number of closed universes (with no boundaries).
The naive quantisation of a phase space which is a disjoint union results in a Hilbert space that is the direct sum over sectors, with each sector corresponding to a connected component of phase space.
However, AdS/CFT requires that these different components \emph{not} give rise to orthogonal sectors in the Hilbert space \cite{Jafferis:2017tiu}; we will return to this point in section \ref{sec:gauge}.

\paragraph{The Two-Boundary Wormhole}

Let us see these considerations explicitly in the case of a two-boundary wormhole.
The spacetime is well-known to be given by the metric
\begin{equation}
  \dd{s}_{3}^{2} = - \frac{(r^{2} - r_{+}^{2}) (r^{2} - r_{-}^{2})}{r^{2}} \dd{t}^{2} + \frac{r^{2} \dd{r}^{2}}{(r^{2} - r_{+}^{2}) (r^{2} - r_{-}^{2})} + r^{2} \left( \dd{\phi} + \frac{r_{+} r_{-}}{r^{2}} \dd{t} \right)^{2}, \qquad s = \pm.
  \label{eqn:rot-btz-metric}
\end{equation}
Remember that in our conventions $r_{-}$ can have either sign.
At fixed $t$ the hypersurface has metric and extrinsic curvature
\begin{equation}
  h =
	\begin{pmatrix}
  \frac{r^{2}}{(r^{2} - r_{+}^{2}) (r^{2} - r_{-}^{2})} & 0 \\
  0 & r^{2} \\
  \end{pmatrix}
	, \qquad 
	K_{ij} = \frac{r_{+} r_{-}}{ \sqrt{(r^{2} - r_{+}^{2}) (r^{2} - r_{-}^{2})}}
	\begin{pmatrix}
	0 & 1 \\
	1 & 0 \\
	\end{pmatrix}
	.
  \label{eqn:rot-btz-hyper}
\end{equation}
This gives for the Mess metrics $h^{\pm}$,
\begin{align}
	\dd{s}_{\pm}^{2} &=  \frac{(r^{2} \mp r_{+} r_{-})^{2}}{r^{2} (r^{2} - r_{+}^{2}) (r^{2} - r_{-}^{2})} \dd{r}^{2} + \frac{(r^{2} \pm r_{+} r_{-})^{2}}{r^{2}} \dd{\phi}^{2} \nonumber\\
	&= \dd{\rho}^{2} + (r_{+} \pm r_{-})^{2} \cosh^{2} \rho \dd{\phi}^{2}, \qquad r \pm \frac{r_{+} r_{-}}{r} = (r_{+} \pm r_{-}) \cosh \rho.
  \label{eqn:rot-btz-mess}
\end{align}
We see that these are formally identical to the $t = 0$ slices of two non-rotating BTZ black holes with radii $r_{+} \pm r_{-}$.
In the final form, $\rho \in \mathds{R}$ covers both sides of the horizon.
The geodesic length and angular velocity of the original rotating black hole are
\begin{equation}
  \ell = 2 \pi r_{+} = \frac{\ell_{+} + \ell_{-}}{2}, \qquad \Omega = \frac{r_{-}}{r_{+}} = \frac{\ell_{+} - \ell_{-}}{\ell_{+} + \ell_{-}}.
  \label{eqn:rprm-ls}
\end{equation}

This example allows us also to understand the role of the marked points mentioned above.
The angular momentum is encoded in the $K_{r\phi}$ component of the extrinsic curvature, which is conjugate to $g_{r \phi}$.
Imagine turning on an infinitesimal $g_{r\phi} (r)$. 
This changes the geodesics on $\Sigma$.
An orthogonal geodesic shot from the left boundary at $\phi = 0$ now reaches the right boundary at $\phi \neq 0$.
However, we can also absorb $g_{r\phi}$ into a redefinition of $\phi$, by using $\dd{\phi'} = \dd{\phi} + g_{r \phi} \dd{r}$; this is a large diffeomorphism and changes the solution, since this is the symplectic flow generated by the angular velocity which is a physical parameter.
A simple way to see how much $g_{r\phi}$ has been turned on is to mark $\phi = 0$ on both boundaries, and compare $\phi = 0$ on the right boundary with the endpoint of an orthogonal geodesic shot out from $\phi = 0$ on the left boundary.
So a rotation of the marked points is a variation conjugate to changing the angular velocity of the corresponding horizon.

\paragraph{A Different Gauge}
While the description of the phase space in the maximal slicing gauge is extremely clean, it will turn out that a modification is more suited to quantisation and our holographic map.
We call this the fixed-area gauge, since it is closely related to how fixed-area states are defined \cite{Dong:2018seb,Akers:2018fow,Dong:2019piw,Marolf:2020vsi,Dong:2023xxe,Dong:2022ilf}.

For $\Sigma \setminus \left\{ \gamma \right\}$, we pick the maximal slicing gauge \eqref{eqn:max-slice-gauge}.
For the geodesics we fix (using the coordinates \eqref{eqn:gamma-nbhd-coords} near each geodesic)
\begin{align}
	\text{Open geodesics}: \quad \text{length } \ell_{\gamma} &\equiv \pi \mathsf{r}_{\gamma}, \nonumber\\
	\text{Closed geodesics}: \quad \text{length } \ell_{\gamma} &\equiv 2\pi \mathsf{r}_{+,\gamma} \quad \text{angular velocity } K_{\rho \phi} \equiv  \mathsf{r}_{-,\gamma} \nonumber\\
	\text{Both}: \quad K_{\phi\phi} &= 0.
  \label{eqn:geodesic-bd-conds-cl}
\end{align}
These conditions ensure that the $\gamma$s are spacetime geodesics, since they are geodesics along $\Sigma$ and also along the normal direction (as ensured by $K_{\phi\phi} = 0$).
In this gauge, $K_{\rho\rho}$ is $\delta$-function supported on $\gamma$, with the coefficient of the $\delta$-function being a relative boost angle between the two sides of the geodesic.
The length is conjugate to the relative boost angle.
We continue to define the lengths $\ell_{\gamma}^{\pm}$ as before.

\begin{figure}[h!]
  \centering
  \includegraphics[width=.3\textwidth]{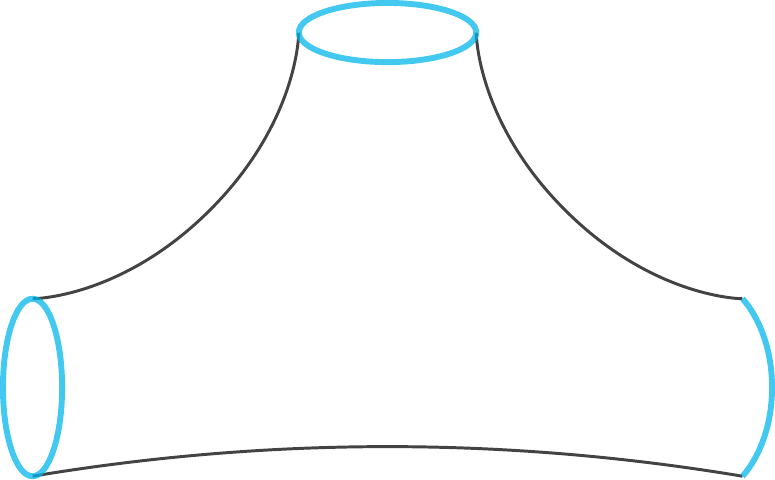}
	\quad
	\includegraphics[width=.3\textwidth]{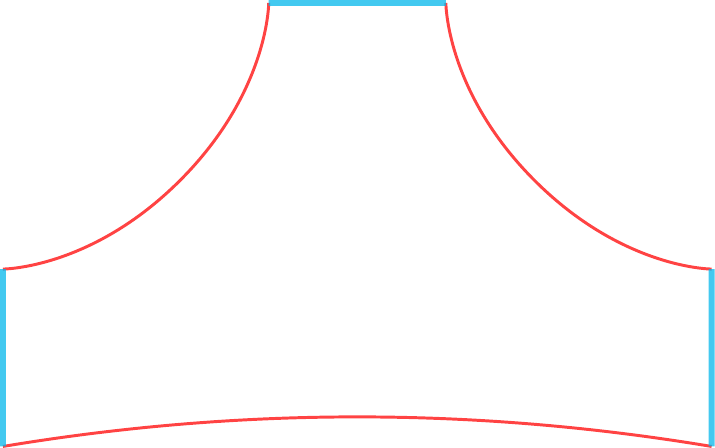}
	\quad
	\includegraphics[width=.3\textwidth]{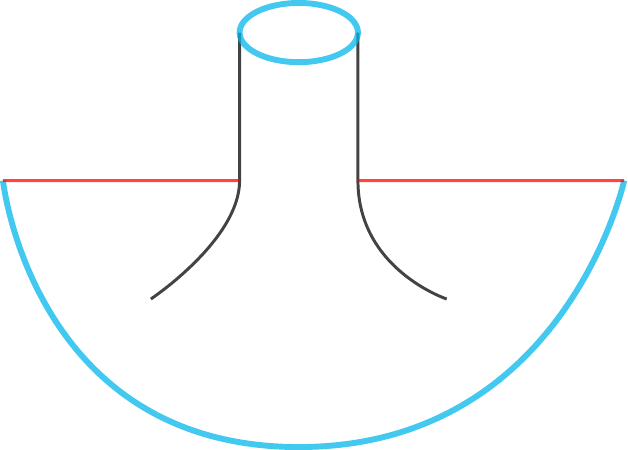}\\
	\includegraphics[width=0.85\textwidth]{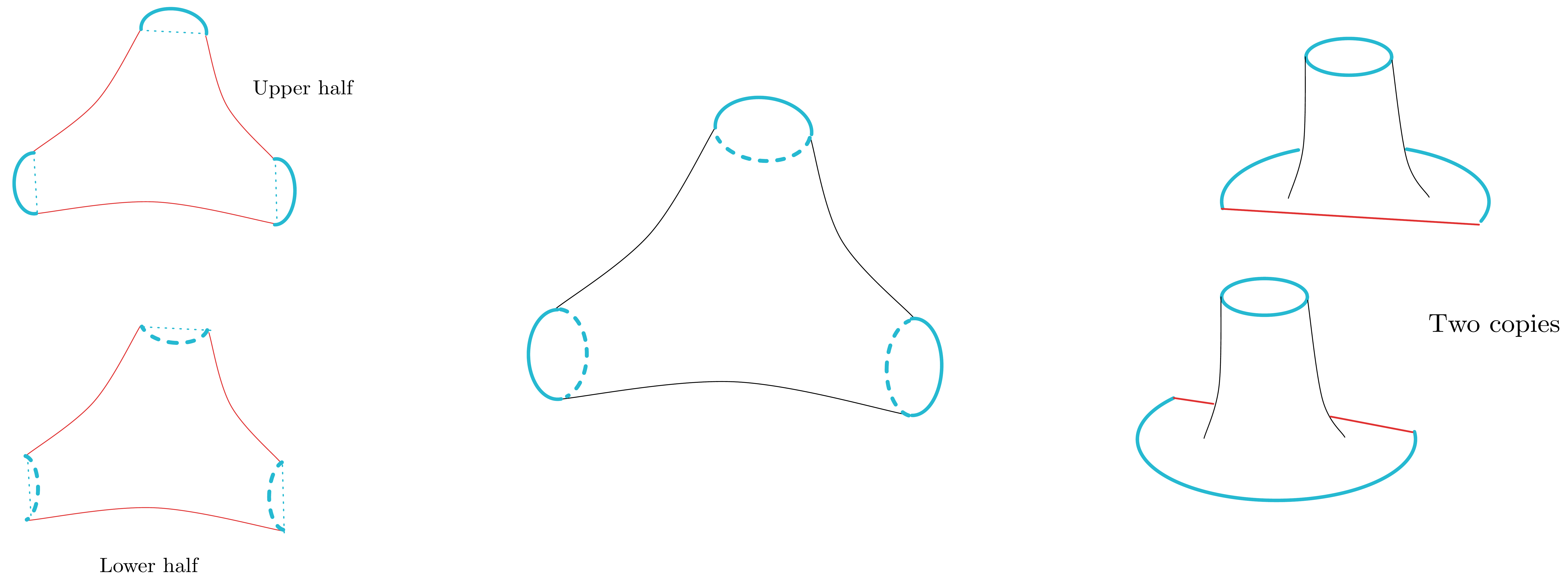}
  \caption{The three elementary pieces out of which any $\Sigma$ is made. The first is a pair-of-pants bounded by three closed geodesics. The second is a half pair-of-pants bounded by three open geodesics and ETW branes (red). The third is a region bounded by a closed geodesic, an open geodesic and an ETW brane.\\
	Bottom line: We can double the second and third piece by taking another copy and gluing it along the ETW brane. In both cases, we get a pair of pants.}
  \label{fig:S-legos}
\end{figure}

The geodesics split $\Sigma$ into three types of elementary pieces, shown in figure \ref{fig:S-legos}.
Each piece is (possibly after using the doubling trick explained in figure \ref{fig:S-legos}) a pair of pants with fixed boundary geodesic lengths.
There is a unique hyperbolic metric on a pair of pants given the boundary lengths, which in turn gives a unique Weyl class of metrics.
We fix the metric on these pieces to lie in the corresponding Weyl class; equivalently, we fix the unique complex structure compatible with the lengths.
In each piece, the diffeomorphism constraints impose holomorphicity of $K$ and antiholomorphicity of $\bar{K}$.
In fact, the boundary conditions $K_{\phi\phi} = K_{\rho\rho} = 0$ and $K_{\rho\phi}$ given fixes $K,\bar{K}$ uniquely.\footnote{
	Uniqueness follows from the following argument.
	Suppose there are two such differentials $K_{1,2}$.
	Pass to the Schottky double of the pair of pants, which is a genus two surface with a reflection symmetry whose fixed-point set is the boundary of the original pair of pants.
	We take $q = K_{1} - K_{2}$, which is also a holomorphic quadratic differential that vanishes on the boundary.
	$q$ can be extended holomorphically to the entire genus two surface.
	Then, the identity theorem for holomorphic sections of a line bundle says that $q = 0$ everywhere.

	This proof was provided to us by Deepseek v4.
}

\begin{figure}[h!]
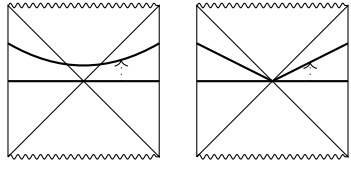

  \centering
  \input{figs/tfd-t-evol.tex} \quad \input{figs/tfd-t-evol-2.tex}
  \caption{The difference in time-evolution between the maximal-slice gauge and the fixed-area gauge, for a two-boundary wormhole. Left: time-evolution moves the entire slice upwards in the Penrose diagram. The length of the slice-geodesic reduces and its extrinsic curvature increases. Right: The slice develops a kink at the bifurcate horizon (whose magnitude is the relative boost angle), so that the length stays the same.}
  \label{fig:t-evol}
\end{figure}

In the maximal slicing gauge, time evolution changes both the lengths and the extrinsic curvatures in a non-trivial way that is tedious to solve.
For example, in the case of the two-sided wormhole, the computation has been done in \cite{Stanford:2014jda}.
Time evolution away from the time-reflection-symmetric slice reduces the length of $\gamma$ and increases $\abs{K^{\rho}_{\rho}} = \abs{K^{\phi}_{\phi}}$.
In this new gauge, however, time evolution is just a change of the boost angle at the external geodesics.
This is illustrated in figure \ref{fig:t-evol}.

\subsection{The Maximally Fixed-Area Basis} \label{ssec:H-bulk}

We quantise this phase space by defining a set of basis elements labelled by a maximal set of Poisson-commuting coordinates on phase space and specifying the inner product between these, as usual.
The coordinates we will pick are the $\ell_{\gamma}^{\pm}$ in \eqref{eqn:symplectic-form-total} for the bulk degrees of freedom, though in the fixed-area gauge.
We will treat the boundary gravitons somewhat cavalierly.
To get a Hilbert space, we will then have to define inner products between these basis elements; we will use the gravitational path integral (GPI) to define this inner product.

\paragraph{The WdW Formalism}

To make this precise, we use the Wheeler-DeWitt (WdW) formalism \cite{DeWitt:1967yk,DeWitt:1967ub,DeWitt:1967uc,Kuchar:1991qf,Isham:1993ha,Araujo-Regado:2022gvw,Held:2025mai}, where basis labels define boundary conditions for semiclassical path integrals.
Canonical quantisation in three dimensions has been studied in depth by Carlip \cite{Carlip:1998uc}, and discussions close to ours in the two-boundary case can be found in \cite{Chua:2023ios,Kaushal:2025rsw}.
In the WdW formalism, the inner product is only defined as a sum over asymptotic expansions in $G_{N}$ and thus the Hilbert space $\mathcal{H}_{\mathrm{bulk}}$ so defined is not a true Hilbert space.
In 3d gravity, the loop corrections can be calculated using Virasoro TQFT (VTQFT) \cite{Collier:2023fwi}, but the convergence of the sum over topologies has not been shown.
We work at the level of classical saddles rather than VTQFT because the subleading corrections of VTQFT are of same or lower order than matter corrections, which we are ignoring; in most other regards, our quantisation agrees with VTQFT
The construction of the holographic map does not require $\mathcal{H}_{\mathrm{bulk}}$ to be non-perturbatively well-defined and so we will soldier on ignoring this subtlety.

Consider a Euclidean path integral (EPI) state $\ket{\Psi} \in \mathcal{H}_{\mathrm{bd}}$.\footnote{
	We use $\mathcal{H}_{\mathrm{CFT}}$ to mean the Hilbert space of a single CFT, defined on either a $S^{1}$ or $I$, and $\mathcal{H}_{\mathrm{bd}}$ as the Hilbert space of many CFTs, one for each connected component of $\partial \Sigma$.
}
By the GKPW dictionary, there is a corresponding boundary condition at the asymptotic boundary for the gravitational path integral (GPI).
Pick another boundary condition on $\Sigma$, which we denote as $h_{\Sigma}$ in this abstract discussion.
The WdW wavefunction is defined as
\begin{equation}
  \Psi [h] = \pip{h}{\Psi} = \int^{*}_{\text{bd conds}} Dg\, e^{- I[g]}.
  \label{eqn:wdw-defn}
\end{equation}
Here, $\int^{*}$ is defined as a sum over loop expansions around saddle-points; we will work at zero loops throughout.
Note that we use rounded kets and bras for bulk states and normal ones for CFT states.
The GKPW dictionary maps an EPI state $\ket{\Psi}$ to a GPI-boundary-condition state $\pket{\Psi}$.
These wavefunctions satisfy the constraints \eqref{eqn:constraints} as differential equations, with $\pi^{ij} \to \delta/\delta h_{ij}$, where products of differential operators are defined as asymptotic expansions in $G_{N}$.
These constraints relate different possible boundary conditions on $\Sigma$, and so we again have to gauge-fix to find a non-redundant basis.

\paragraph{Boundary Conditions for WdW Wavefunctions}

\begin{figure}[h!]
  \centering
  \includegraphics[width=0.8\textwidth]{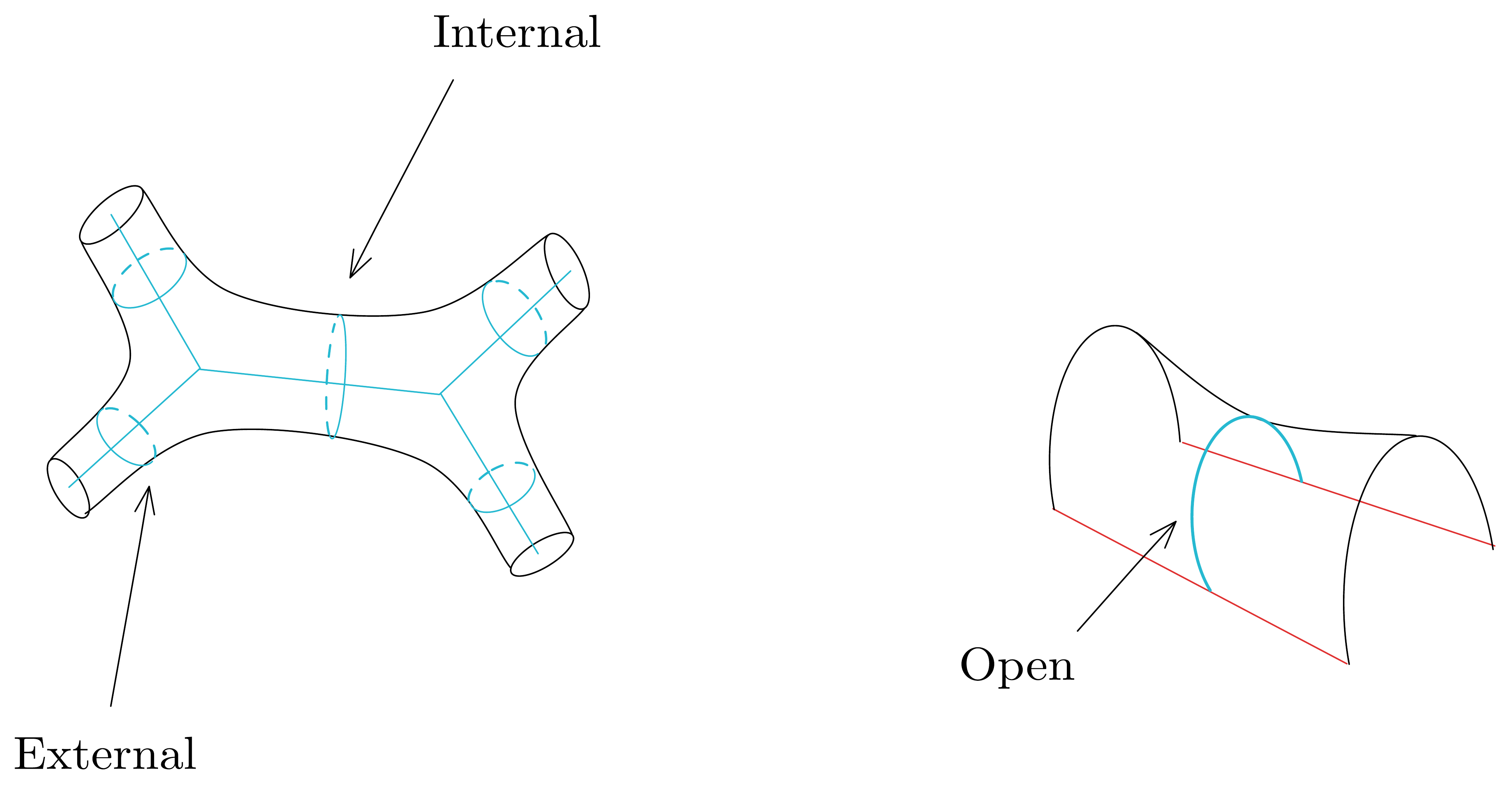}
  \caption{Different types of geodesics}
  \label{fig:geod-types}
\end{figure}

The boundary conditions we choose are as follows.
Pick a maximal non-intersecting set of geodesics $\left\{ \gamma \right\}$.
We will classify these geodesics in two ways.
Firstly, some of these are closed (topology of $S^{1}$) and others are open (those with the topology of $I$ ending on ETW branes).
Secondly, some of these are `external,' by which we mean that they are homotopic to a component of $\partial \Sigma$, and the rest are internal.
Apart from these geodesics, we also pick a `framing' of $\Sigma$, which is a graph that intersects each geodesic orthogonally.
This graph being connected will define the absence of a twist below.
We illustrate these in figure \ref{fig:geod-types}.

We work in the fixed-area gauge introduced around \eqref{eqn:geodesic-bd-conds-cl}.
For $\Sigma \setminus \left\{ \gamma \right\}$, we pick the maximal slicing gauge \eqref{eqn:max-slice-gauge}.
For the geodesics we fix (using the coordinates \eqref{eqn:gamma-nbhd-coords} near each geodesic)
\begin{align}
	\text{Open geodesics}: \quad \text{length } \ell_{\gamma} &\equiv \pi \mathsf{r}_{\gamma}, \nonumber\\
	\text{Closed geodesics}: \quad \text{length } \ell_{\gamma} &\equiv 2\pi \mathsf{r}_{+,\gamma} \quad \text{angular velocity } K_{\rho \phi} \equiv  i \mathsf{r}_{-,\gamma} \nonumber\\
	\text{Both}: \quad K_{\phi\phi} &= 0.
  \label{eqn:geodesic-bd-conds}
\end{align}
An important difference from section \ref{ssec:phase-space} is that we now work in Euclidean conventions where the normal is defined by $n \cdot n = +1$.
With this convention, $K_{\rho \phi}$ becomes imaginary for a real angular velocity.
It is common to analytically continue the angular velocity, but that will lead to further complications down the road for us.
With this choice, $\ell^{\pm}$ become complex and we will refer to them as the complex lengths henceforth.\footnote{
	A different suggestion for complex lengths was given in \cite{Hartman:2025cyj,Hartman:2025ula}.
	There, the boundary conditions on $\Sigma \setminus \gamma$ were taken to be $K_{ij} = 0$.
	We will comment on why we haven't taken this choice in footnote \ref{fn:Kij-2}.
	\label{fn:Kij-1}
}
Further, we take the metric on each elementary piece $\Sigma \setminus \left\{ \gamma \right\}$ to be the one fixed by the lengths of the bounding geodesics.
Finally, we also draw a framing graph on our surface, which is just a graph that intersects each geodesic once.
We do \emph{not} require that the framing graph be connected in the on-shell solution.

We call a choice of non-intersecting geodesics along $\ell^{\pm}$ on each geodesic along with a framing graph a fixed-area network.
We use a simplified graphical notation where, for example,
\begin{align}
  \pbra{\includegraphics[height=3em,valign=c]{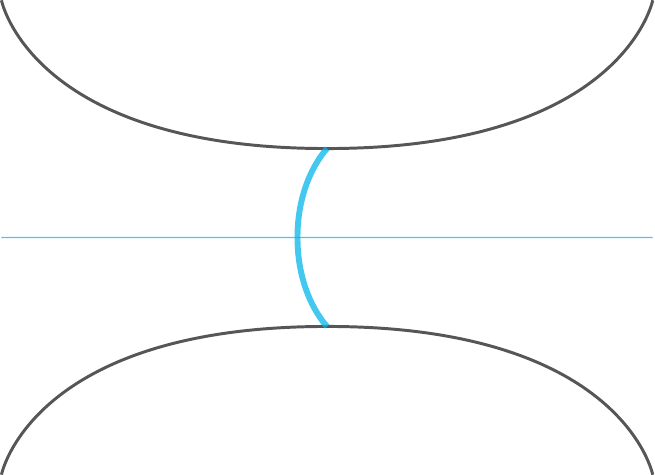}} 
			&= \pbra{ 
\begin{tikzpicture}[baseline=-.2cm]
	\draw[cyan] (0,0) to[out=-90,in=-90] node[pos=.5,below] {\tiny $\ell$} (1,0);
\end{tikzpicture}
 } \nonumber\\
			\pbra{\includegraphics[height=3em,valign=c]{figs/four-bd-wormhole-eg.pdf}}
			&= \pbra{ \input{figs/fa-4-bd-s.tex} } \nonumber\\
			\pbra{\includegraphics[height=3em,valign=c]{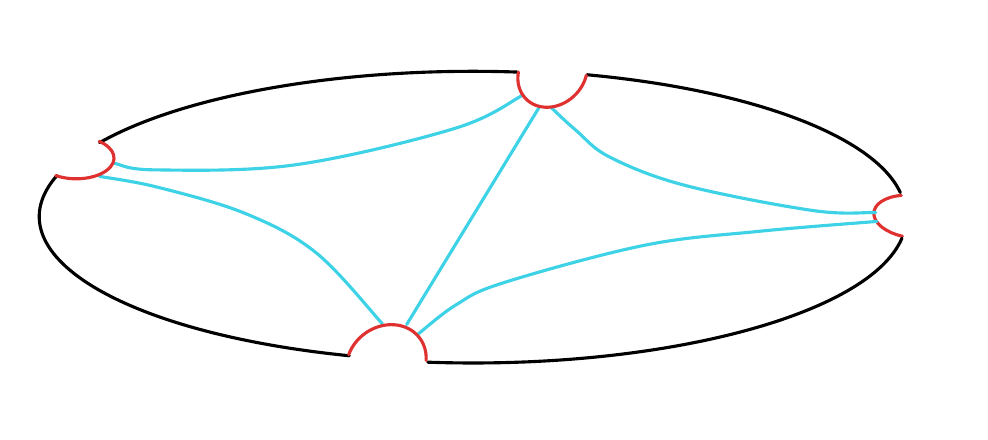}} &= \pbra{
				\begin{tikzpicture}[baseline]
				  \draw[double,cyan] (0,0) -- node[below] {\tiny $\ell_{s}$} ++(  0:1);
				  \draw[double,cyan] (0,0) -- node[pos=1,above] {\tiny $\ell_{1}$} ++(120:.5);
				  \draw[double,cyan] (0,0) -- node[pos=1,above] {\tiny $\ell_{2}$} ++( 60:.5);
				  \draw[double,cyan] (1,0) -- node[pos=1,above] {\tiny $\ell_{3}$} ++(120:.5);
				  \draw[double,cyan] (1,0) -- node[pos=1,above] {\tiny $\ell_{4}$} ++( 60:.5);
				\end{tikzpicture}
			}.
  \label{eqn:fa-nets}
\end{align}

This is not a complete specification of boundary conditions on $\Sigma$, since we have not fixed a basis for the boundary gravitons.
Thus, our fixed-area wavefunctions are actually states in the Hilbert space of boundary gravitons,
\begin{equation}
  \pip{\ell_{i}}{\Psi_{B}} \in \mathcal{H}_{\mathrm{bdgrav}}.
  \label{eqn:fa-wavefn-as-state}
\end{equation}
For this to work, it needs to be the case that $\mathcal{H}_{\mathrm{bdgrav}}$ admits a basis where $K_{i}^{i} = 0$ near the boundary.
It was shown in \cite{Scarinci:2011np} that this gauge allows for the entire phase space, and therefore it must also allow for such a basis.
The WdW wavefunction of $\pket{\Psi_{B}}$ is given by the GPI with boundary conditions $\Psi$ at the asymptotic boundary and the above boundary conditions on $\Sigma$.
Schematically,
\begin{equation}
  \includegraphics[width=0.55\linewidth,valign=c]{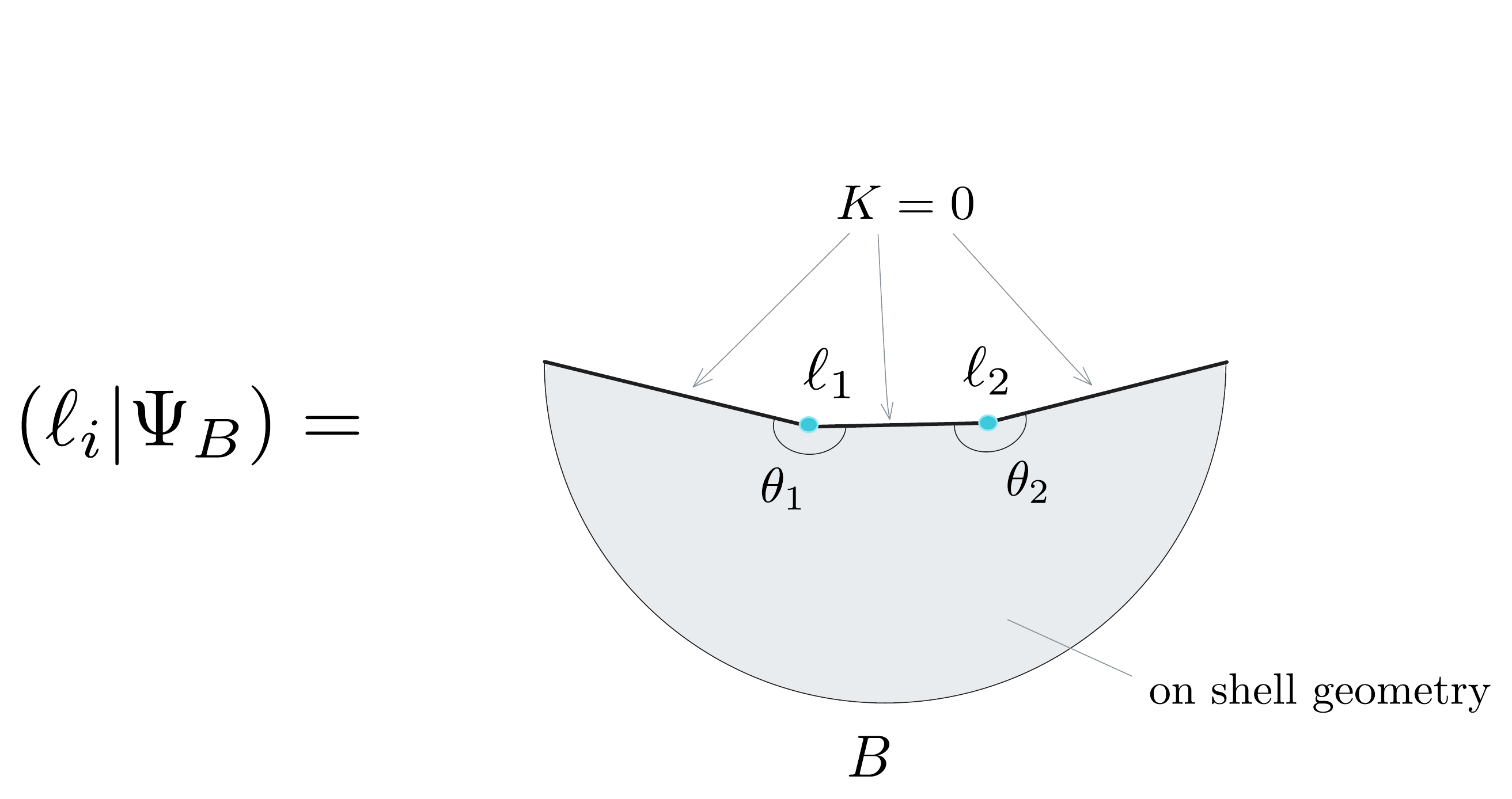} \in \mathcal{H}_{\mathrm{bdgrav}}.
 \label{eqn:wdw-fa}
\end{equation}

\begin{figure}[h!]
  \centering
  \includegraphics[width=.5\textwidth]{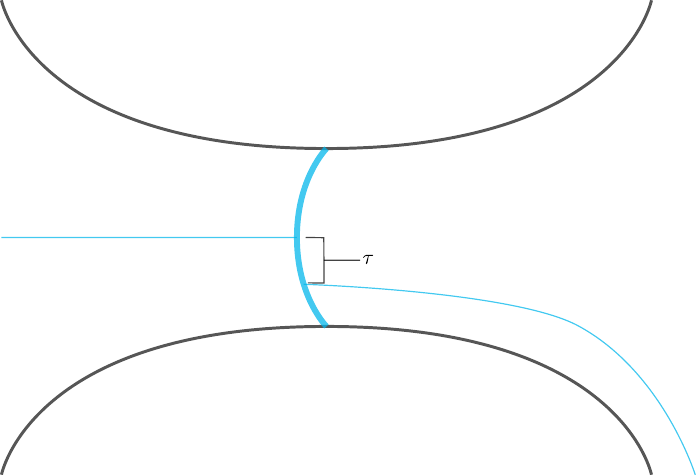}
  \caption{We measure the twist by the discontinuity in the framing graph at a geodesic.}
  \label{fig:twist-eg}
\end{figure}

On $\Sigma$, we have fixed the lengths of the geodesics and their angular velocities.
By the uncertainty principle, there should be two unfixed quantities conjugate to these two; in the semiclassical evaluation, these quantities will be determined by equations of motion.
The conjugate to the length is a $\delta$-function in $K^{\rho\rho} \propto \pi^{\phi\phi}$, which is a relative boost angle between the two sides of a geodesic.
This property is not intrinsic to the surface $\Sigma$.
We have fixed $K^{\rho \phi}$ on a codimension-one submanifold of $\Sigma$, so the conjugate is a $\delta$-function in $h_{\rho \phi}$.
Such a $\delta$-function contribution can be geometrically visualised by cutting $\Sigma$ at the geodesic, rotating one side relative to the other and re-pasting it, as shown in figure \ref{fig:twist-eg}.
This creates a discontinuity in the framing graph.
We will define the twist $\tau$ as the angular distance on the geodesic between the ends of the graph.

To summarise, our basis elements are determined by a maximal set of non-intersecting homotopy classes on $\Sigma$ and the complex lengths of their geodesics.
The choice of gauge is (a) in the $K^{i}_{i} = 0$ condition on the smooth parts and (b) in the choice of set of homotopy classes.
The different basis elements in this gauge are labelled by different complex lengths.
Below, we will find that, since our geodesics are restricted to being spacetime geodesics, the choice of homotopy classes actually fixes the Hamiltonian gauge symmetry.

\paragraph{The Boundary Term}

Now, we need a boundary term for the action on $\Sigma$ that gives a good variational principle.
Defining $\Sigma_{\mathrm{sm}} \equiv \Sigma \setminus \left\{ \gamma \right\}$, we take
\begin{equation}
	8 \pi G_{N} I_{\Sigma} = -  \int_{\Sigma_{\mathrm{sm}}} \sqrt{h} K - \frac{1}{2} \sum_{\gamma} \left[ (\pi - \vartheta_{\gamma}) \ell_{\gamma} + (\pi - \bar{\vartheta}_\gamma) \bar{\ell}_{\gamma} \right]  -  \int_{\partial \Sigma} \left( \frac{\pi}{2} - \theta \right).
  \label{eqn:Sigma-action}
\end{equation}
Here, $\theta$ is the internal angle between $\partial \Sigma$ and $B$ and $\theta_{\gamma} \equiv \vartheta_{\gamma} + \bar{\vartheta_{\gamma}}$ is the internal angle at $\gamma$.
$\tau_{\gamma} = (\vartheta_{\gamma} - \bar{\vartheta}_\gamma)/4 r_{+}$ is the twist at $\gamma$, defined by the angular distance between the endpoints of the framing graph.\footnote{
	This provides a natural physical interpretation for the two independent angles suggested in \cite{Hartman:2025cyj,Hartman:2025ula} for defining conformal Turaev-Viro theory including angular momentum.
}
The imaginary part of the angles are relative boosts, and the twist is real always.
The corner terms at $\partial \Sigma$ will identically vanish on-shell due to our boundary conditions, but they are important for loop corrections (which we will not calculate).
The part of the corner term proportional to $r_{+}$ at the geodesics vanishes if the internal angle is $\pi$; this corresponds to a conical angle of $2\pi$, i.e. no conical singularity, in the norm.

To see that this provides a good variational principle, we first use the well-known fact \cite{Hayward:1993my}, see \cite{Lehner:2016vdi} for a pedagogical discussion, that
\begin{equation}
	\delta \left[ - \int \sqrt{g} (R + 2) - 2 \int_{\Sigma_{\mathrm{sm}}} \sqrt{h} K \right] = \text{eoms} + \int_{\Sigma_{\mathrm{sm}}} \pi^{ij} \delta h_{ij} - 2\pi \sum_{\gamma} \delta \theta_{\gamma} \mathsf{r}_{+,\gamma}.
  \label{eqn:smooth-part-varn}
\end{equation}
Focusing on one geodesic, we use the coordinates \eqref{eqn:gamma-nbhd-coords} and \eqref{eqn:ang-vel-relns} to expand the corner term as
\begin{equation}
	- \frac{1}{2} \left[ (\pi - \vartheta_{\gamma}) \ell_{\gamma} + (\pi - \bar{\vartheta}_\gamma) \bar{\ell}_\gamma \right] = - (\pi - \theta_{\gamma}) 2\pi \mathsf{r}_{+,\gamma} + 2 r_{+} \tau_{\gamma} \int_{\gamma} \dd{\phi} K_{\rho \phi}.
  \label{eqn:corner-term-phys}
\end{equation}
The variation of the first term converts the $\delta \theta_{\gamma} \cdot \mathsf{r}_{+}$ in \eqref{eqn:smooth-part-varn} to $\theta_\gamma \delta \mathsf{r}_{+}$; this is known as the Hayward corner term.
The corner term at $\partial \Sigma$ is also a Hayward corner term.
The variation of the second term gives $\tau \int \delta K_{\rho \phi}$.
To see this, imagine smoothing out the $\delta$-function twist a little bit to
\begin{equation}
  \tau_{\gamma} \int_{\gamma} K_{\rho \phi} \to \int \sqrt{h} h^{\rho \phi} K_{\rho \phi} = 16 \pi G_{N} \int \pi^{\rho \phi} h_{\rho \phi}, \quad h_{\rho \phi} = \tau_{\gamma} f_{\epsilon}(\rho),
  \label{eqn:ang-mom-term-smooth}
\end{equation}
where $f_{\epsilon} (\rho)$ is a smoothing of the $\delta$-function.
The variation of this term converts the $\int \pi^{\rho x} \delta h_{\rho x}$ in \eqref{eqn:smooth-part-varn} to $\int \delta \pi^{\rho x} h_{\rho x} \xrightarrow{\epsilon \to 0} \tau \delta J$ \cite{Krishnan:2016mcj}.
Thus, we see that we get a good variational principle from \eqref{eqn:Sigma-action} using the boundary conditions \eqref{eqn:geodesic-bd-conds}.
Further, note that the term on $\Sigma_{\mathrm{sm}}$ vanishes on-shell.

\paragraph{A Convenient Generalisation}

Naively, $\ell$ takes values in the region of $\mathds{C}$ where $\Re \ell > 0$ and $\Im \ell < \Re \ell$.
$\Im \ell = \Re \ell$ is the extremal limit and our results do not generalise straightforwardly to this limit.
We will also allow $\ell = \bar{\ell} = 2\pi i$ for later convenience.
This corresponds to the case where $\gamma$ belongs to a contractible homotopy class.
To see this, notice that taking $r_{+} \to i$ in the non-rotating BTZ metric $[- (r^{2} - r_{+}^{2})] \dd{t}^{2} + \dd{r}^{2}/(r^{2} - r_{+}^{2}) + r^{2} \dd{\phi}^{2}$ gives $[- (r^{2} + 1)] \dd{t}^{2} + \dd{r}^{2}/(r^{2} + 1) + r^{2} \dd{\phi}^{2}$ --- which is the global AdS metric.
This choice will allow us to treat different numbers of connected components of $\Sigma$ on the same footing and simplify many statements.
We expect that our results will also generalise to the case where $\ell \in (\pi i, 2\pi i)$, which corresponds to heavy particles in the bulk, as in \cite{Chandra:2022bqq}, but we don't consider this for simplicity.

\paragraph{A Non-Perturbative Gauge Condition}

Notice that there may be multiple solutions in principle with these boundary conditions.
Following \cite{Iliesiu:2024cnh}, we also impose a `non-perturbative gauge choice:' for the WdW wavefunction, we only allow solutions where $\Sigma$ is homotopic to $B$ through the bulk solution.
In this we allow some of the lengths to be $2\pi i$.
All non-trivial topology will be taken into the inner product, so that physical boundary quantities will be correctly calculated in our Hamiltonian formalism.

\paragraph{Kets and Bras}
There is an important difference between the boundary conditions for ket and bra basis elements.
Till now, we have dealt exclusively with bras, which are relevant for calculating WdW wavefunctions.
For a ket, since the outward normal points in the opposite direction, the boundary condition at each geodesic should be modified to
\begin{equation}
  K_{\rho\phi} = - i r_{-,\gamma}.
  \label{eqn:ket-ang-mom}
\end{equation}

\subsection{Bulk Inner Products} \label{ssec:inner-products}

Now we have fixed a basis for our Hilbert space, it is time to define an inner product.
The inner product is given by a rigging map, where the rigging map is defined by the gravitational path integral.
This type of inner product automatically deals with issues of residual gauge symmetry of perturbative as well as non-perturbative type, and also allows us to work with different gauges at the same time.

Consider two slices $\Sigma$, $\Sigma'$ such that $\partial \Sigma = \partial \Sigma'$.
We define three inner products on our basis elements.
The first is the kinematical inner product.
Traditionally, the kinematical inner product is defined before gauge-fixing as $\pip{\Sigma,h} {\Sigma',h'} = \delta_{\Sigma,\Sigma'} \delta(h-h')$ where $h,h'$ are metrics on $\Sigma,\Sigma'$ and $\delta_{\Sigma,\Sigma'}$ is one only when $\Sigma,\Sigma'$ have the same topology.
For notational convenience, we also define a renormalised version of the kinematical inner product after our gauge-fixing:
\begin{equation}
	\pip{\Sigma',\left\{ \ell_{\gamma'}' \right\}}{\Sigma, \left\{ \ell_{\gamma} \right\}} = \delta_{\Sigma_{\gamma},\Sigma'_{\gamma'}} \prod_{\gamma} \delta^{(2)} (\ell_{\gamma} - \ell'_{\gamma}).
  \label{eqn:kinematical-ip}
\end{equation}
where $\delta_{\Sigma_{\gamma}, \Sigma'_{\gamma'}}$ is $1$ if and only if $\Sigma$ and $\Sigma'$ have the same topology and the fixed-area geodesics lie in the same homotopy class.
This inner product would be the physical inner product if we chose a specific set of homotopy classes to fix for each topology, and if the gauge had completely been fixed.

In practice, we would like the freedom to fix areas for different sets of homotopy classes for a given topology.
We define the perturbative inner product using the gravitational path integral on manifolds $\mathcal{M}$ such that $\Sigma,\Sigma'$ are homotopic through $\mathcal{M}$,
\begin{equation}
	\pmel{\Sigma, \left\{ \ell_{\gamma'} \right\}}{\eta_{\mathrm{pert}}}{\Sigma, \left\{ \ell_{\gamma} \right\}} = \delta_{\Sigma,\Sigma'} \int_{\substack{\partial \mathcal{M} = \Sigma \cup \Sigma' \\ \Sigma,\Sigma' \text{ homotopic}}}^{*} Dg\, e^{- I_{\mathrm{full}} [g, \Sigma,\Sigma']}, \quad I_{\mathrm{full}} = I_{0} + I_{\Sigma} + I_{\Sigma'}.
  \label{eqn:kinematical-ip-1}
\end{equation}
This prescription was originally proposed in \cite{Teitelboim:1981ua}, and we refer the reader to that work for an explanation of why this choice is appropriate.\footnote{
	Our inner product differs from \cite{Teitelboim:1981ua} in that we don't fix the relative causal order of $\Sigma$ and $\Sigma'$ within $\mathcal{M}$.
	See \cite{Teitelboim:1981ua,DiazDorronsoro:2017hti,Araujo-Regado:2022gvw} for a conceptual explanation of this choice.
	At a more lowbrow level, our holographic map won't work if we fix the causal order as advocated in \cite{Teitelboim:1981ua}.
}
$\eta_{\mathrm{pert}}$ is known as the rigging map; it is similar to the classical limit of the group-averaging inner product for the Hamiltonian and diffeomorphism constraints \cite{Marolf:1995cn,Marolf:2000iq}.

However, it is well-known that the perturbative inner product does not reproduce the CFT inner product, see e.g. \cite{Jafferis:2017tiu,Marolf:2020xie,Penington:2019kki}.
As we will see, states with different topologies are not orthogonal in the CFT.
So, we define our physical (or non-perturbative) inner product using the full gravitational path integral as
\begin{equation}
\pmel{\Sigma', \left\{ \ell_{\gamma}' \right\}}{\eta} {\Sigma, \left\{ \ell_{\gamma} \right\}} = \int^{*}_{ \partial \mathcal{M} = \Sigma \cup \Sigma'} Dg e^{- I_{\mathrm{full}} [g,\Sigma,\Sigma']}, \quad I_{\mathrm{full}} \equiv I_{0} + I_{\Sigma} + I_{\Sigma'}.
  \label{eqn:physical-ip}
\end{equation}
Here, $\eta$ is known as the rigging map.
The non-trivial topologies that contribute to this inner product but not to the perturbative ones are often known as `non-perturbative gauge transformations;' we will use this phrase below, but we are only using the phrase for convenience and not claiming that these are the same sort of thing as the usual gauge transformations.
We denote the Hilbert space defined by \eqref{eqn:kinematical-ip-1} as $\mathcal{H}_{\mathrm{pert}}$ and that by \eqref{eqn:physical-ip} as $\mathcal{H}_{\mathrm{bulk}}$.

In the case of the four-boundary wormhole, the inner product is schematically
\begin{equation}
	\pmel{
		\begin{tikzpicture}[baseline,scale=.75]
			\draw[cyan] (0,0) -- node[below] {\tiny $\ell'_{s}$} ++(  0:1);
			\draw[cyan] (0,0) -- node[pos=1,above] {\tiny $\ell'_{1}$} ++(120:.5);
			\draw[cyan] (0,0) -- node[pos=1,above] {\tiny $\ell'_{2}$} ++( 60:.5);
			\draw[cyan] (1,0) -- node[pos=1,above] {\tiny $\ell'_{3}$} ++(120:.5);
			\draw[cyan] (1,0) -- node[pos=1,above] {\tiny $\ell'_{4}$} ++( 60:.5);
		\end{tikzpicture}
	}{\eta}{
		\begin{tikzpicture}[baseline,scale=.75]
			\draw[cyan] (0,0) -- node[below] {\tiny $\ell_{s}$} ++(  0:1);
			\draw[cyan] (0,0) -- node[pos=1,above] {\tiny $\ell_{1}$} ++(120:.5);
			\draw[cyan] (0,0) -- node[pos=1,above] {\tiny $\ell_{2}$} ++( 60:.5);
			\draw[cyan] (1,0) -- node[pos=1,above] {\tiny $\ell_{3}$} ++(120:.5);
			\draw[cyan] (1,0) -- node[pos=1,above] {\tiny $\ell_{4}$} ++( 60:.5);
		\end{tikzpicture}
	} =
	\int^{*} Dg \exp{ - I_{\mathrm{full}} \left[ 
			\includegraphics[width=.2\textwidth,valign=c]{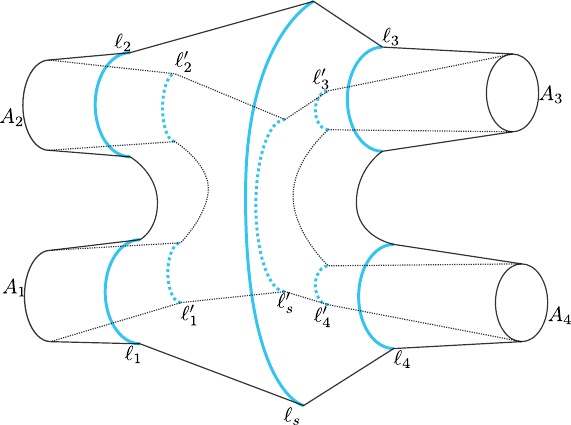}
	\right]}
\end{equation}
Here, the black circles are the asymptotic boundary and the blue ones are the fixed-area surfaces, and we have not drawn the framing graph in the RHS.
We have placed $\Sigma'$ inside $\Sigma$ in the figure to indicate that the boundary conditions are given by $\Sigma,\Sigma'$; the actual bulk solution may have a topology distinct from the one indicated in the figure.

\begin{figure}[h!]
  \centering
	\begin{subfigure}[c]{.3\textwidth}
		\centering
		\begin{tikzpicture}
			\node[cyan,fill,circle,inner sep=.5mm,label=below:{\small $\gamma$} ] (g)	at (0,-.5) {};
			\node[cyan,fill,circle,inner sep=.5mm,label=above:{\small $\gamma'$}] (gp) at (0, .5) {};
			\draw (0,-.5) -- node[below] {\small $\Sigma$} node[pos=1,right] {\small $B$} (1,0) -- node[above] {\small $\Sigma'$} (0,.5);
			\draw[dotted,->] (0,-.5) .. controls (.8,-.1) and (.8,.1) .. (0,.5);
		\end{tikzpicture}
		\caption{Two geodesics homotopic to the same boundary are homotopic to each other through the bulk.}
		\label{fig:hom-arg}
	\end{subfigure}
	\quad
	\begin{subfigure}[c]{.5\textwidth}
		\centering
		\begin{tikzpicture}[baseline]
			\node[fill,cyan,inner sep=.5mm,circle,label=above right:{\tiny $\gamma'$}] (gp) at (0, .5) {};
			\node[fill,cyan,inner sep=.5mm,circle,label=below right:{\tiny $\gamma$ }] (g ) at (0,-.5) {};

			\draw (gp) -- node[above] {\tiny $\Sigma'$} ++(1,0);
			\draw (g ) -- node[below] {\tiny $\Sigma $} ++(1,0);
			\draw (gp) -- ++(120:1);
			\draw (g ) -- ++(260:1);
			\draw[dashed] (gp) -- ++(-1,0);
			\draw[dashed] (g ) -- ++(-1,0);

			\draw (gp) ++( .2,0) arc(360:120:.2) node[pos=.85,left] {\tiny $\pi + \theta'$};
			\draw (g ) ++( .2,0) arc(	0:260:.2) node[pos=.85,left] {\tiny $\pi + \theta $};
		\end{tikzpicture}
		$\quad \xrightarrow{\text{identify}}	\quad $
		\begin{tikzpicture}[baseline]
			\node[fill,cyan,inner sep=.5mm,circle,label=below right:{\tiny $\gamma$ }] (g) at (0,0) {};

			\draw (g) -- ++(1,0);
			\draw (g) -- node[right] {\tiny $\Sigma'$} (120:1);
			\draw (g) -- node[right] {\tiny $\Sigma$} (260:1);

			\draw (g) ++ (120:.2) arc(120:260:.2) node[left] {\tiny $\theta + \theta'$};
		\end{tikzpicture}
		\caption{When we identify $\gamma,\gamma'$ along with a section of $\Sigma,\Sigma'$ ending on them, the internal angles add up to $2\pi + \theta + \theta'$, where $\theta + \theta'$ is the internal angle in the geometry where we excise the identified parts of $\Sigma,\Sigma'$.}
		\label{fig:I-half-pic}
	\end{subfigure}
	\caption{We can remove the external trumpets while calculating inner products, and consequently simplify the action as well.}
\end{figure}
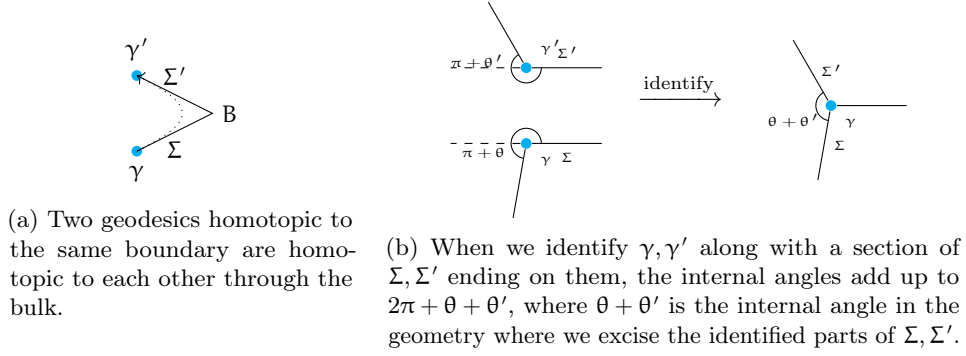

Suppose that $B_{1}$ is homotopic to a non-zero length geodesic $\gamma_{1}$ in $\Sigma$ and also to $\gamma_{1}'$ in $\Sigma'$.
Then, $\gamma_{1}$ and $\gamma_{1}'$ are homotopic to each other in the boundary condition for the inner product.
The simplest choice for a homotopy surface is $\Sigma \cup \Sigma'$ itself, but as long as there are a finite number of handles there will also be a homotopy surface through the bulk, as shown in figure \ref{fig:hom-arg}.
In $\text{AdS}_{3}$ as well as $\mathds{H}^{3}$ (i.e. Euclidean $\text{AdS}_{3}$), there cannot be two distinct geodesics in the same homotopy class.
Therefore, in the solution, $\gamma_{1}$ must be identified with $\gamma_{1}'$.
Thus, the inner product simplifies to
\begin{equation}
	\pmel{
		\input{figs/fa-net-Sigp.tex}
	}{\eta}{
		\input{figs/fa-net-Sig.tex}
	} =
	\prod_{i=1}^{4} \delta(\ell_{i} - \ell'_{i}) \int^{*} \exp{ - I_{\mathrm{half}} \left[ \includegraphics[width=.2\textwidth,valign=c]{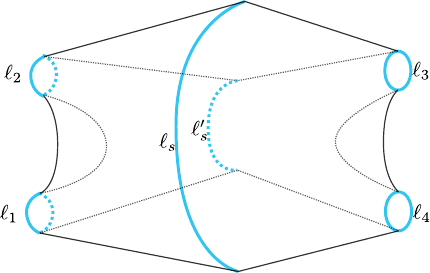} \right]}
	\label{eqn:ip-core}
\end{equation}
Here, we have removed the `trumpet' regions connecting the external geodesics to the corresponding asymptotic boundaries.
Now the boundary conditions for the gravitational path integral are on codimension-one slices and codimension-two geodesics.
Here $I_{\mathrm{half}}$ is
\begin{equation}
  I_{\mathrm{half}} = I_{\mathrm{EH}} + I_{\mathrm{GHY}} + I_{\mathrm{ETW}} - \sum_{\gamma \text{ internal}} (\pi - \vartheta_{\gamma}) \frac{\ell_{\gamma}}{8 \pi G_{N}} + \sum_{\gamma \text{ external}} \vartheta_{\gamma} \frac{\ell_{\gamma}}{8 \pi G_{N}} + c.c.\,,
  \label{eqn:core-action}
\end{equation}
where $\tau_{i}$ is the sum of the internal angle and twist at $\gamma_{i}$ in this `core' geometry and $c.c.$ means that we also add the corresponding $\bar{\tau}_{i} \bar{\ell}_{i}$ terms.\footnote{
	We can think of c.c. as an abbreviation for `chiral conjugate' instead of `complex conjugate.'
}
The reason for this modification of the action is that the internal angle being $\tau_{i}$ here is the same as the sum of the internal angles in the geometry of \eqref{eqn:physical-ip} being $2\pi + \tau_{i}$, as shown in figure \ref{fig:I-half-pic}.
Note that the boundary condition shown in \eqref{eqn:ip-core} does not distinguish between internal and external geodesics, whereas $I_{\mathrm{half}}$ does; for symmetry, we define $I$ by
\begin{equation}
  I \equiv I_{\mathrm{half}} + \sum_{\gamma \text{ internal}} \pi \frac{\ell_{\gamma}}{8 \pi G_{N}}.
  \label{eqn:I-fin}
\end{equation}

\eqref{eqn:ip-core} can be niftily represented using our schematic line diagrams,
\begin{align}
	\pmel{
		\input{figs/fa-4-bd-sp.tex}
	}{\eta}{
		\input{figs/fa-4-bd-s-no-l.tex}
	} = 
	\prod_{i=1}^{4} \delta(\ell_{i} - \ell'_{1}) e^{\frac{\ell_{s} + \ell'_{s}}{8 G_{N}}} \int^{*} Dg \exp{ - I \left[ \input{figs/fa-net-ip.tex} \right]}.
  \label{eqn:ip-line-diag}
\end{align}
The closed line diagram denotes a closed boundary condition with fixed areas, and from now on we will label the lines by $\gamma$ instead of $\ell_{\gamma}$.
From now onwards, we simplify the notation by using the closed line diagrams to denote both boundary conditions as well as the value the GPI takes with those boundary conditions.
\begin{equation}
 \input{figs/fa-net-ip.tex} \equiv \int^{*} Dg \exp{ - I \left[ \input{figs/fa-net-ip.tex} \right]}.
  \label{eqn:simp-act}
\end{equation}

While we have focused on the case without ETW branes in the above discussion, it goes through essentially unmodified when we include them as well.
We just have to remember that we are identifying only the open geodesics; the ETW branes are allowed to have a non-zero area in our saddles.
We will see examples below and in appendix \ref{app:bulk-ip-egs}.

In the general case, the prescription is as follows.
Define $\Sigma_{\mathrm{int}}$ to be the region of $\Sigma$ bounded by the external geodesics.
The boundary condition is given by $\Sigma_{\mathrm{int}} \# \Sigma_{\mathrm{int}}'$, where $\#$ is the union of the two slices where corresponding external geodesics are identified.
The external geodesic lengths must be the same for this to be possible, leading to a factor of $\prod_{\gamma \text{ external}} \delta(\ell_{\gamma} - \ell'_{\gamma})$.
A more abstract description of this prescription is to take the line diagrams corresponding to $\Sigma$ and $\Sigma'$ and join the open ends as in \eqref{eqn:ip-line-diag}.
This line diagram is related to the one that appears in Virasoro TQFT \cite{Collier:2023fwi}, as described above and pointed out in \cite{Hartman:2025cyj,Hartman:2025ula}.

We can think of the connection between the diagram in \eqref{eqn:ip-core} and that in \eqref{eqn:ip-line-diag} as follows (see \cite{Belin:2026pko}, specifically the section on the geometric interpretation of the fixed-$P$ partition function, for a more detailed explanation).
On the one hand, take the diagram in \eqref{eqn:ip-core} and invert it so that the inside is now outside; on the other hand, thicken each line in \eqref{eqn:ip-line-diag} into a cylinder and each vertex into a pair-of-pants.
These two diagrams are identical,
\begin{equation}
	\includegraphics[height=20mm,valign=c]{figs/s-ch-ip.pdf} \xrightarrow{\text{inside-out}} \includegraphics[height=20mm,valign=c]{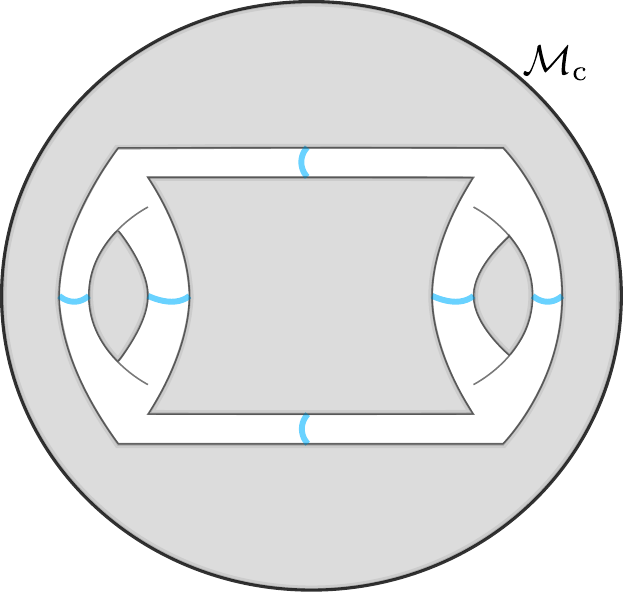} \,.
  \label{eqn:inside-out}
\end{equation}
The rule that we have to fill in the boundary conditions in \eqref{eqn:ip-core}, after inversion, becomes the rule that we have to embed the inverted diagram/thickened line diagram into a closed 3d manifold $\mathcal{M}_{\mathrm{c}}$ like $S^{3}, S^{2} \times S^{1}, T^{3}$ etc.
The bulk solution is the closed 3d manifold with the interior of the thickened line diagram excised.
This manipulation is useful for (a) drawing the boundary conditions, (b) describing some saddles and (c) connecting to the literature on Virasoro TQFT/Conformal Turaev-Viro theory in e.g. \cite{Collier:2023fwi,Hartman:2025cyj,Hartman:2025ula,Belin:2026pko}.
These uses will become apparent below.

If the line diagram contains double lines (i.e. the boundary condition contains open geodesics), we have to embed it into an open 3d manifold with ETW brane boundaries, such that the ETW branes in the solution end on the ETW branes in the boundary condition.

Notice that we have not said anything about boundary gravitons in the above rules.
The reason is that our fixed-area basis acts as the identity on that part of the Hilbert space, so the inner product between genuine states is
\begin{equation}
	\pmel{\Psi'}{\eta} {\Psi} = \sum_{\Sigma,\Sigma'} \int \dd{\ell_{\gamma}} \dd{\ell'_{\gamma}} \pmel{\Sigma',\left\{ \ell'_{\gamma} \right\}}{\eta} {\Sigma, \left\{ \ell_{\gamma} \right\}} \pmel{\Psi'}{\, \left\{ \, \pket{\Sigma', \left\{ \ell'_{\gamma} \right\}} \pbra{\Sigma, \left\{ \ell_{\gamma} \right\}} \, \right\} \,}{\Psi},
  \label{eqn:fa-ip-states}
\end{equation}
where the last factor is an inner product in the boundary graviton Hilbert space.
In the CFT, we will interpret the first factor as statistics of OPE coefficients; for Euclidean path integral (EPI) states, the second factor will become a conformal block.

\subsubsection*{How to Calculate Inner Products}

Let us see how to calculate these inner products.
One can simply use the rules of Virasoro TQFT \cite{Collier:2023fwi,Belin:2026pko} to do so.
Here, we will review the major building blocks of classical solutions to these fixed-area boundary conditions, and their actions.
Readers can consult appendix \ref{app:bulk-ip-egs} for a more detailed exposition.

\paragraph{The Degenerate Double Torus}
Consider the inner product
\begin{equation}
	\pmel{
\begin{tikzpicture}[baseline=.5cm]
	\draw[cyan] (0,0) to[out=120,in=-90] node[pos=.9,left]  {\small $1'$} (-.5,1);
	\draw[cyan] (0,0) to[out= 90,in=-90] node[pos=.9,left]  {\small $2'$} (  0,1);
	\draw[cyan] (0,0) to[out= 60,in=-90] node[pos=.9,right] {\small $3'$} ( .5,1);
\end{tikzpicture}
}{\eta} {
\begin{tikzpicture}[baseline=.5cm]
	\draw[cyan] (0,0) to[out=120,in=-90] node[pos=.9,left]  {\small $1$} (-.5,1);
	\draw[cyan] (0,0) to[out= 90,in=-90] node[pos=.9,left]  {\small $2$} (  0,1);
	\draw[cyan] (0,0) to[out= 60,in=-90] node[pos=.9,right] {\small $3$} ( .5,1);
\end{tikzpicture}
} = \prod_{\gamma = 1}^{3} \delta^{(2)} (\ell_{\gamma} - \ell_{\gamma}') \input{figs/3-bd-cis-ip.tex}.
  \label{eqn:3-bd-cis-ip}
\end{equation}
The leading solution is when $\Sigma$ and $\Sigma'$ are entirely identified and so there is no interior.
In this case, all internal angles are $0$, and we find
\begin{equation}
	\pmel{
\begin{tikzpicture}[baseline=.5cm]
	\draw[cyan] (0,0) to[out=120,in=-90] node[pos=.9,left]  {\small $1'$} (-.5,1);
	\draw[cyan] (0,0) to[out= 90,in=-90] node[pos=.9,left]  {\small $2'$} (  0,1);
	\draw[cyan] (0,0) to[out= 60,in=-90] node[pos=.9,right] {\small $3'$} ( .5,1);
\end{tikzpicture}
}{\eta} {
\begin{tikzpicture}[baseline=.5cm]
	\draw[cyan] (0,0) to[out=120,in=-90] node[pos=.9,left]  {\small $1$} (-.5,1);
	\draw[cyan] (0,0) to[out= 90,in=-90] node[pos=.9,left]  {\small $2$} (  0,1);
	\draw[cyan] (0,0) to[out= 60,in=-90] node[pos=.9,right] {\small $3$} ( .5,1);
\end{tikzpicture}
} \supset \prod_{\gamma=1}^{3} \delta^{(2)} ( \ell_{\gamma} - \ell'_{\gamma}).
  \label{eqn:3-bd-cis-ip-leading}
\end{equation}
The boundary condition for this inner product is a closed manifold with three geodesics --- i.e. a genus-$2$ double torus.
The solution `fills in' these boundary conditions with a zero-volume manifold.
We will call this solution the degenerate double torus.

This degenerate double torus will also appear in more general inner products.
If there are two pairs of pants such that all the geodesics bounding them are identified in the bulk solution, then we find that that part of the solution is a degenerate double torus.
Consider, for example, the inner product between two four-boundary wormholes with no angular velocities:
\begin{equation}
  \pmel{\input{figs/fa-4-bd-sp.tex}}{\eta} {\input{figs/fa-4-bd-s-no-l.tex}} = \prod_{\gamma=1}^{4} \delta(\ell_{\gamma} - \ell_{\gamma}') e^{\frac{\ell_{s} + \ell'_{s}}{8 G_{N}}} \input{figs/fa-4-bd-ss-ip.tex}.
  \label{eqn:4-bd-ip-main}
\end{equation}

One solution, which reproduces the kinematical inner product, is where $s$ gets identified with $s'$.
The topology of the solution is given by filling in the gaps between the surfaces in \eqref{eqn:ip-core}; in this topology, the $s,s'$ geodesics are homotopic and get identified,
\begin{equation}
	 \input{figs/fa-4-bd-ss-ip.tex} \supset \exp{ - I \left[ \includegraphics[height=20mm,valign=c]{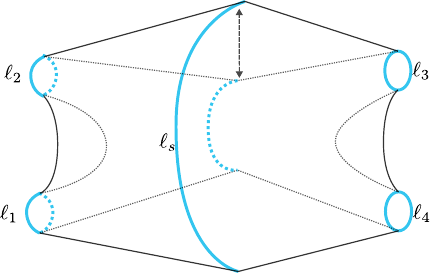} \right]} = e^{- \frac{\ell_{s}}{4 G_{N}}} \delta (\ell_{s} - \ell'_{s}).
  \label{eqn:4-bd-ip-s3-1}
\end{equation}
Notice that this forms two degenerate double tori --- bounded by $1,2,s$ and $s,3,4$ respectively --- joined at the $s$ geodesics.
The factor comes from the fact that the internal angles at $s,s'$ add up to $2\pi$ by an argument similar to the one in figure \ref{fig:I-half-pic}.
The `inside-out' perspective on this solution is that it arises from embedding the graph in $S^{3}$:
\begin{equation}
  \input{figs/fa-4-bd-ssp-ip-hom.tex} \propto \delta (\ell_{s} - \ell'_{s}) \ \input{figs/fa-4-bd-ssp-ip-hom-pinch.tex}
  \label{eqn:4-bd-ip-s3-hom}
\end{equation}
The dotted arrows indicate the homotopy between the geodesics and in the RHS we have identified the $s,s'$ geodesics to make it look like two copies of \eqref{eqn:3-bd-cis-ip}.

Another solution fills in the boundary conditions like
\begin{equation}
	\exp{ -I \left[ \includegraphics[height=40mm,valign=c]{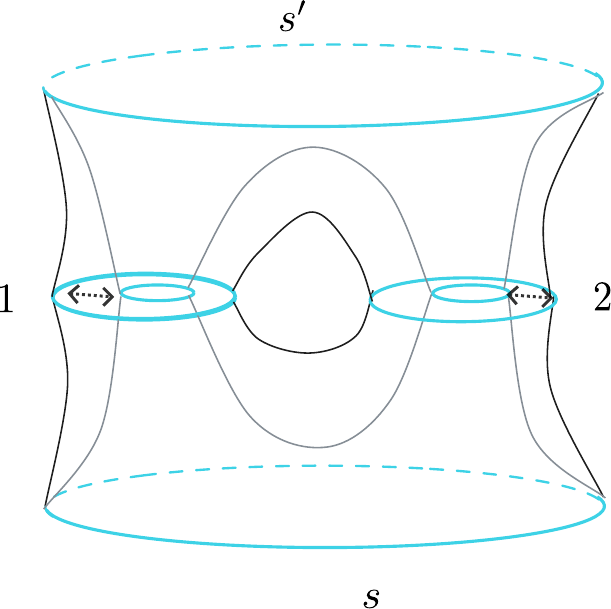} \right] } = \delta(\ell_{1} - \ell_{4}) \delta (\ell_{2} - \ell_{3}) e^{- \frac{\ell_{1} + \ell_{2}}{4 G_{N}}}.
	\label{eqn:4-bd-ip-s2s1-1}
\end{equation}
In this topology, the $1$ geodesic is homotopic to the $4$ one and the $2$ to the $3$, leading to the $\delta$-functions.
The internal angles at identified geodesics add up to $\pi$ as before and those at $s,s'$ vanish.
Notice that this is also the union of two degenerate double tori, one containing $1,2,s$ and another containing $1,2,s'$.
The inside-out perspective is that this is the embedding of the graph in $S^{2} \times S^{1}$.\footnote{
	We thank Tom Hartman for explaining this to us.
}
Representing $S^{2}$ as a circle and taking the $S^{1}$ to be the union of two intervals, this looks like\footnote{
	Alternatively, one can draw an $\Omega$-loop surrounding the $2$ and $3$ lines in the boundary conditions.
	An $\Omega$-loop in a given homotopy class enforces contractibility of that homotopy class.
}
\begin{equation}
  \input{figs/4-bd-ip-s2s1.tex}.
  \label{eqn:4-bd-ip-s2s1-2}
\end{equation}
The geodesics corresponding to the lines hitting the $S^{2}$s are homotopic along the respective $S^{2}$.

As a shorthand, we will simply say that some pair or pairs-of-pants collapses into a degenerate double torus.
So the $S^{3}$ contribution corresponds to the vertices $1,2,s$ and $3,4,s$ collapsing into degenerate double tori and the $S^{2} \times S^{1}$ contribution corresponds to the vertices $1,2,s$ and $1,2,s'$ collapsing into degenerate double tori.

Another important appearances of the degenerate double torus is when there are two copies of tori with holes that are identified,
\begin{equation}
  
\begin{tikzpicture}[baseline]
	\draw[cyan] (0,1) -- (0,0) arc(90:450:.5);
	\draw[cyan] (2,1) -- (2,0) arc(90:450:.5);
	\draw[<->,dotted,thin] (0,.5) -- (2,.5);
	\draw[<->,dotted,thin] (.5,-.5) -- (1.5,-.5);
\end{tikzpicture}

  \label{eqn:deg-2-tori}
\end{equation}
This gives a degenerate double torus with two ends identified.

We can simplify the calculations of actions of general solutions by removing degenerate double tori in the following way.
When the boundary condition graph is embedded in some $\mathcal{M}_{\mathrm{c}}$, some geodesics might become homotopic to some other geodesics.
Suppose there are two vertices $123, 1'2'3'$ in the graph such that the $1,2,3$ geodesics get identified with $1',2',3'$, then the corresponding pairs-of-pants collapse into a degenerate double torus.
We can then remove this pair of vertices from the graph, taking into account the contribution from the degenerate double torus.
Since this region of the solution has zero volume, the contribution is confined to the geodesics.
Focusing on the $1,1'$ geodesics, there are two distinct cases: either (a) they correspond to the same line in the graph or (b) they do not.
If they correspond to the same line, then the internal angle is $0$ and there is no contribution from this geodesic to the action.
Otherwise, by an argument similar to that in figure \ref{fig:I-half-pic}, then we get a factor of $\delta(\ell_{1} - \ell_{1'}) \exp{- (\ell_{1} + \bar{\ell}_{1})/8 G_{N}}$.
The same holds for the other two pairs.
There may also be braiding phases as we explain below; these factors are just the ones we get from the internal angles.
After taking into account these factors, we can remove the two vertices from the graph and end up with a simpler problem.
In the CFT, this operation is known as a Gaussian contraction.

The degenerate double torus is two copies of a pair-of-pants that get identified.
Similarly, we can take two copies of the other two elementary pieces and identify them also.
These play a similar role.

\paragraph{Dealing with Angular Momentum}
With our boundary conditions, the inner product with non-zero angular velocity is simply related to that with zero angular velocity.
Remember that $h^{\pm}$ defined in \eqref{eqn:mess-map} are related to the $PSL(2, \mathds{R})$ gauge fields.
The gravitational action in the Chern-Simons description takes the form $I[\mathds{A}] - I[\bar{\mathds{A}}]$, and therefore the inner product factorises between the part dependent on $h^{+}$ and that dependent on $h^{-}$.
Denoting by $\eta_{\mathcal{M}}$ the contribution to the inner product due to the solution $\mathcal{M}$, we can write
Therefore,
\begin{equation}
  \pmel{\ell',\bar{\ell}'}{\eta} {\ell,\bar{\ell}} = \sum_{\mathcal{M}} \sqrt{\pmel{\ell',\ell'}{\eta_{\mathcal{M}}}{\ell,\ell} \pmel{\bar{\ell}', \bar{\ell}}{\eta_{\mathcal{M}}} {\bar{\ell}, \bar{\ell}} }.
  \label{eqn:rot-trick}
\end{equation}
So, for the most part, we need to only deal with non-rotating geodesics and then the rotating case is given by a straightforward analytic continuation, as anticipated in \cite{Hartman:2025cyj,Hartman:2025ula}.\footnote{
	As pointed out in footnote \ref{fn:Kij-1}, our boundary conditions differ from those suggested in \cite{Hartman:2025cyj,Hartman:2025ula}.
	It is not clear how to generalise this trick to the case of $K_{ij} = 0$ boundary conditions.
	\label{fn:Kij-2}
}

The main novelty of including rotation is that there can be non-trivial braidings.
In practice, it is useful to calculate the contributions of these braidings separately and \emph{then} use \eqref{eqn:rot-trick}.
Consider the boundary condition
\begin{equation}
  \input{figs/3-bd-xing-ip.tex}.
  \label{eqn:3-bd-trans-ip}
\end{equation}
This also becomes a degenerate double torus, but one pair-of-pants needs to be rotated relative to the other to identify them,
\begin{equation}
  \includegraphics[height=20mm,valign=c]{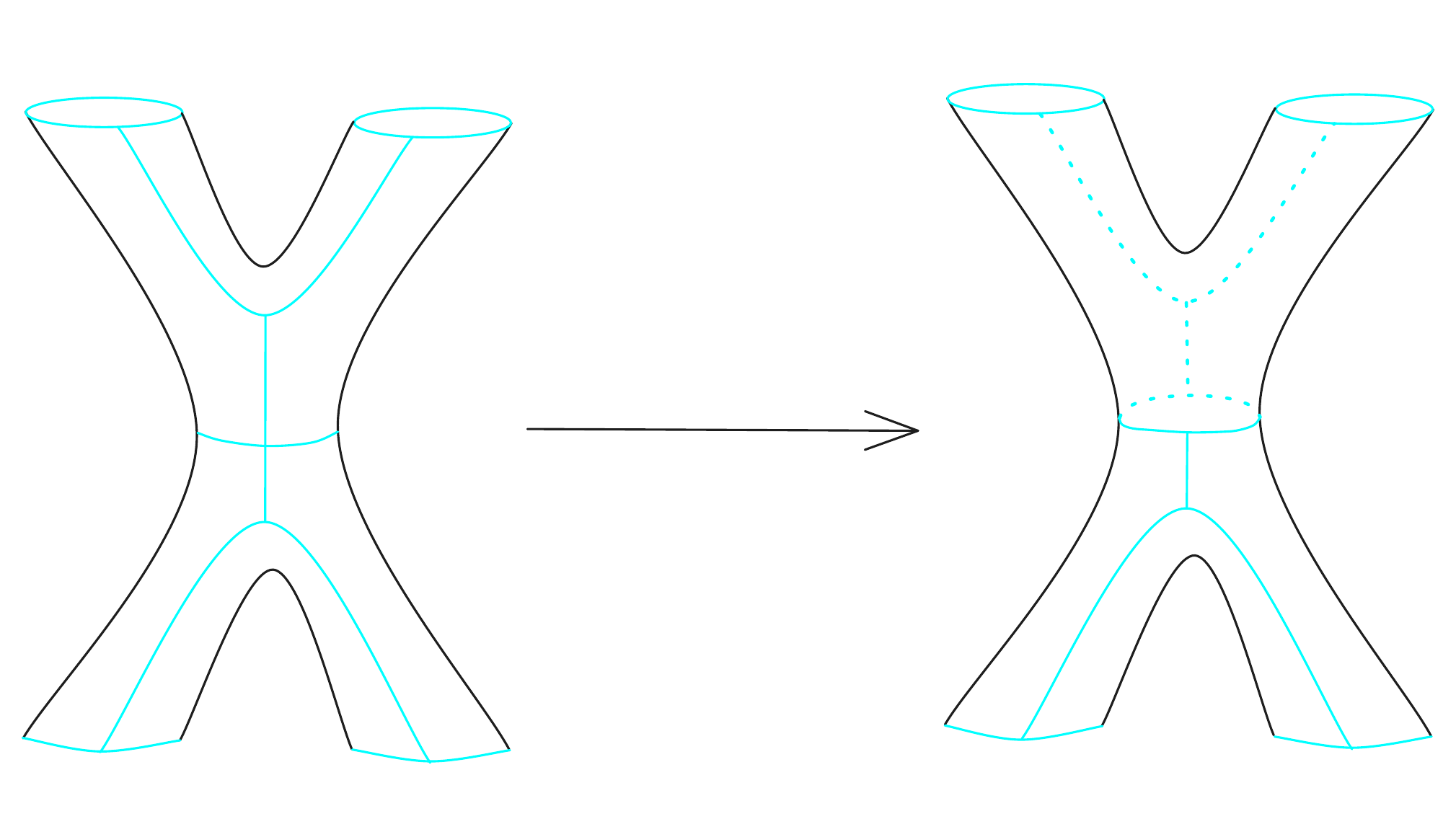}.
  \label{eqn:braiding}
\end{equation}
The cost is that the framing graph on one of the pairs is diametrically opposite to that on the other, giving
\begin{equation}
  \input{figs/3-bd-xing-ip.tex} = (-1)^{J_{1} + J_{2} - J_{3}}.
  \label{eqn:3-bd-trans-ip-ans}
\end{equation}
The relative sign in the exponent is entirely conventional, since, as we will see presently, angular momenta are integer-quantised.

In general, we get a factor of the braiding phase $(-1)^{J_{1} + J_{2} - J_{3}}$ whenever we rotate a pair-of-pants by $\pi$.
The direction of rotation and the relative signs in the exponent are not important only because the angular momenta are integer-quantised.
In particular, these details become important when we consider the terms inside the square root of \eqref{eqn:rot-trick}.
This is why it is useful to calculate the braiding phases separately.

\paragraph{Quantisation of Angular Momentum}
Since the twist is a periodic variable, its conjugate --- the angular momentum --- must be quantised.
This can be seen also in the following way.
For any saddle, we have an infinite class of saddles where one side is rotated by $2\pi n$ relative to the other, as pointed out in \cite{deBoer:2025rct,Belin:2026pko,Goker:2026tct}.
This gives a factor of
\begin{equation}
  \sum_{n \in \mathds{Z}} e^{2\pi i n \frac{r_{+} r_{-}}{2 G_{N}}} = \sum_{J \in \mathds{Z}} \delta \left( \frac{r_{+} r_{-}}{2 G_{N}} - J \right).
  \label{eqn:spin-quant-factor}
\end{equation}
The right side forces the angular momentum to be an integer.

\paragraph{The Generalised Tetrahedron}
All non-degenerate parts of the inner product can be built out of generalised tetrahedra \cite{Hartman:2025cyj,Hartman:2025ula}, which can be visualised as
\begin{equation}
  \input{figs/gen-tet.tex} = 
	\begin{Bmatrix}
	1 & 2 & s \\
	3 & 4 & t \\
	\end{Bmatrix}
	\,,
  \label{eqn:gen-tet}
\end{equation}
where each face is a surface with $K_{ij} = 0$ and the RHS is known as the Virasoro $6j$-symbol \cite{Ponsot:1999uf,Eberhardt:2023mrq}.
The Virasoro $6j$-symbol exhibits all the symmetries of the generalised tetrahedron.
The arguments for the $6j$-symbol are $\ell_{i}/2\pi b$.
Here, we only consider non-rotating geodesics since \eqref{eqn:rot-trick} ensures us that this is enough.

The first set of examples are the duals of F transformations in CFT bootstrap.
For example,
\begin{align}
  \pmel{\input{figs/fa-4-bd-tp.tex}}{\eta} {\input{figs/fa-4-bd-s-no-l.tex}} &=  \prod_{i=1}^{4} \delta(\ell_{i} - \ell'_{i}) \includegraphics[height=20mm,valign=c]{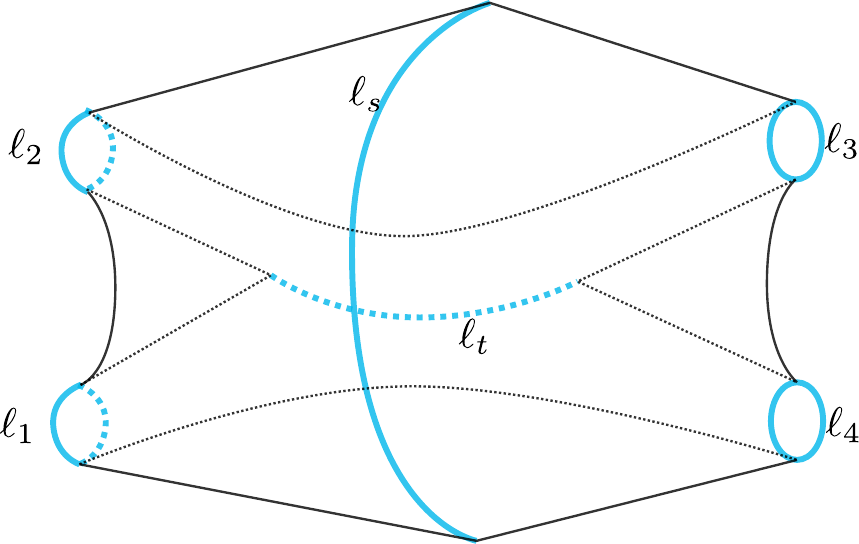} \nonumber\\
  &= \prod_{i=1}^{4} \delta (\ell_{i} - \ell'_{i}) \ 
	\input{figs/double-gen-tet.tex} \,,
  \label{eqn:fa-st-xing-main}
\end{align}
where the dotted arrows represent identification.
A similar inner product where all the geodesics are open and $T = 0$ is given by a single generalised tetrahedron, where all the small triangles are ETW branes:
\begin{equation}
  \input{figs/fa-tet-open.tex} = \input{figs/gen-tet-etw.tex}
  \label{eqn:fa-st-xing-open}
\end{equation}
The case of non-zero tension is easily handled as explained below.

The third example is the modular S-transformations, which is a diagram similar to \eqref{eqn:deg-2-tori} but where the circles are not identified but rather the two geodesics on the circles are related by an S-transformation,
\begin{equation}
  
\begin{tikzpicture}[baseline]
	\draw[cyan] (0,1) -- (0,0) arc(90:450:.5);
	\draw[cyan] (2,1) -- (2,0) arc(90:450:.5);
	\node at (2,-.5) {\small $S$};
	\draw[<->,dotted,thin] (0,.5) -- (2,.5);
\end{tikzpicture}
 = \input{figs/gen-tet-S.tex} \,.
  \label{eqn:S-trans}
\end{equation}
If the two tori are related by a general $PSL(2,\mathds{Z})$ transformation, we have to tile the region in between with many generalised tetrahedra and integrate over the lengths of all extra geodesics introduced in this tiling \cite{Hartman:2025cyj,Hartman:2025ula}.

There are further non-degenerate pieces when you have ETW branes, but they can be rewritten in terms of generalised tetrahedra by the doubling trick.
Take the saddle with ETW branes (with $T = 0$), then reflect about the ETW branes to get a closed manifold.
The inner product is given by the action of a $PSL(2,\mathds{R})$ Chern-Simons theory (with only one gauge field $\mathds{A}$) on this closed manifold.
Since we have a chiral theory on the double, the direction in which we move the framing graph matters and the calculation can get quite delicate.
In this work, we will either do simple calculations or work at an abstract level where it is enough to know that there is a correct answer.
The precise inner products are described in \cite{Jafferis:2026gzn}.

All of these pieces can be used to remove vertices and simplify diagrams to calculate on-shell actions as in the case of the degenerate double torus.
In the CFT, these are known as non-Gaussian contractions.

\paragraph{Non-Zero Tension}
The case of non-zero tension is quite simple \cite{Wang:2025bcx,Jafferis:2026gzn}.
The point is that the action of the region between a $K_{ij} = 0$ surface and a $K_{ij} = T h_{ij}$ surface is topological, and just gives a contribution of $n \log g$ in the action, where $n$ is the number of connected components of the ETW brane,
\begin{equation}
  I [T] - I[T=0] = \frac{n}{4 G_{N}} \tanh^{-1} T.
  \label{eqn:action-tension}
\end{equation}

\section{The CFT Hilbert Space} \label{sec:H-CFT}

In this work, we would like to consider a CFT whose dual theory is general relativity coupled to some matter fields in an appropriate semiclassical limit.
In this section, we review some relevant properties of these CFTs that are either known or believed to be true.
We use the following parametrisation for the central charge and conformal weights,
\begin{equation}
  h =  \frac{c-1}{24} + P^{2}, \qquad b \approx \sqrt{ \frac{6}{c-1}} = \sqrt{4 G_{N}}.
  \label{eqn:liouville-param}
\end{equation}

The Hilbert space of a generic irrational 2d CFT on a circle decomposes into a sum over primary Hilbert spaces,
\begin{equation}
  \mathcal{H}_{S^{1}} = \bigoplus_{(P, \bar{P}) \in \operatorname{spec}} \mathcal{H}_{(P,\bar{P})} \cong \mathcal{H}_{\mathds{1}} \oplus \mathcal{H}_{\mathrm{prim}} \otimes \mathcal{H}_{\mathrm{desc}}.
  \label{eqn:H-S1}
\end{equation}
Here, $\operatorname{spec}$ denotes the set of primaries in the theory, $\mathcal{H}_{(P,\bar{P})}$ is the Hilbert space of descendants of the primary with weights $P,\bar{P}$ and $\mathds{1}$ denotes the identity operator.
For a generic irrational CFT, all the descendant Hilbert spaces are isomorphic and so we can decompose the Hilbert space as in the third expression.\footnote{
	One has to be careful, since the inner product in $\mathcal{H}_{(P,\bar{P})}$ depends on $b,P,\bar{P}$.
	One has to do a Gram-Schmidt orthogonalisation at each level of the Verma modules and identify the corresponding orthonormal bases.
	This means that the stress tensor is not an operator purely on $\mathcal{H}_{\mathrm{desc}}$.
	For more details, see \cite{Soni:2025qau}.
}
Since any closed curve in a Euclidean path integral is conformally related to a circle, we may apply this decomposition to any closed curve.
Similarly, the Hilbert space on an interval with boundary conditions $\sigma$ on both ends can be written as
\begin{equation}
  \mathcal{H}_{\sigma\sigma} = \bigoplus_{P \in \operatorname{spec}_{\sigma\sigma}} \mathcal{H}_{P} = \mathcal{H}_{\mathds{1}} \otimes \mathcal{H}_{\mathrm{prim}} \otimes \mathcal{H}_{\mathrm{desc}}.
  \label{eqn:H-I}
\end{equation}
$\mathcal{H}_{\mathrm{desc}}$ on an interval is only one copy of a Verma module whereas that on a circle is two Verma modules; we will not overburden our notation by differentiating these since it will be clear from context.

\subsection{Holographic $\text{CFT}_{2}$s: A Review} \label{ssec:cft-review}

A holographic $\text{CFT}_{2}$ is a unitary compact CFT with a large central charge $c \gg 1$. 
$c$ is related to Newton's constant in the bulk by the Brown-Henneaux relation $c = 3/2 G_{N}$.
A holographic 2d CFT has a sparse spectrum of primaries below the black hole threshold $h, \bar{h} < c/24$ (i.e. $P, \bar{P} \in i \mathds{R}$), and a dense spectrum sufficiently above.
By dense, we mean that the density of primaries is \cite{Hartman:2014oaa,Mukhametzhanov:2019pzy,Pal:2019zzr,Dey:2024nje}
\begin{equation}
  \log \rho_{\mathrm{p}} (P,\bar{P}) = 2\pi \frac{P + \bar{P}}{b} - \log c + \mathcal{O} (c^{0}),
  \label{eqn:prim-dos}
\end{equation}
where $b,P$ are defined in \eqref{eqn:liouville-param}.
By density, we mean that this is the number of primaries in an $\mathcal{O} (c^{0})$ window in $h,\bar{h}$ space, which we will call a microcanonical window henceforth.
The corrections to this formula depend on the details of the window, but the exhibited terms are believed to be universal.\footnote{
	A rigorous derivation for $h \neq \bar{h}$ is lacking as of now.
	\eqref{eqn:prim-dos} has been proved for $h, \bar{h} \gg c$ in \cite{Pal:2019zzr} and $h = \bar{h} \sim c \gg 1$ in \cite{Mukhametzhanov:2019pzy}.
	This formula agrees with the large $c$ expansion of the modular transformation of the vacuum Virasoro character.
}

Holographic $\text{CFT}_{2}$s are also believed to exhibit similar universality in the OPE coefficients.
For three primaries, if one has $h, \bar{h} \gg c >1$, then the average of the OPE coefficients over a microcanonical window over the heavy operators is \cite{Collier:2019weq} 
\begin{equation}
	\mathds{E}_{\mu c} \left[ C_{123} C_{123}^{*} \right] \approx C_{0} (P_{1}, P_{2}, P_{3}) \cdot  C_{0} (\bar{P}_{1}, \bar{P}_{2} , \bar{P}_{3} ) \equiv \mathfrak{C}_{123},
  \label{eqn:ope-universality}
\end{equation}
where $C_{0}$ is proportional to the DOZZ formula for OPE coefficients in Liouville theory.\footnote{
	Liouville CFT plays the same role for Virasoro representation theory that a particle on a sphere plays for $SO(3)$ representation theory.
}
The detailed form of the function $C_{0}$ will not be used in this work.
There are corrections to this formula from terms proportional to $\delta^{(2)} (P_{1} - P_{2})$ etc.
We use the standard convention that
\begin{equation}
  C_{213} = C_{123}^{*}.
  \label{eqn:ope-convention}
\end{equation}
We also sometimes use a diagrammatic notation for the OPE coefficients, e.g.
\begin{equation}
  C_{123} C_{123}^{*} = \input{figs/g2-ope.tex}.
  \label{eqn:ope-diagram}
\end{equation}
In these diagrams, each vertex is an OPE coefficient and a line joining two vertices indicates that those OPE coefficients share an index. In \eqref{eqn:ope-diagram}, the bottom vertex gives $C_{123}$, following the convention that sitting at the vertex, the lines labelled 1, 2, 3 appear in clockwise order. Reverse holds for the top vertex, giving $C^*_{123} = C_{213}$.

While \eqref{eqn:ope-universality} has been shown for $h,\bar{h} \gg c$, the question of whether there is an extended range of validity of this formula in holographic CFT, to the regime where the heavy operators have $h,\bar{h} \sim c$, is a lot more subtle. 
These universal results are derived by assuming the existence of a surface $M$ such that $Z(M)$ is dominated by $C_{123} C_{123}^{*}$ in one channel and the identity in another, as we review in appendix \ref{app:ope-statistics}.\footnote{
	There are further contributions that cannot be derived in such a way but which are required for consistency with physical principles motivated by eigenstate thermalisation hypothesis, see \cite{Belin:2026pko} for more details.
	However, the leading pieces usually follow from identity block domination.
}
It was found in \cite{Belin:2017nze,Dong:2018seb} that in some regimes of moduli space a sufficiently light bulk field can condense and therefore, in the boundary, no channel is dominated by the identity conformal block.
In the bulk, this means that the corresponding boundary conditions are such that GR is not a good effective theory.

We further assume that the CFT of interest has at least one Cardy boundary condition, which we denote $\sigma$, and that the corresponding boundary state has non-zero overlap with the vacuum,
\begin{equation}
  g \equiv \obra{\sigma} \mathds{1} \rangle \neq 0.
  \label{eqn:g-defn}
\end{equation}
The analogues of \eqref{eqn:prim-dos} and \eqref{eqn:ope-universality} in the BCFT, using upper/lower indices for boundary/bulk primaries, are \cite{Kusuki:2021gpt,Numasawa:2022cni}
\begin{align}
  \log \rho_{\mathrm{p}} (g, P) &= 2\pi \frac{P}{b} - \log c + 2 \log g,
  \label{eqn:bcft-prim-dos} \\
	\mathds{E} \left[ C^{123} C^{213} \right] &= \mathds{E} \left[ \input{figs/g2-bcft-ope.tex} \right] = \frac{\mathfrak{C}^{123}}{g^{2}}, \nonumber\\
	\mathds{E} \left[ C^{1}_{2} C^{1}_{2} \right] &= \mathds{E} \left[ 
\begin{tikzpicture}[baseline=0cm]
	\draw[dashed] (-.5,0) arc(-90:-270:.5) node[pos=.5,left] {\small 2};
	\draw[dashed,double=white] (-.5,0) arc(-90:90:.5) node[pos=.5,right] {\small 1};
\end{tikzpicture}
 \right] = \frac{\mathfrak{C}^{1}_{2}}{g^{2}},
	\label{eqn:bcft-ope-universality}
\end{align}
where
\begin{equation}
  \mathfrak{C}^{123} = C_{0} (P_{1}, P_{2}, P_{3}), \quad \mathfrak{C}^{1}_{2} = C_{0} (P_{1}, P_{2}, \bar{P}_{2}).
  \label{eqn:Cfk-bcft}
\end{equation}
We use double dashed lines to denote boundary primaries in these diagrams.
Yet again, we assume these formulas to be valid when at least one primary has $h \sim c$, and there are subleading corrections to these formulas.

\subsection{Euclidean Path Integral (EPI) States} \label{ssec:vtft-ip}

One class of states of particular interest is states prepared by Euclidean path integral with possible stress tensor insertions, which we refer to as EPI states.
These states are usefully described in terms of objects known as Virasoro OPE blocks \cite{Czech:2016xec,Fitzpatrick:2016mtp,Besken:2018zro,DHoker:2019clx,Chandra:2023dgq}.
Here, we describe what an OPE block is and point out some interesting facts about the inner product in terms of OPE blocks.

Consider an open surface $\Sigma_{k}$ with $n_{1}$ circular boundaries, some number of $\sigma$ boundaries and $n_{2}$ interval boundaries.
A Euclidean path integral on this surface prepares a state in $\mathcal{H}_{S^{1}}^{\otimes n_{1}} \otimes \mathcal{H}_{I}^{\otimes n_{2}}$.
We refer to states prepared by Euclidean path integrals as Euclidean path integral (EPI) states.
The name is somewhat misleading, since we can allow for $\Sigma_{k}$ to have regions where the metric is Lorentzian without affecting the discussion below.

For concreteness, we discuss the case where $n_{1} = 4$ and $n_{2} = 0$; the story generalises straightforwardly.
This state can be expanded in terms of OPE blocks \cite{Chandra:2023dgq}, which are maps from $\mathcal{H}_{\mathrm{prim}}^{\otimes n} \to \mathcal{H}_{\mathrm{CFT}}^{\otimes n}$, as follows
\begin{equation}
  \ket{\Sigma} = \sum_{(P_{i}, \bar{P}_{i}) \in \operatorname{spec}, i = 1\dots 4,s} \abs*{ \quad \input{figs/ope-block-dot.tex}\ (\Sigma_{k})\ \ket{P_{1},P_{2},P_{3},P_{4}} }^{2}
  \label{eqn:ope-block-exp}
\end{equation}
Let us explain this equation.
The $\abs{\,}^{2}$ indicates that this is the holomorphic half of the story and one should also add the anti-holomorphic one.
$\ket{P_{i}} \ket{ \bar{P}_{i}}$ lives in $\mathcal{H}_{\mathrm{prim}}^{\otimes 4}$.
The diagram in the middle is an unconventionally normalised OPE block, which describes the state of the descendants created once we fix the primary flowing through every non-intersecting cycle of $\Sigma_{k}$.
Remember that any circle in a path integral hosts a copy of $\mathcal{H}_{S^{1}}$; so we can draw any circle and project onto a fixed primary there.
Denoting such a projector by a grey circle,
\begin{equation}
	\abs*{\input{figs/ope-block-dot.tex}}^{2}
	= C_{12s} C_{s34} \abs*{\input{figs/ope-block.tex}}^{2}
	\quad \propto \quad 
  \input{figs/ope-block-M.tex}
	\quad ,
  \label{eqn:prim-cuts}
\end{equation}
where the proportionality constant is the Weyl anomaly which is independent of the conformal weights, and a grey circle with label $i$ means that we have projected onto a specific primary in the Hilbert space \eqref{eqn:H-S1} of the corresponding circle.
A label $i$ means that the Liouville momentum of the primary is $P_{i}$.
The OPE block is a function of the moduli of $\Sigma$, and in \eqref{eqn:ope-block-exp} we have shown this dependence explicitly; we will drop it soon.

We can also allow for stress tensor insertions on $\Sigma$, as well as real-time evolution on all the $B_{i}$s.
The representation of the final state $\ket{\Psi; \Sigma_{k}}$ similar to \eqref{eqn:ope-block-exp} is as follows.
First, we absorb any real-time evolution into an analytic continuation of $\Sigma_{k}$.
The stress tensor insertions, via the conformal Ward-Takahashi identities, do two things.
First, they may act as derivatives on the moduli of $\Sigma_{k}$.\footnote{
	This can be seen for example by writing a $T(z)$ insertion in \eqref{eqn:prim-cuts} in terms of Virasoro generators on some cycle and performing contour deformations.
	The details will not be relevant for us.
}
Second, they may act as descendant operators on the external states.
So, we denote the OPE blocks contributing to $\ket{\Psi}$ as
\begin{equation}
	\ket{\Psi; \Sigma_{k}} = \sum_{(P_{i}, \bar{P}_{i})} \abs*{ \mathds{D}_{\Psi} \cdot\,   \input{figs/ope-block-dot.tex}\ (\Sigma_{\mathrm{k}})\ \ket{P_{1},P_{2},P_{3},P_{4}} }^{2}.
  \label{eqn:ope-block-Psi}
\end{equation}
The font of the $\mathds{D}$ represents the fact that it is also a matrix.
When we take the norm, the OPE blocks join up to become vanilla conformal blocks and $\mathds{D}_{\Psi}$ simplifies to an ordinary differential operator acting on the moduli of $\mathcal{M}$; for example,
\begin{equation}
  \braket{\Psi;\Sigma_{b}}{\Psi;\Sigma_{k}} = \sum \abs*{ D_{\Psi,\Psi^{*}} \input{figs/s-channel-conf-block.tex}\ (\Sigma_{k} \cup \Sigma_{b})}^{2}.
  \label{eqn:norm-conf-block}
\end{equation}

We can similarly define OPE blocks for a BCFT.
Parallel to the above, for $\Sigma$ a disk topology manifold that prepares a state on four intervals,
\begin{equation}
  \ket{\Sigma} = \sum_{P_{i}}
	\input{figs/bcft-ope-block-s.tex}
	(\Sigma) \ket{P_{1}\dots P_{4}}
  \label{eqn:bcft-ope-bloc}
\end{equation}

It is well-known that EPI states on $\Sigma_{\mathrm{k}}$ construct bulk states with Cauchy slices $\Sigma$ that are homeomorphic to $\Sigma_{\mathrm{k}}$ for some values of the moduli \cite{Skenderis:2009ju,Krasnov:2001va,Balasubramanian:2014hda,VanRaamsdonk:2010pw,VanRaamsdonk:2018zws}.
We will call these EPI states classical, for reasons that will be explained in section \ref{sec:gauge}.
Our work also applies to semiclassical states, which can be constructed from EPI state by integrating over $\mathcal{O}(1)$ ranges of moduli, see \cite{Marolf:2018ldl} for an example.

While EPI states are the most classical and also the most convenient for AdS/CFT via the GKPW dictionary, they are often more tedious to work with in boundary calculations.
The reason is simply that the probability distribution over primaries on any cycle is determined by a conformal block.
For example, the probability of finding the weight $P_{s}$ on the $s$ line is
\begin{equation}
  \mel{\Sigma_{\mathrm{k}}}{\hat{P}_{s}}{\Sigma_{\mathrm{k}}} = \frac{1}{\norm{\Sigma_{\mathrm{k}}}^{2}} \sum_{P_{i}} P_{s} \input{figs/s-channel-conf-block.tex}.
  \label{eqn:epi-ps}
\end{equation}
In other words, we need to calculate a conformal block even if we are interested only in the primary Hilbert space.
Another way to say this is that the map from moduli to the probability distribution over primaries is quite complicated.
For example, if we are interested in calculating entanglement entropy, the leading contribution comes from the entanglement in the primary space but we still need the conformal blocks to find its value.
Therefore, it is common to truncate EPI states \cite{Chandra:2023dgq}: we replace the conformal block by an explicit wavefunction $\Psi(P_{i})$ over primaries that smears each line over a microcanonical window with slowly varying phases.

\subsection{OPE Statistics from Identity Block Domination} \label{ssec:id-block-dom-ope}

In appendix \ref{app:ope-statistics}, we give a short introduction to calculating OPE statistics from the assumption of identity block domination as developed by \cite{Collier:2019weq,Kusuki:2021gpt,Numasawa:2022cni,Belin:2021ryy} and other works.
To calculate the microcanonical average of a series of OPE coefficients, we assume the existence of a surface $\Sigma$ such that, in one channel, the partition function $Z(\Sigma)$ is dominated by the microcanonical window in question and, in another channel, it is dominated by the identity primary on all the cuts in the channel.
This provides the duals of handlebody contributions to the sum over topologies, but misses the contributions of non-handlebodies \cite{Belin:2026pko}.
Here, we list some basic facts about the results one obtains in this way, paralleling the discussion in section \ref{ssec:inner-products}.

\paragraph{Angular Momentum}
The chiral factorisation shown in \eqref{eqn:rot-trick} is dual to the fact that, since the OPE densities are all crossing kernels of the conformal bootstrap, they factorise holomorphically.
The angular momentum $h - \bar{h}$ is quantised in any local CFT.

\paragraph{Gaussian OPE Statistics}
First, consider
\begin{equation}
	\mathds{E}_{\mu c} \left[ \input{figs/g2-ope.tex} \right] \approx \mathfrak{C}_{123}.
  \label{eqn:3-bd-ope}
\end{equation}
This is obtained by dualising any of the legs to the identity.
In larger OPE coefficient networks, we can pair any two vertices to get a similar answer along with appropriate $\delta$-functions \cite{Chandra:2022bqq}.
In the case that the two vertices are connected in the OPE coefficient network, this result can be obtained by dualising one of the lines joining the vertices,
\begin{align}
	\mathds{E}_{\mu c} \left[ \input{figs/ope-gauss-1line.tex} \right] &\supset   \frac{\delta^{(2)} (P_{1} - P_{5}) \delta^{(2)} (P_{2} - P_{4}) + (-1)^{J_{3} - J_{4} - J_{5}} \delta^{(2)} (P_{1} - P_{3}) \delta^{(2)} (P_{2} - P_{5})}{\rho_{\mathrm{p}} (P_{1}, \bar{P}_{1}) \rho_{\mathrm{p}} (P_{2}, \bar{P}_{2})}  \mathfrak{C}_{123} \nonumber\\
	\mathds{E}_{\mu c} \left[ 
\begin{tikzpicture}[baseline]
	\draw[dashed] (0,0) -- node[above] {\small $1$} (1,0);
	\draw[dashed] (2,0) -- node[above] {\small $4$} (3,0);
	\draw[dashed] (1,0) arc(180:  0:.5) node[pos=.5,above] {\small $2$};
	\draw[dashed] (1,0) arc(180:360:.5) node[pos=.5,below] {\small $3$};
\end{tikzpicture}
 \right] & \supset \frac{\delta^{(2)} (P_{1} - P_{4})}{\rho_{\mathrm{p}} (P_{1}, \bar{P}_{1})} \mathfrak{C}_{123}.
  \label{eqn:gaussian-statistics}
\end{align}
We get $\delta$-functions despite the spectrum being discrete, since the microcanonical window function can be peaked at any point in a continuum of values.
The contributions one finds from using these relations (and their BCFT analogs) are known as Gaussian statistics in the literature, see e.g. \cite{Chandra:2022bqq,Bao:2024ixc,Bao:2025plr,Geng:2025efs,Bao:2025nbz}.
Using these Gaussian contractions gives answers similar to pairing off pairs of pairs of pants into degenerate double tori.

As an example, consider the case of the four vertex diagram,
\begin{align}
	\mathds{E}_{\mu c} \left[ \input{figs/s-ch-dashed.tex} \right]  \xrightarrow{\text{pairing vertices on the same line}} & \frac{\mathfrak{C}_{12s} \mathfrak{C}_{21s'}}{\rho_{\mathrm{p}} (P_{1}, \bar{P}_{1}) \rho_{\mathrm{p}} (P_{2}, \bar{P}_{2})} \left[ \delta^{(2)} (P_{1} - P_{4}) \delta^{(2)} (P_{2} - P_{3}) \right. \nonumber\\
																																 & \qquad \qquad  + \left. (-1)^{J_{s} + J'_{s}} \delta^{(2)} (P_{1} - P_{3}) \delta^{(2)} (P_{2} - P_{4}) \right] \nonumber\\
																																 \xrightarrow{\text{pairing vertices vertically}} & \frac{\mathfrak{C}_{12s} \mathfrak{C}_{s34}}{\rho_{\mathrm{p}} (P_{s}, \bar{P}_{s})} \delta^{(2)} (P_{s} - P'_{s}).
  \label{eqn:ope-4-vertex}
\end{align}
Restricting to scalars, these two contraction patterns are dual to \eqref{eqn:4-bd-ip-s2s1-1} and \eqref{eqn:4-bd-ip-s3-1} respectively.

\paragraph{Non-Gaussian Statistics}
Gaussian statistics are not enough, since they only give you the contributions of zero-volume solutions.
Non-Gaussian statistics are given by non-trivial crossing kernels.
As a result, they can all be generated by a few moves.

The first set of rules is given by the crossing move or F-move \cite{Belin:2021ryy,Hartman:2025cyj,Hartman:2025ula}
\begin{align}
	\input{figs/ope-tet.tex} &\supset \sqrt{\mathfrak{C}_{12s} \mathfrak{C}_{s34} \mathfrak{C}_{41t} \mathfrak{C}_{t23}} \abs*{
		\begin{Bmatrix}
		1 & 2 & s \\
		3 & 4 & t \\
		\end{Bmatrix}
	}^{2} \nonumber\\
	\input{figs/ope-b-tet.tex} &\supset g^{4} \sqrt{\mathfrak{C}^{12s} \mathfrak{C}^{s34} \mathfrak{C}^{41t} \mathfrak{C}^{t23}}
	\begin{Bmatrix}
	1 & 2 & s \\
	3 & 4 & t \\
	\end{Bmatrix}
  \label{eqn:F-stat}
\end{align}
These correspond to \eqref{eqn:fa-st-xing-main} and \eqref{eqn:fa-st-xing-open} respectively.
The contribution corresponding to the S-transformation in \eqref{eqn:S-trans} is
\begin{equation}
	\mathds{E}_{\mu c} \left[ 
\begin{tikzpicture}[baseline]
	\draw[dashed] (0,0) -- node[above] {\small $2$} (1,0);
	\draw[dashed] (0,0) arc(  0:360:.5) node[pos=.5,left ] {\small $1$};
	\draw[dashed] (1,0) arc(180:540:.5) node[pos=.5,right] {\small $3$};
\end{tikzpicture}
 \right] \supset \sqrt{\mathfrak{C}_{112} \mathfrak{C}_{233}} \abs*{\widehat{\mathds{S}_{13}} [2]}^{2}.
  \label{eqn:S-stat}
\end{equation}
The classical limit of this kernel agrees with \eqref{eqn:S-trans} \cite{Hartman:2025cyj,Hartman:2025ula}.

Similarly, there are contributions corresponding to the different crossing transformations for BCFTs.
These can also be written in terms of the braiding phases and Virasoro $6j$-symbols by passing to a chiral CFT on the Schottky double.
The explicit OPE densities can be found in \cite{Jafferis:2025yxt}.

\paragraph{$\delta$-Functions in OPE Densities}
OPE densities often have $\delta$-functions; since OPE coefficients are chaotic objects, they often average to $0$ unless some of the primaries are identified during averaging.
Each contribution to the OPE density corresponds to a different topology for the bulk.
If a particular contribution is non-zero only when two a priori distinct primaries are identified, the corresponding bulk topology is such that the corresponding geodesics are homotopic.

Here, it is important to distinguish between coarse-grained equality and fine-grained equality.
A coarse-grained equality is when two primaries are \emph{independently} microcanonically smeared around the same weights.
For example, denoting by $f_{\ell} (P)$ a microcanonical smearing function
\begin{equation}
  \sum \abs{f_{\ell_{1}} (P_{1}) f_{\ell_{1}} (P_{2}) f_{\ell_{3}} (P_{3})}^{2} \input{figs/g2-ope.tex} = \mathfrak{C}_{113}.
  \label{eqn:cg-eqty}
\end{equation}
This corresponds in the bulk to two \emph{disntinct} geodesics with the same length.
However, we can smear two primaries together as in
\begin{equation}
  \sum \abs{f_{\ell_{1}} (P_{1}) f_{\ell_{3}} (P_{3})}^{2} \input{figs/g2-ope-112.tex}.
  \label{eqn:fg-eqty}
\end{equation}
It is not obvious how to calculate this in the CFT, since such a joint smearing does not arise in any path integral.
A bulk solution that contributes to this is where we embed the graph in $S^{2} \times S^{1}$,
\begin{equation}
  \input{figs/3-bd-ip-s2s1.tex}.
  \label{eqn:fg-eqty-bulk}
\end{equation}
The two geodesics are homotopic around the $S^{2}$ and therefore identified in the solution.
The $3$ geodesics is contractible around the $S^{2}$ and therefore its length must be $2\pi i$ (i.e. the primary must be the identity).
\cite{Belin:2026pko} also studies a different solution where the first two geodesics are homotopic but the third one is not contractible.

\section{The Holographic Map} \label{sec:V}

Now that we have set up the bulk as well as the boundary Hilbert spaces, we are finally ready to state the central proposal of this work, which is the holographic map mapping the one to the other.
This map is non-isometric, but it is approximately isometric on many semiclassical states.
We also list some other sources of failure of the map, and argue that these are physical.

Taking the example of a four-boundary wormhole for concreteness, the map is
\begin{align}
	&V \pket{ 
		\input{figs/fa-4-bd-s-no-l.tex}
	} \otimes \pket{\psi}_{\mathrm{bdgravs}} \nonumber\\
		&=  \sum_{(P_{I}, \bar{P}_{I}) \in \mathrm{prim}} \prod_{I = 1\dots 4,s} \frac{\tilde{\delta}^{(2)} (\ell_I, \bar{\ell}_I; P_I,\bar{P}_I)}{ \sqrt{\rho_{\mathrm{p}} (P_{I}, \bar{P}_{I})}} 
		\; \cdot \;   \frac{\input{figs/ope-coeff-diag.tex}}{ \sqrt{\mathfrak{C}_{12s} \mathfrak{C}_{s34}}}  \; \cdot \;\abs*{\ket{P_{1} \dots P_{4}}}^{2} \;\;\otimes \;\; \ket{\psi}_{\mathrm{desc}},
		\label{eqn:hol-map-1}
\end{align}
where we have defined
\begin{equation}
  \tilde{\delta}^{(2)} (\ell_I, \bar{\ell}_I; P_I,\bar{P}_I) \equiv \delta[(\ell_{I} + \bar{\ell}_{I}) - 2 \pi b (P_{I} + \bar{P}_{I})] \,\cdot\,  \delta [-i (\ell_{I} - \bar{\ell}_{I}) - 2\pi b (P_{I} - \bar{P}_{I})].
  \label{eqn:ell-realisation}
\end{equation}
Remember that the cyan line diagram on the LHS is a fixed-area network and the dashed line diagram on the RHS just means $C_{12s} C_{s34}$.
The state of the boundary gravitons becomes the state of the descendants in a straightforward manner.

For a general fixed-area network, the map
\begin{enumerate}
    \item Projects every line into the primary spectrum of the CFT, through the delta function \eqref{eqn:ell-realisation}, with a normalisation coefficient of $\rho_{\mathrm{p}} (P_{i}, \bar{P}_{i})^{-1/2}$ for every line.

  \item Converts every vertex to an OPE coefficient as follows: if the lines meeting the vertex have labels $\ell_i, \ell_j, \ell_k$, we get a $C_{ijk} /\sqrt{\mathfrak{C}_{ijk}}$ factor in the wavefunction.
  \item Creates a state proportional to $\otimes_{i \text{ external}} \ket{P_{i}, \bar{P}_{i}}$ (or just $\ket{P_{i}}$ for an open geodesic) in the primary Hilbert space.
  
\end{enumerate}
Notice that the last point implies the internal geodesics only turn up in the wavefunction.

There are two ingredients to this map.
The first is the identification between lengths and Liouville momenta of boundary primaries, and the second the identification of a pair of pants with a normalised OPE coefficient.
The first identification can be motivated by the relation between Wilson loops in the bulk and Verlinde loops in the boundary \cite{Jackson:2014nla}.
The bulk Wilson loop $P \exp{\oint_{\gamma} \tr \mathds{A}}$ in the homotopy class $\gamma$ evaluates to $\cosh \ell_{\gamma}$.\footnote{
	More precisely, it evaluates to $\cosh 2\pi (r_{+} + r_{-})$ without the $i$.
}
When deformed to a homotopic component of the boundary, it becomes a Verlinde loop (which in particular commutes with the Virasoro algebra), and it evaluates to $\cosh 2\pi b P_{\gamma}$.
For external geodesics, the homotopic component of the boundary is where the actual Hilbert space lives and the Verlinde loop is a self-adjoint operator --- so the identification is proved for these geodesics.
For internal geodesics, however, the homotopic component is either a part of the Euclidean path integral preparing the state or may not even exist depending on the bulk solution.
In this case, the operator is not self-adjoint and so this argument is not sufficient to prove the identification.\footnote{
	The most explicit way to see this is to try and use the CFT coproduct \cite{Gaberdiel:1993td} to deform the operator to the open ends.
	This coproduct explicitly doesn't preserve self-adjointness.
}
The Wilson loop for open geodesics has been formulated in \cite{Takayanagi:2020njm}.
We will give a full justification below that will justify both ingredients in our proposal.

Before proceeding to this justification, we need to address an important property of the map, which is that it is not isometric, $V^{\dagger} V \neq \mathds{1}$.
This is immediately obvious from the $\delta$-functions above: fixed-area basis elements that have infinite norm can map to $0$ if the lengths are not in the spectrum.
The other important reason for non-isometry is that the OPE coefficients can deviate from the OPE statistics.
The isometry domain of this map includes semiclassical states, where the fluctuations from both of these sources average out.

An important example of semiclassical states are EPI states, states created by a 2d surface $\Sigma_{\mathrm{k}}$ with fixed (and generic) moduli.
In the boundary, such states take the form \eqref{eqn:ope-block-Psi}.
In the bulk, the wavefunctions of these states are WdW wavefunctions like \eqref{eqn:wdw-fa}, with Dirichlet (or conformal Dirichlet) boundary conditions on $B$.
We expect that in these states (for generic moduli), all lengths are smeared over an $\mathcal{O} (\sqrt{G_{N}})$ window by the wavefunction.
Whenever a given geodesic is the minimal one in its homology class, this can be shown from the behaviour of the Renyi entropies \cite{Bao:2018pvs}.

Another example of semiclassical states are those where all the lengths are smeared over microcanonical windows, with standard deviation $\sigma_{\ell} \sim G_{N}$ (so that the standard deviation in the conformal weights $\sim G_{N}^{0}$).
This is a wavefunction $\psi(\ell_{\gamma})$ that satisfies
\begin{align}
	\forall \gamma, \qquad \qquad \qquad \qquad \qquad  \ \int \dd[m]{\ell} \ell_{\gamma} \abs{\psi(\ell_{\gamma})}^{2} &\sim G_{N}^{0}, \nonumber\\
	\int \dd[m]{\ell} \ell_{\gamma}^{2} \abs{\psi(\ell_{\gamma})}^{2} - \left[ \int \dd[m]{\ell} \ell_{\gamma} \abs{\psi(\ell_{\gamma})}^{2} \right]^{2} &\sim G_{N}, \nonumber\\
	\int \dd[m]{\ell} \psi^{*} \overset{\leftrightarrow} {\partial}_{\ell_{\gamma}} \psi &\lesssim G_{N},
  \label{eqn:mucan-wavefn}
\end{align}
where $m$ is the number of geodesics in $\Sigma$.
Note that we are defining the variances and expectation values with respect to the kinematical inner product.
EPI states are superpositions over such states with slowly varying phases.

The claim is that these states are within the isometry domain of our holographic map.
To see this, consider two wavefunctions $\psi,\phi$ satisfying \eqref{eqn:mucan-wavefn}.
The inner product between their images under $V$ is given by a sum over OPE coefficients.
For a four-boundary example, suppressing the antiholomorphic weight,
\begin{align}
	\int \dd[10]{\ell} \phi^{*} (\ell') \psi (\ell) &\pmel{\input{figs/fa-4-bd-sp.tex}}{V^{\dagger} V} {\input{figs/fa-4-bd-s.tex}} \nonumber\\
   &= \sum_{P_{1\dots 4,s,s'}} \frac{ \phi^{*} (P_{1\dots 4}, P_{s'}) \psi(P_{1\dots 4}, P_{s}) \input{figs/ss-ope-cont.tex}}{\prod_{I=1}^{4} \rho_{\mathrm{p}} (P_{I}) \sqrt{\rho_{\mathrm{p}} (P_{s}) \rho_{\mathrm{p}} (P_{s'})} \sqrt{\mathfrak{C}_{12s} \mathfrak{C}_{s34} \mathfrak{C}_{12s'}^{*} \mathfrak{C}_{s'34}^{*}}} \nonumber\\
																																																																				&\approx \int \dd[6]{P} \phi^{*} \psi \sqrt{\rho_{\mathrm{p}} (P_{s}) \rho_{\mathrm{p}} (P_{s'})} \frac{\mathds{E}_{\mu c} \left[ \input{figs/ss-ope-cont.tex} \right]}{\sqrt{\mathfrak{C}_{12s} \mathfrak{C}_{s34} C_{12s'}^{*} C_{s'34}^{*}}}.
  \label{eqn:four-bd-wormhole-V-ip}
\end{align}
The last line follows from the semiclassicality conditions, which ensure that the wavefunctions are slowly varying functions and so the OPE coefficients are the only rapidly varying parts of the sum.
We have also abused notation to convert the arguments of $\psi,\phi$ to Liouville momenta without taking care of the detailed conversion from lengths to Liouville momenta.
The conversion of the sum to the integral in the last line is $\sum_{P} \to \int \dd{P} \rho_{\mathrm{p}} (P)$ to account for the number of primaries in a microcanonical window.
By the conjecture of \cite{Hartman:2025cyj,Hartman:2025ula,Belin:2026pko},
\begin{equation}
	\mathds{E} \left[ \input{figs/ss-ope-cont.tex} \right] = \sqrt{\mathfrak{C}_{12s} \mathfrak{C}_{s34} \mathfrak{C}_{12s'}^{*} \mathfrak{C}_{s'34}^{*}} \input{figs/fa-4-bd-ss-ip.tex}
  \label{eqn:ope-conj}
\end{equation}
so that \eqref{eqn:four-bd-wormhole-V-ip} has the same transseries as $\pmel{\phi} {\eta} {\psi}$.
An analogous analysis holds for arbitrary inner products of semiclassical states.

The conjecture \eqref{eqn:ope-conj} in known to fail in some cases.
\begin{enumerate}
  \item The path integral is not dominated by identity in any channel.
		This happens for example when a bulk field condenses \cite{Belin:2017nze,Dong:2018esp}.
		In this case, GR is not a good effective theory and so the map is correct to break down.

  \item As mentioned previously, OPE statistics are calculated using the assumption of identity dominance, which is dual to the bulk solution being a handlebody.
		There are non-handlebody contributions which are important for the OPE statistics to have some properties related to the eigenstate thermalisation hypothesis \cite{Belin:2026pko}, and we do not yet have a CFT derivation of these contributions.
		So this is another possible way for \eqref{eqn:ope-conj} to fail.
		If these contributions can indeed be found in a single CFT, this is not a failure of our map.
		If they can't, then GR with the rule of summing over all topologies is again not a good effective theory.
  
  \item There are off-shell topologies, topologies without any solutions, which do contribute to OPE statistics \cite{Cotler:2020ugk,DiUbaldo:2023qli}.
		Here the problem is in how we are calculating the GPI, as a sum over saddle-points.
  
\end{enumerate}
Importantly, all the causes of non-isometry, both non-semiclassicality as well as these reasons, are physical: they are either failures of GR to be the correct bulk EFT or they are a mistake in defining the sum over topologies.
Thus, these are features rather than bugs of this map.

While the map we will study throughout is \eqref{eqn:hol-map-1} which maps the bulk state to a single holographic CFT, let us note that we can also construct a map to an ensemble of theories as follows.
Denote the members of the ensemble by $\alpha$.
Then, we construct a map to $\oplus_{\alpha} \mathcal{H}_{\mathrm{bd},\alpha}$ as
\begin{equation}
  V_{\mathrm{ens}} = \bigoplus_{\alpha} V_{\alpha}.
  \label{eqn:hol-map-ens}
\end{equation}
Inner products with this map become ensemble averages of sums over OPE coefficients.
This $\alpha$ label can be thought of as different $\alpha$ states a la \cite{Marolf:2020xie}.

\subsection{Tensor Network Representation} \label{ssec:tns}
We can represent the holographic map as a tensor network as follows.
For each closed geodesic in the fixed-area network, we introduce the operator
\begin{equation}
  \Pi_{\mathrm{spec}} = \int \dd[2]{\ell} \sum_{(P,\bar{P}) \in \operatorname{spec}} \frac{\tilde{\delta}^{(2)} (\ell - 2\pi bP)}{\sqrt{\rho_{\mathrm{p}} (P,\bar{P})}} \pket{\ell} \pbra{\ell}.
  \label{eqn:Pi-spec}
\end{equation}
We overload the symbol $\Pi_{\mathrm{spec}}$ to also mean the similar operator for open geodesics.
Next, we define the fanout operator on any Hilbert space as
\begin{equation}
  \mathrm{FANOUT} = \sum_{i} \ket{i} \ket{i} \bra{i},
  \label{eqn:fanout}
\end{equation}
where $i$ indexes a basis for the Hilbert space.
Then, define the tensors
\begin{align}
	\bra{C_{abc}} &= \sum_{(P_{i}, \bar{P}_{i})} \frac{C_{123}}{\sqrt{\mathfrak{C}_{123}}} \bra{(P_{1}, \bar{P}_{1}), (P_{2}, \bar{P}_{2}), (P_{3}, \bar{P}_{3})} \nonumber\\
	\bra{C_{a}^{A}} &=  \sum_{(P_{1}, \bar{P}_{1}, P_{2})} \frac{C_{1}^{2}}{\sqrt{\mathfrak{C}^{2}_{1}}} \bra{P_{1}, \bar{P}_{1}} \bra{P_{2}} \nonumber\\
	\bra{C^{ABC}} &= \sum_{P_{1,2,3}} \frac{C^{123}}{\sqrt{\mathfrak{C}^{123}}} \bra{P_{1}, P_{2}, P_{3}}.
  \label{eqn:C-tensor}
\end{align}
The indices on the LHS are abstract indices similar to those used in GR; their only role is to tell us what sort of OPE coefficient is appearing.

A fixed-area network is a graph with five types of vertices and two types of edges.
We denote the sets of vertices as $V_{abc}, V_{a}^{A}, V^{ABC}, V_{a}, V^{A}$, where the indices are again abstract indices that agree with the OPE coefficients; the single index vertex sets are where the CFT lives.
All the vertices are denoted $V$ and the type of vertex is denoted $t (V)$.
The two edge sets are $E_{\mathrm{c}}, E_{\mathrm{o}}$, where the $\mathrm{c}$ and $\mathrm{o}$ subscripts means closed and open respectively.
There are obvious adjacency rules in the graph arising from matching the type of geodesic to the type of vertex.
For an edge $e$, $\partial e = v_{1} \cup v_{2}$ and $\mathrm{FANOUT}_{e}: \mathcal{H}_{\mathrm{prim}} (e) \to \mathcal{H}_{\mathrm{prim}} (e,v_{1}) \otimes \mathcal{H}_{\mathrm{prim}} (e,v_{2})$.
We will denote the set of edges that end at a vertex $v$ as $E(v)$ and also define $\mathcal{H}_{\partial E(v)} \equiv \otimes_{e \in E(v)} \mathcal{H}_{\mathrm{prim}} (e,v)$.
With these definitions, the holographic map is
\begin{equation}
  V = \left( \otimes_{v \in V} \bra{C_{t(v)}}_{\partial E(v)} \right) \left( \otimes_{e \in E} \mathrm{FANOUT}_{e} \Pi_{\mathrm{spec}} \right).
  \label{eqn:hol-map-tn-eqn}
\end{equation}

Diagrammatically, we denote $\Pi_{\mathrm{spec}}$ by a dashed line, each $FANOUT$ by a trivalent vertex, and bras by triangular nodes with edges below.
For the four-boundary wormhole case, our holographic map is
\begin{equation}
  V = \input{figs/tn.tex}
  \label{eqn:hol-map-tn}
\end{equation}

This tensor network is qualitatively similar to random tensor network models of multiboundary wormholes that have appeared before \cite{Akers:2022zxr}.
The main difference is that the random tensors are replaced by pseudo-random OPE coefficients.
This network is more closely related to that in \cite{Chandra:2023dgq}, except that we are not restricting attention to an $\mathcal{O} (1)$ code subspace.
Notice that there is no link in this network whose bond dimension measures the value of the $t$ geodesic; this is a manifestation of the fact that intersecting lengths do not commute \cite{Bao:2018pvs}.

\section{Bulk Gauges and Emergent Bases} \label{sec:gauge}
Another interesting output of our story is that it allows us to study certain Hamiltonian gauge transformations in the bulk.
Our main finding is a refinement of the observation in \cite{Parrikar:2025xmz}, where they related Hamiltonian gauge transformations to changes of basis in the physical Hilbert space.
The first refinement is that we study this story in the full non-perturbative CFT Hilbert space.
The second is that, here, we find that it is dual not to a change of exact basis but to a change of an emergent basis.

\paragraph{A Story of Gauge and Basis}

Let us first summarise the major insight of \cite{Parrikar:2025xmz}.
The usual perspective on the Hilbert space of gauge theories is as follows.
There is an extended Hilbert space $\mathcal{H}_{\mathrm{ext}}$ that arises from quantising both physical as well as redundant degrees of freedom.
The physical Hilbert space $\mathcal{H}_{\mathrm{phys}}$ is isomorphic to the subspace $\mathcal{H}_{\mathrm{inv}} \subset \mathcal{H}_{\mathrm{ext}}$ that is annihilated by the gauge constraints.
However, completely fixing gauge\footnote{
	We are assuming that neither are there residual gauge freedoms nor are there physical states excluded by the gauge condition.
	Finding such a gauge is hard.
} gives a different subspace $\mathcal{H}_{\mathrm{fix}} \subset \mathcal{H}_{\mathrm{ext}}$ that is also isomorphic to $\mathcal{H}_{\mathrm{phys}}$.
Each gauge, as well as smearing over different gauge conditions, provides such a representative of the physical Hilbert space.
\cite{Parrikar:2025xmz} turned this around: they pointed out that each gauge provides a different \emph{parametrisation} of the physical states.
A basis of $\mathcal{H}_{\mathrm{fix}}$ is after all also a basis for $\mathcal{H}_{\mathrm{phys}}$ because of the isomorphism; thus, the natural bases in different gauges give different bases for $\mathcal{H}_{\mathrm{phys}}$.
This concurs with the oft-repeated slogan that a gauge is merely an aspect of description: a basis is also an aspect of description so this is a relation between like and like.
Of course, the more precise statement gives a little more than that: relational observables are naturally understood as those that are close to diagonal in a particular basis (or, more generally, observables that are defined using a certain basis).

\cite{Parrikar:2025xmz} used this perspective to analyse the perturbative Hilbert space of JT gravity, where $\mathcal{H}_{\mathrm{inv}}$ was defined as the one invariant under the Hamiltonian and diffeomorphism constraints, with $\Sigma$ having the topology of an interval.
There are also non-perturbative gauge transformations \cite{Jafferis:2017tiu} which relate slices of different topologies (or relate slices with the same topology in a topologically non-trivial way); these play an important role in the unreasonable effectiveness of semiclassical gravity \cite{Marolf:2020xie}.
In particular, invariance under these is required for GPI answers to match coarse-grained boundary answers, see e.g. \cite{Iliesiu:2024cnh}.
Here, we will analyse perturbative and non-perturbative gauge transformations in the bulk as basis changes and map them to the boundary using our holographic map.\footnote{
	One could imagine applying the holographic map of \cite{Iliesiu:2024cnh} to the results of \cite{Parrikar:2025xmz}, but there is an important technical issue: the York evolution of \cite{Parrikar:2025xmz} introduces an infinite phase to the wavefunction, making comparison between different gauges tricky.
	We will not run into this issue in our bases, though on the other hand we also don't have a continuous one-parameter family of Hamiltonian gauge transformations.
}

\paragraph{Perturbative Gauge Transformations}

\begin{figure}[h!]
  \centering
  \includegraphics[width=\linewidth]{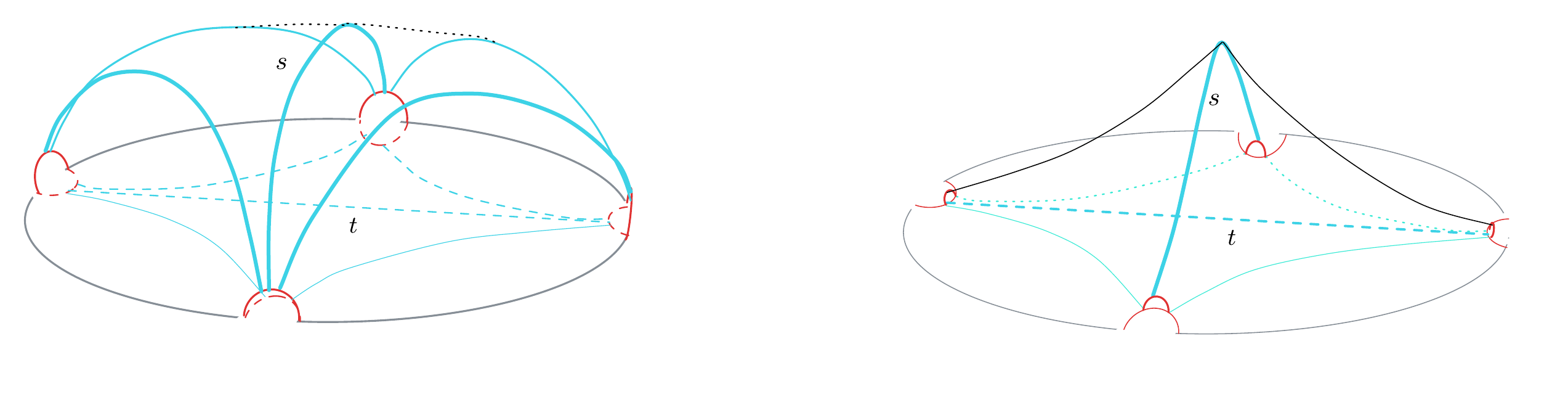}
  
  \caption{An $s-t$ crossing transformation changes the Cauchy slice.}
  \label{fig:F-cauchy}
\end{figure}

Let us begin with the observation, already mentioned in section \ref{sec:H-bulk}, that changing the set of geodesics whose lengths we fix is a gauge transformation.
We again focus on a four-boundary case, but with four interval boundaries instead of circles.
An output of our analysis in section \ref{sec:H-bulk} is that
\begin{equation}
  \pip{\input{figs/fa-4-bd-I-t.tex}}{\Psi} = \int \dd{\ell_{t}} e^{\frac{\ell_{s} - \ell_{t}}{8 G_{N}}} \input{figs/fa-tet-open.tex}
	\pip{\input{figs/fa-4-bd-I-s-no-l.tex}}{\Psi}.
  \label{eqn:basis-change}
\end{equation}
The basis transformation is given in the GPI by a codimension-zero region of spacetime, which changes the final slice, as shown in figure \ref{fig:F-cauchy}.
Thus, this is an example of a Hamiltonian gauge transformation in the bulk.

We are not fixing the lapse function between the slices, and so the equations of motion can determine it to be of any value.
Depending on the various lengths, the lapse being real, imaginary or $0$, corresponding to the region can be Euclidean, Lorentzian or even codimension-one \cite{Liu:2025tzv}.
Let us fix $\ell_{1\dots 4,s}$.
For small $\ell_{t}$ the codimension-one region is Euclidean.
The volume of this region keeps decreasing till it reaches a critical value, where it vanishes.
In this case, both the $s$ and the $t$ geodesics live on the same slice.
As we increase $t$ further, we find that the region becomes Lorentzian and thus the lapse becomes real.

The question now arises: what is the boundary dual of this gauge transformation?
The first point to note is that there is no dual at the level of individual basis elements, since
\begin{equation}
  V \pket{\input{figs/fa-4-bd-I-s-no-l.tex}} \propto V \pket{\input{figs/fa-4-bd-I-t.tex}}.
  \label{eqn:hol-map-b-el-prop}
\end{equation}
(Sometimes the proportionality constant is $0$.)
This follows from the fact that $\ell_{s,t}$ appear only in the wavefunctions.
We should not be worried about this failure, since these are not semiclassical states.

So let us take an overcomplete basis of semiclassical states, as is familiar from the text book harmonic oscillator example.
For this topology, one choice of overcomplete basis is given by $s$-channel fixed-area networks where all lengths are smeared over microcanonical windows.
Denoting this smearing by an overline over the fixed-area network, the image of this set is
\begin{equation}
	\left\{ V \pket{ \overline{ \input{figs/fa-4-bd-I-s-no-l.tex}}} \right\},
  \label{eqn:mucan-basis}
\end{equation}
This forms an overcomplete basis in the boundary as well, since all the states in $\mathcal{H}_{\mathrm{prim}}$ appear many times in the set.
Further, this overcomplete basis also has the property of approximate orthogonality, since the leading-order contribution to an inner product between different values of $\ell_{s}$ is given by
\begin{equation}
	\mathds{E}_{\mu c} \left[ \input{figs/ss-bcft-ope-cont.tex} \right] \approx \delta (P_{s} - P'_{s}) \mathfrak{C}_{12s} \mathfrak{C}_{s34}.
  \label{eqn:mucan-diag-ip-ope}
\end{equation}
We will call the set \eqref{eqn:mucan-basis} an \emph{emergent} basis.
The gauge transformation shown in figure \ref{fig:F-cauchy} can then be understood as a transformation between emergent bases.

Semiclassical states are superpositions over these emergent basis elements with slowly varying wavefunctions.
A natural set of conditions is given in \eqref{eqn:mucan-wavefn}.
If we smear each length over an $\mathcal{O} (G_{N})$ window, we land on the fixed-area states discussed in \cite{Dong:2018seb,Akers:2018fow,Dong:2019piw,Marolf:2020vsi,Dong:2023xxe,Dong:2022ilf}.
An element of the set \eqref{eqn:mucan-basis} is in fact the boundary dual of such a fixed-area state.
For a fixed-area state where $\ell_{s}$ is fixed, $\ell_{t}$ has larger fluctuations \cite{Bao:2018pvs,Kaplan:2022orm}, since the two do not commute.

There is a special subclass of `classical' states where both $\ell_{s}$ as well as $\ell_{t}$ have the same $\mathcal{O} (\sqrt{G_{N}})$ uncertainty.
By our terminology, the usual coherent states of the harmonic oscillator are classical states whereas squeezed coherent states are semiclassical states.
The paradigmatic examples of these classical states are EPI states with fixed moduli.
In this case, the state has parametrically similar spread in both $\ell_{s}$ as well as $\ell_{t}$.
For fixed-area states, the state is more classical in one basis compared to the other, whereas EPI states are equally classical in both bases.
These are the gauge transformations of the classical theory: basis transformations where the spread of the wavefunction is preserved.

\paragraph{Non-Perturbative Gauge Transformations}

Let us now focus on these classical states and consider the case of non-perturbative gauge transformations \cite{Jafferis:2017tiu}.\footnote{
	The example of the non-perturbative gauge transformation explored in \cite{Jafferis:2017tiu} is not actually accessible in our formulation, see around \eqref{eqn:jafferis-case}.
	However, we are able to reproduce any non-perturbative gauge transformations where both ket and bra have same lengths for external geodesics.
}
These non-perturbative gauge transformations are bulk manifestations of the fact that inequivalent classical limits are not orthogonal.

To set this up, we harken back to \eqref{eqn:phase-space-disjoint-union}, which states that the phase space of solutions with a fixed number of boundaries is a disjoint union of many connected components.
While we initially quantised each of the components separately in the fixed-area basis, non-perturbative contributions to the inner product ensure that these do not give orthogonal sectors of $\mathcal{H}_{\mathrm{bulk}}$.
Our holographic map maps all of these connected components to a single Hilbert space $\mathcal{H}_{\mathrm{bd}}$, where they are not orthogonal.
Thus, the non-perturbative gauge transformations are bulk manifestations of the fact that all these different phase spaces emerge from the same Hilbert space.

Let us see this in more detail.
Much as in the case of the harmonic oscillator, a set of classical states provides us with a classical limit.
This classical limit is given by a set of almost orthogonal states such that time-evolution leaves you within that set \cite{klauder1963continuous1,klauder1963continuous2}.
The fact that the set is invariant under time-evolution then allows us to interpret it as a phase space \cite{Ashtekar:1997ud}.
Thus, the phase space that emerges in the classical limit is nothing but a submanifold of Hilbert space.
The fact that the classical phase space has many connected components is the fact that there are many such non-intersecting sets.
While they don't intersect, they are also not orthogonal.

For example, with four boundaries we can consider either a connected Cauchy slice $\Sigma_{4}$ with four-boundaries, or a two-component Cauchy slice $\Sigma_{2} \cup \Sigma_{2}$ with two boundaries each.
The corresponding states map to
\begin{align}
  V \int \dd[5]{\ell} \psi_{4} \pket{\input{figs/fa-4-bd-I-s-no-l.tex}} &= \sum_{P_{1,\dots 4,s}} \psi_{4} N(P) \input{figs/4-bd-I-s.tex} \ket{P_{1} \dots P_{4}} \nonumber\\
  V \int \dd[2]{\ell} \psi_{2,2}  \pket{
\begin{tikzpicture}[baseline=-.5cm]
	\draw[double,cyan] (  0,0) to[out=-90,in=-90] node[pos=.5,above] {\small $1$} (  1,0);
	\draw[double,cyan] (-.5,0) to[out=-90,in=-90] node[pos=.5,below] {\small $2$} (1.5,0);
\end{tikzpicture}
} &=  \sum_{P_{1}, P_{2}} \psi_{2,2} N(P) \ket{P_{2}, P_{1}, P_{1}, P_{2}}.
  \label{eqn:two-tops-map}
\end{align}
We have again collected various factors in \eqref{eqn:hol-map-1} into the factor $N(P)$.
We see that classical states in the $\Sigma_{2} \cup \Sigma_{2}$ topology are ones where the $1,4$ and the $2,3$ boundaries have the same primary.
In the bulk, this is implemented by a gauge constraint (or, equivalently, the fact that both boundaries share a homotopic geodesic).
This gauge constraint, that the ADM masses on both boundaries agree, is thus emergent (see \cite{Akers:2024ixq} for a tensor network model).
On the other hand, the classical states in the $\Sigma_{4}$ topology do not exhibit this emergent gauge constraint.
This validates the assertion above that these are two different sets of classical states in the same Hilbert space.
In the $G_{N} \to 0$ limit, these become two different connected components of phase space.
Notice also that these components have different dimensionality: the emergent phase space is only four-dimensional for $\Sigma_{2} \cup \Sigma_{2}$ whereas it is ten-dimensional for $\Sigma_{4}$.

\paragraph{Summary}

To summarise this discussion, perturbative gauge transformations are transformations between overcomplete `emergent' bases that parametrise a given classical limit; whereas non-perturbative gauge transformations are transformations between inequivalent classical limits, and they encode the fact that all these different connected components of phase space are non-orthogonal in the Hilbert space.

The boundary dual of this change of basis corresponding to a perturbative gauge transformation is \emph{not} a change of basis, since $P_{s,t}$ never appear in the boundary basis elements, only in the wavefunctions.
Further, the bulk basis change is not even approximately a basis change in the boundary, since $V \pket{s}$ and $V \pket{s'}$ (for example) are not even linearly independent and so are very far from forming a basis.

Finally, as an aside, let us note that we can also define a notion of classicality of a basis using the isometry of the action of the holographic map.
The holographic map is highly non-isometric when acting on our fixed-area basis elements, as already pointed out.
But if we go to a basis for $\mathcal{H}_{\mathrm{bulk}}$ that has appreciable spread in the fixed-area basis, then we might find that the holographic map acts more isometrically on them.
For example, in JT gravity, the action of $V$ on fixed-energy basis states is very similar to \eqref{eqn:hol-map-1}, in that it annihilates most of them \cite{Iliesiu:2024cnh}.
However, in the fixed-length basis, the map doesn't annihilate any of them, and even orthogonality is well-preserved.
\cite{Basu:2024tgg,Basu:2025mmm} explores a similar notion using Wigner negativity, and it would be interesting to understand the connection.

\section{Interior Operators} \label{sec:rel-ops}
We now study semiclassical gauge-invariant interior operators using our map.
The natural operators to reconstruct are non-perturbatively relational operators, which measure a length given a slice topology.
The reconstruction is defined as a boundary operator that approximately agrees with the bulk operator on semiclassical states.
There is no unique reconstruction, but we find one prescription that has the right behaviour at our level of approximation, assuming that OPE densities satisfy typicality \cite{Foini:2018sdb}.

We first begin with an argument that the length operators of interior geodesics that are naturally defined using the bulk description are not linear operators in the boundary.
Consider again the four-boundary wormhole.
The natural bulk operator that measures $\ell_{s}$ is one that acts as
\begin{equation}
  \hat{\ell}_{s} \pket{\input{figs/fa-4-bd-s-no-l}} = \ell_{s} \pket{\input{figs/fa-4-bd-s-no-l}},
  \label{eqn:ell-s-op-bulk}
\end{equation}
But one can also measure $\ell_{t}$, where $t$ denotes the homotopy class surrounding $B_{2,3}$.
It is easy to show that $\left[ \hat{\ell}_{s}, \hat{\ell}_{t} \right] \neq 0$, since they have different eigenstates (the inner product is given by the double tetrahedron as discussed in \eqref{eqn:fa-st-xing-main}).
However, all eigenstates of $\hat{\ell}_{s}$ and all eigenstates of $\hat{\ell}_{t}$ --- the fixed-area networks where the corresponding lengths are fixed --- map to the \emph{same} state in the boundary, differing only in overall factors.
Thus, there is no linear operator on the boundary that has a similar action as these length operators on all of $V \mathcal{H}_{\mathrm{bulk}}$.\footnote{
	A non-linear operator is allowed to have different actions on $\ket{\psi}$ and $\alpha \ket{\psi}$.
}
However, we will only need operators that agree with $\hat{\ell}_{s}$ on semiclassical states.

\paragraph{Bulk Length Operators}

First, let us define the bulk length operator more carefully.
In terms of fixed-area basis elements, we define it as
\begin{equation}
  \hat{\ell}_{s} = \int \dd[5]{\ell} \ell_{s} \pket{\input{figs/fa-4-bd-s.tex}} \pbra{\input{figs/fa-4-bd-s.tex}} \eta.
  \label{eqn:ls-defn}
\end{equation}
This is the analogue of writing $\hat{x} = \int \dd{x} x \ket{x} \bra{x}$ in quantum mechanics, apart from the presence of the rigging map.
The rigging map makes the action of this operator well-defined on four-boundary wormholes of different topologies; since different topologies are not orthogonal, we don't want the operator to annihilate any other topologies.
Now let us see that this definition reproduces \eqref{eqn:ell-s-op-bulk}:
\begin{align}
  \hat{\ell}_{s} \pket{\input{figs/fa-4-bd-s.tex}} &= \int \dd[5]{\ell'} \prod_{\gamma=1}^{4} \delta(\ell_{\gamma} - \ell'_{\gamma}) \ell'_{s} e^{\frac{\ell_{s} + \ell'_{s}}{8 G_{N}}} \pket{\input{figs/fa-4-bd-sp.tex}} \input{figs/fa-4-bd-ss-ip.tex} \nonumber\\
																									&\approx \ell_{s} \pket{\input{figs/fa-4-bd-s.tex}},
  \label{eqn:ls-ent-fact-just}
\end{align}
where we have used \eqref{eqn:4-bd-ip-s3-hom} as the leading contribution in the last line.

More generally, for two slices $\Sigma_{\mathrm{k}}, \Sigma_{\mathrm{b}}$, both containing four boundaries, we have the matrix elements
\begin{align}
	\pmel{\input{figs/Sigma-b.tex}}{\eta \hat{\ell}_{s}}{\input{figs/Sigma-k.tex}} = \prod_{i=1}^{4} \delta (\ell_{i} - \ell'_{i}) \int \dd{\ell_{s}} &e^{\frac{2\ell_{s} + \sum_{\gamma \text{ internal}} \ell_{\gamma}}{8 G_{N}}} \ell_{s} \input{figs/Sigma-o.tex}\,,
  \label{eqn:l-op-fa}
\end{align}
where all geodesics with the same label are identified across the two diagrams, so that there is only one copy of the $1,2,3,4,s$ geodesics (but the internal angles are counted twice, as per the diagram).\footnote{
	In the language of \cite{Belin:2026pko}, we have to perform cylinder surgery across the lines with identical labels.
}

As an example, consider $\Sigma_{\mathrm{k}}, \Sigma_{\mathrm{b}}$ to both be four-boundary wormholes in the $t$-channel.
Then, we find at leading order
\begin{align}
  \pmel{\input{figs/fa-4-bd-tp.tex}}{\eta \hat{\ell_{s}}}{\input{figs/fa-4-bd-t.tex}} &\approx  \prod_{i=1}^{4} \delta(\ell_{i} - \ell'_{i}) \int \dd{\ell_{s}} \ell_{s} \input{figs/fa-tetp.tex} \input{figs/fa-tet.tex} \nonumber\\
  &= \prod_{i=1}^{4} \delta(\ell_{i} - \ell'_{i}) \int \dd{\ell_{s}} \ell_{s} \abs*{
		\begin{Bmatrix}
		1 & 2 & s \\
		3 & 4 & t' \\
		\end{Bmatrix}
	}^{2}
	\abs*{
		\begin{Bmatrix}
		1 & 2 & s \\
		3 & 4 & t \\
		\end{Bmatrix}
	}^{2}
  \label{eqn:l-op-fa-t-s}
\end{align}
The solution is given by two double tetrahedrons with the $23s$ and $14s$ faces identified between them.

More generally, we might want to calculate the matrix element of some function of the lengths on a slice $\Sigma_{\mathrm{o}}$, which we call $F(\ell_{\mathrm{o}})$.
Then, we identify the external geodesics of $\Sigma_{\mathrm{o}}$ to those of $\Sigma_{\mathrm{k}}$ and also to those of $\Sigma_{\mathrm{b}}$, use the resulting network as `boundary conditions' and integrate over the lengths of internal geodesics in $\Sigma_{\mathrm{o}}$ with the weight $\exp(\sum_{\gamma_{\mathrm{o}} \text{ internal}} \ell_{\gamma_{\mathrm{o}}}/4 G_{N})$.
We put boundary conditions in quotes since $\Sigma_{\mathrm{o}}$ can be in the interior of the solution.
In the language of \cite{Hartman:2025cyj,Hartman:2025ula}, we are summing over all manifolds whose large triangulations admit the network as a subgraph.

One might wonder why one needs all of $\Sigma_{\mathrm{o}}$ in the middle; the answer is that this projects onto the classical limit described by $\Sigma_{\mathrm{o}}$ and our length operators are only defined in this classical limit.
One can think of this as the operator being relational --- ``what is the length of this geodesic \emph{given} the slice topology?''
There's an important difference from usual relational operators, however.
Relational definitions are useful because they allow us to define operators that are invariant under some gauge constraint; in this case, the addition of all of $\Sigma_{\mathrm{o}}$ is allowing us to define it in a way invariant under \emph{non-perturbative} gauge transformations.

\paragraph{Boundary Reconstruction}

Now, let us turn to the reconstruction of this operator in the boundary theory.
To reconstruct $\hat{F}(\ell_{\mathrm{o}})$, we need to find an operator $\mathcal{R}^{*}_{g,n} (\hat{F})$ (here, $g$ and $n$ are the genus and the number of boundaries of $\Sigma_{\mathrm{o}}$) such that \cite{Akers:2025ynh}
\begin{equation}
  \forall \phi,\psi\ s.t.\ \pmel{\phi} {\eta} {\psi} \approx \pmel{\phi} {V^{\dagger} V} {\psi} \quad \implies \quad \pmel{\phi} {\eta \hat{F} (\ell_{o})}{\psi} \approx \pmel{\phi} {V^{\dagger} \mathcal{R}_{g,n}^{*} (\hat{F}) V} {\psi}.
  \label{eqn:rec-map-defn}
\end{equation}

One way to do this uses a trick developed in \cite{Soni:2025qau}.
We will restrict to the case of $\hat{\ell}_{s}$ in a four-boundary wormhole for simplicity, but the arguments below generalise straightforwardly.
We first take a partition of unity $f_{\ell} (P)$ such that
\begin{equation}
  \sum_{P} f_{\ell} (P) = 1 \ \forall \ell, f_{\ell} (P) > 0 \quad \text{and} \quad f_{\ell} (P) \text{ is peaked at } P = \frac{\ell}{4 \pi b}  \text{ with width of order } G_{N}^{0}.
  \label{eqn:partition-of-unity}
\end{equation}
Then define the microcanonically smeared states
\begin{equation}
  \ket{\overline{\ell_{I}}} = \sum_{P_{I}} N (P_{I}) \input{figs/ope-coeff-diag.tex} \sqrt{\prod_{I = 1\dots4,s} f_{\ell_{I}} (P_{I})} \ket{P_{1}, \dots P_{4}}.
  \label{eqn:mic-states}
\end{equation}
Here, $N(P_{I})$ collects the density of states and OPE coefficient factors in \eqref{eqn:hol-map-1}.
The operator we define is
\begin{equation}
	\mathcal{R}^{*}_{0,4} (\hat{\ell}_{s}) = \int \dd[5]{\ell} \ell_{s} \sum_{P_{I}} \abs{N(P_{I}) }^{2} \input{figs/ope-coeff-op.tex} \left( \prod_{I} f_{\ell_{I}} (P_{I}) \right)  \ket{P_{I}} \bra{P_{I}}.
  \label{eqn:rec-map-explicit}
\end{equation}

The reason \eqref{eqn:rec-map-explicit} satisfies \eqref{eqn:rec-map-defn} is that the factor of $\ell_{s}$ inside the integral in \eqref{eqn:rec-map-explicit} varies slowly and therefore doesn't change the saddle-point, which ends up reproducing \eqref{eqn:l-op-fa}.
Denoting by $\Sigma_{\mathrm{k},\mathrm{b},\mathrm{o}}$ patterns of contracting OPE coefficients (i.e. the cyan lines become black dashed lines) and collecting all OPE coefficient and density of states factors in $\tilde{N}$, we find
\begin{align}
	& \sbox0{$\overline{\input{figs/Sigma-b-ope.tex}}$} \mathopen{\resizebox{1.2\width}{\ht0}{$\Bigg\langle$ }} \usebox{0} \mathclose{\resizebox{1.2\width}{\ht0}{$\Bigg|$ }}
	\mathcal{R}^{*}_{0,4} (\hat{\ell}_{s})
	\sbox0{$\overline{\input{figs/Sigma-k-ope.tex}}$} \mathopen{\resizebox{1.2\width}{\ht0}{$\Bigg|$ }} \usebox{0} \mathclose{\resizebox{1.2\width}{\ht0}{$\Bigg\rangle$ }}
 \nonumber\\
	& \qquad \qquad \approx \prod_{i=1}^{4} \delta (P_{i} - P_{i}') \sum_{P_{s}} 4 \pi b P_{s} \tilde{N} \mathds{E}_{\mu c} \left[ \input{figs/Sigma-b-o-ope.tex} \input{figs/Sigma-k-o-ope.tex} \right]
	\label{eqn:l-op-exp-bd}
\end{align}
To evaluate this, we need to know how to calculate OPE densities when a primary appears more than once in the indices.
It is currently not known how to calculate these for a single CFT.

However, if we assume that a single holographic CFT satisfies the typicality requirement of \cite{Foini:2018sdb}, then we must have
\begin{equation}
	\mathds{E}_{\mu c} \left[ \input{figs/Sigma-bok-ope-2.tex} \right] = \frac{1}{\prod\limits_{I=1\dots4,s}\rho(P_{I})} \mathds{E}_{\mu c} \left[ \input{figs/Sigma-b-o-ope.tex} \right] \mathds{E}_{\mu c} \left[ \input{figs/Sigma-k-o-ope.tex} \right] \,.
  \label{eqn:typicality}
\end{equation}
This is exactly what the bulk path integral gives \cite{Belin:2026pko}.
In the example \eqref{eqn:l-op-fa-t-s}, the $e^{- \frac{\ell_{s}}{4 G_{N}}}$ cancels a factor in \eqref{eqn:l-op-fa}.
So we find that (under the assumption of typicality) \eqref{eqn:rec-map-defn} is satisfied by \eqref{eqn:rec-map-explicit}.

While we have focused on one specific operator for a specific slice topology, the argument generalises.
Thus, we have found a candidate reconstruction for length operators.
We are not claiming that it is the optimal one; there could be other reconstructions that satisfy \eqref{eqn:rec-map-defn} for a larger class of semiclassical states.\footnote{
	\cite{Akers:2025ynh} did find an optimal reconstruction in this sense, though this part of their claim is brought into question by the results of \cite{Preskill:2026guv}.
	Indeed, the analysis of \cite{Preskill:2026guv} shows that the set of semiclassical states can depend on highly non-universal features of the full theory, and therefore an optimisation of the sort attempted in \cite{Akers:2025ynh} would require an engagement with such details.
	We thank Shreya Vardhan for comments on this.
}
Further, we have not defined a reconstruction map for the conjugate of the length, i.e. the relative boost angle.
Due to the non-linearity of reconstruction, the map for the conjugate does not directly follow from the one above.
We postpone this to future work.

\section{Closed Universes} \label{sec:cus}
We now apply our holographic map to closed universes.
We will find a large breakdown of semiclassicality for closed universes without matter, which will get partially lifted when we add bulk matter.

As usual, we work with a simple example involving no angular momentum.
A closed universe is given by a fixed-area network with no open ends.
A simple closed universe state is
\begin{equation}
  \pket{\psi} = \int \dd[3]{\ell} \psi (\ell_{\gamma}) \pket{\input{figs/3-bd-cis-ip.tex}}.
  \label{eqn:cu-state}
\end{equation}
Its boundary dual is
\begin{equation}
  V \pket{\psi} = \sum_{P_{I}} \psi(P_{I}) N(P) \input{figs/g2-ope.tex}.
  \label{eqn:cu-state-V}
\end{equation}
Note that this is a number not a state in a Hilbert space; alternatively, the Hilbert space dimension is $1$, in agreement with arguments about closed universes \cite{Usatyuk:2024mzs,Usatyuk:2024isz,Harlow:2025pvj} and references therein.
This follows from the lack of open ends in this network, and therefore this is equally true for any closed universe topology.

The bulk inner product of two such states can be written in terms of boundary conditions for the GPI as
\begin{equation}
  \pmel{\phi} {\eta} {\psi} = \int \dd[3]{\ell} \dd[3]{\ell'} e^{\frac{\sum_{i} \ell_{i} + \ell'_{i}}{8 G_{N}}} \phi^{*} (\ell') \psi(\ell) \input{figs/3-bd-cis-ip-p.tex} \input{figs/3-bd-cis-ip.tex}.
  \label{eqn:cu-ip-bulk}
\end{equation}
The leading solution is that both the bra and the ket collapse into separate degenerate double tori.
Since all internal angles are $0$, we find
\begin{equation}
  \pmel{\phi} {\eta} {\psi} = \left( \int \dd[3]{\ell} e^{\frac{\sum_{i} \ell_{i}}{8G_{N}}} \phi(\ell) \right) \left( \int \dd[3]{\ell} e^{\frac{\sum_{i} \ell_{i}}{8 G_{N}}} \psi(\ell) \right) + \dots
  \label{eqn:cu-ip-bulk-leading}
\end{equation}
where the $\dots$ denote subleading terms.
We see that the leading order term factorises as expected (assuming that neither wavefunction integrates to $0$).
This has also been observed in \cite{Belin:2025ako}.
If we just take this leading answer, then we find that the bulk Hilbert space is one-dimensional, since after normalisation we find that the inner product between any two such states is a phase.
This lines up with the fact that the boundary dual is just a number.

But let us look at the first subleading term.
This is the one where the ket and bra slices get identified with each other, giving
\begin{equation}
  \pmel{\phi} {\eta} {\psi} \supset \int \dd[3]{\ell} \phi^{*} (\ell) \psi(\ell) = \pip{\phi} {\psi}.
  \label{eqn:cu-ip-bulk-subleading}
\end{equation}
This is exactly the same as the kinematical inner product!
The fact that the kinematical inner product is subleading in this case signals a breakdown of semiclassicality.

Let us try to make this more precise.
Take the states $\psi = \phi$ to be coherent states for the phase space of $\Sigma_{2,0}$ centred on some lengths $\bar{\ell}_{1,2,3}$ and the internal angles $\theta_{1,2,3} = \pi$ (this corresponds to a smooth slice).
In a semiclassical limit, we would expect the norm to be dominated by a geometry where the angles are $\pi$.
However, the internal angles in the leading solution are all $0$!
Thus, even for coherent states, the classical solution is unrelated to the `momentum' label.
This implies that there are no coherent states which form the classical phase space, and so there is no classical limit.

Further the subleading term that agrees with the kinematical inner product actually contradicts \eqref{eqn:cu-state-V}, since it gives non-trivial contributions to the inner product.
This is related to the typicality condition on ETH \cite{Foini:2018sdb}, which in the context of 2d CFT implies that
\begin{equation}
	\mathds{E}_{\mu c} \left[ \input{figs/g2-ope.tex} \input{figs/g2-ope-p.tex} \right] \supset \prod_{i = 1}^{3} \frac{\delta(P_{i} - P'_{i})}{\rho(P_{i})} \mathds{E}_{\mu c} \left[ \input{figs/g2-ope.tex}^{2} \right].
  \label{eqn:typicality-2}
\end{equation}
While the full inner product of course factorises, keeping only the averages can retain these sorts of contributions.
This implies that the bulk non-perturbative inner product agrees with a large-$N$ filter applied on the product of numbers \cite{Liu:2025ikq}.

Note that the leading term factorising and the existence of non-trivial subleading contributions is independent of the topologies of the bra and the ket.
In particular, they need not even have the same topology.
Thus, we have found that for any two states in the Hilbert space of closed universes, the bulk inner product is exponentially close to $1$.
This has also been found recently in \cite{Abdalla:2026mxn}.

There are at least three ways of restoring semiclassicality in the above sense.
The first is to consider bulk \emph{mixed} states,
\begin{equation}
  \rho_{\mathrm{bulk}} = \int \dd[3]{\ell} \rho(\ell) \pket{\input{figs/3-bd-cis-ip.tex}} \pbra{\input{figs/3-bd-cis-ip.tex}} \eta
  \label{eqn:bulk-mixed-state}
\end{equation}
This was explored in \cite{Belin:2025ako} and we refer the reader to that work for more details.
The second is to restrict attention to wavefunctions that integrate to $0$ in the leading term.

The third way of restoring semiclassicality is to include massive probes with mass below the black hole threshold.\footnote{
	This sort of object has been called an observer in some recent literature.
	However, there is no measurement theory associated to such a probe and so we refrain from using this word.
}
Each such probe generates a conical defect $\Delta \phi$ (the total conical angle is $2\pi + \Delta \phi$).
Based on \cite{Chandra:2022bqq,Hartman:2025cyj,Hartman:2025ula}, the natural extension of the holographic map to include these is to identify a defect with a primary whose conformal weight is
\begin{equation}
  h = \frac{c}{24} \left( 1 - \frac{\Delta\phi^{2}}{4\pi^{2}} \right).
  \label{eqn:hol-map-conical-defect}
\end{equation}

Denoting such a massive probe by a line that points horizontally, a simple such state is
\begin{equation}
  \pket{\psi} = \int \dd{\ell_{1}} \dd{\Delta \phi_{2}} \psi(\ell_{1}, \Delta\phi_{2}) 
\begin{tikzpicture}[baseline]
	\draw[cyan] (.5,0) arc(0:360:.5) node[pos=.5,left] {\small $1$};
	\draw[cyan] (0,0) -- node[above] {\small $2$} (.5,0);
\end{tikzpicture}
.
  \label{eqn:cu-probe}
\end{equation}
This denotes a slice whose topology is a once-punctured torus.
When we consider the inner product of two such states, there is no disconnected solution similar to \eqref{eqn:cu-ip-bulk-leading}, since there is only one conical defect on the surface and so that point cannot be identified with any other on the same surface.
This means that the kinematical term is the leading one.\footnote{
	This disagrees with the main conclusion of \cite{VanRaamsdonk:2026tnv}.
}
Semiclassicality follows.
The argument here remains true with multiple massive probes for generic masses and topologies.\footnote{
	If pairs of probes have the same mass, then this might not be true.
}

This can be seen in the boundary as follows.
The image of \eqref{eqn:cu-probe} under the holographic map is
\begin{equation}
  V \pket{\psi} = \sum_{P_{i}} \psi(P_{i}) N(P) C_{112}.
  \label{eqn:cu-probe-V}
\end{equation}
The microcanonical average of $C_{112}$ vanishes, meaning that
\begin{equation}
	\mathds{E}_{\mu c} \left[ C_{112} C^{*}_{1'1'2'} \right] = \frac{\delta(P_{1} - P'_{1}) \delta(P_{2} - P'_{2})}{\rho(P_{1}) \rho(P_{2})} \mathfrak{C}_{112} + \dots
  \label{eqn:cu-probe-ope}
\end{equation}
This agrees with the kinematical inner product.
Thus, we find that light probes can restore semiclassicality to closed universes.
It would be interesting to understand the connections between this observation and the observer rules recently proposed in \cite{Abdalla:2025gzn,Harlow:2025pvj,Akers:2025ahe}.

Note here that by semiclassicality, we mean the existence of coherent states that approximate the classical phase space.
If we consider multiple copies, then by the same arguments as in \cite{Usatyuk:2024mzs,Usatyuk:2024isz}, we will find non-trivial contributions to the inner product where closed universes get exchanged at the same order as we find the kinematical inner product.

\section{Discussion} \label{sec:disc}

Using recent progress in the bootstrap of holographic CFTs, we have proposed a holographic map from the semiclassical Hilbert space of pure GR in $\text{AdS}_{3}$ to the Hilbert space of holographic $\text{CFT}_{2}$.
This map is approximately isometric for semiclassical states where pure GR is the correct bulk EFT, but highly non-isometric in general.
We have also applied it to the study of the boundary dual of certain bulk gauge transformations, the reconstruction of interior operators, and closed universes.

We have focused on pure GR with one type of ETW brane in the bulk.
The generalisation to include sufficiently heavy point particles and more types of ETW branes should be straightforward using the techniques of  \cite{Chandra:2022bqq,Jafferis:2026gzn}.
Further, we also expect that our bulk Hilbert space can be upgraded to that of Virasoro TQFT without changing the main story (though matter corrections live at the same order as the VTQFT corrections).
However, the convergence of the sum over topologies has not been shown, which in turn implies that our inner product may diverge.
In examples where the sum over topologies has been performed \cite{Maloney:2007ud}, the sum diverges but can be appropriately regulated to give a convergent answer.
It is important to understand how this affects the validity of our proposal.

Another interesting direction is to compare the holographic map approach with the maximum ignorance principle approach of \cite{deBoer:2023vsm}.
\cite{deBoer:2023vsm} would map a pure state in the bulk to a mixed state in the boundary.
The physical differences between the two proposals remain to be understood.

It should also be possible to extend the tensor network construction in section \ref{ssec:tns} to construct a tensor network that allows us to find the entropies of arbitrary subregions by adding multipartite edge modes \cite{Akers:2024wab}.
First, we need to write $\mathcal{H}_{\mathrm{pert}}$ as a Levin-Wen-like model using the ideas in \cite{Chen:2024unp,Takahashi:2024ukk,Bao:2024ixc}.
Then, we need to add $SL^{+}_{q} (2, \mathds{R})$ edge modes \cite{Mertens:2022ujr,Wong:2022eiu,Krishnan:2026mpa}, using the techniques developed in \cite{Balasubramanian:2025rcr} to deal with the non-compactness of the quantum semigroup.
Finally, this will allow us to define a multipartite factorisation map.

There are many other directions forward.
On the CFT side, many assumptions about the OPE statistics we have made are yet unproven.
On the bulk side, it is important to also allow for the bulk EFT to contain quantum fields.
This will change the very definition of a fixed-area basis element or state, due to renormalisation effects \cite{Dong:2019piw}, and it will correct the results here in an important way.
Perhaps the developments in the Wilson spool program \cite{Castro:2023dxp,Bourne:2025azc,Bourne:2026jbx} will be relevant.

Further, there are many more things to do to flesh out the analyses in sections \ref{sec:gauge}, \ref{sec:rel-ops} and \ref{sec:cus}, as indicated in the respective sections.

\section*{Acknowledgements}
We thank
Arhum Ansari,
Stefano Antonini,
Jeevan Chandra,
Abhijit Gadde,
Anirban Ganguly,
Thomas Hartman,
Yuya Kusuki,
Alok Laddha,
Shiraz Minwalla,
Pranjal Nayak,
Sridip Pal,
Onkar Parrikar,
Suvrat Raju,
Pratik Rath,
Shashank Sengar,
Shreya Vardhan,
Tanmoy Sengupta,
and the authors of \cite{Belin:2026pko} for useful discussions.
We also thank Pratik Rath for comments on a draft.

This work has been presented at
14th Joburg Workshop on String Theory and 1st India/South Africa Joint String Meeting,
UC Berkeley \cite{bctptalk},
Stanford,
Chennai Strings Meeting 2025,
\href{https://sites.google.com/view/holographicuniverse2026}{``Holographic Universe 2026,'' YITP, Kyoto U},
and
\href{https://sites.iiserpune.ac.in/~suneeta/webpage.html}{``Advances in Black Hole Theory,'' IISER Pune}.
We thank the organisers and audiences of all these talks.

\appendix
\section{Examples of the Bulk Inner Product} \label{app:bulk-ip-egs}

Here we illustrate various examples of bulk inner products for slices with closed and open geodesics.

Below, $\Sigma_a^b$, through the subscript and superscript, will represent slices with $a$ closed and $b$ open geodesics respectively.
For example, $\Sigma_2$ will label a slice with 2 boundaries and 1 fixed length.
Likewise, $\Sigma^2$ will label a slice with 1 fixed length open geodesic with 2 interval boundaries.
$\Sigma_1^1$ will label a slice with 1 CFT and 1 BCFT boundary.   

Inner products involving slices with only closed geodesics will be our first focus.
Cases involving both open and closed geodesics will follow next.
A more systematic treatment of part of our discussion can be found in \cite{Belin:2026pko}.
Here we will tread some of the same ground, except at the level of classical saddles.

\subsection{Closed Geodesics} \label{ssec:egs-closed}

Here we will go through illustrations in the order of increasing number of CFT boundaries. In non-trivial cases, we will give multiple illustrations.

\paragraph{One Boundary}

In the case of one CFT boundary, the simplest example of a Cauchy slice which can be used to represent a network is a disk, $\Sigma_1 = D^{2}$.
However, this case has no closed geodesics.
We represent this topology by a point $\tikz{\node[cyan,fill,circle,inner sep=.5mm] (D) at (0,0) {};}$ in our line diagrams.
Its inner product with itself is trivially $1$.
More complicated examples of single boundary elements can arise when  genus $g > 0$, and when we have open geodesics.
We will see this later. 

\paragraph{Two Boundaries}

Consider now the two-boundary case with $g=0$.
In this case, there are two types of Cauchy slices, the connected wormhole $\Sigma_2 = S^{1} \times \mathds{R}$ and two disconnected disks $\Sigma_1 \cup \Sigma_1 = D^{2} \cup D^{2}$.
Therefore, the types of inner products are simple to enumerate. 

Consider first the case where both ket and bra are two-boundary wormholes, with no angular velocity.
Then there is only one external geodesic in each, so the core is forced to be empty and we find simply
\begin{equation}
  \pmel{
\begin{tikzpicture}[baseline=-.2cm]
	\draw[cyan] (0,0) to[out=-90,in=-90] node[pos=.5,below] {\tiny $\ell'$} (1,0);
\end{tikzpicture}
}{\eta} {
\begin{tikzpicture}[baseline=-.2cm]
	\draw[cyan] (0,0) to[out=-90,in=-90] node[pos=.5,below] {\tiny $\ell$} (1,0);
\end{tikzpicture}
} = \delta (\ell - \ell').
  \label{eqn:2-bd-ip}
\end{equation}

A more non-trivial example is the rotating case.
As discussed in section \ref{sec:H-bulk}, now there is an infinity of saddles due to the twist $\tau \in 2\pi \mathds{Z} r_{+}$.
 While this doesn't change anything about the saddle, it adds a phase factor due to the $i \tau r_{-}$ term in the action.
The sum over these quantises the angular momentum, since
\begin{equation}
  \pmel{
\begin{tikzpicture}[baseline=-.2cm]
	\draw[cyan] (0,0) to[out=-90,in=-90] node[pos=.5,below] {\tiny $\ell'$} (1,0);
\end{tikzpicture}
}{\eta}{
\begin{tikzpicture}[baseline=-.2cm]
	\draw[cyan] (0,0) to[out=-90,in=-90] node[pos=.5,below] {\tiny $\ell$} (1,0);
\end{tikzpicture}
}_{r_+ \neq  r_-} = \delta^{(2)} (\ell - \ell') \cdot  \sum_{n \in \mathds{Z}} e^{2\pi i n r_{+} r_{-}} = \delta^{(2)} (\ell - \ell') \cdot  \delta (r_{+} r_{-} \in 2 G_{N} \mathds{Z}).
  \label{eqn:2-bd-ip-J}
\end{equation}
There will be a similar sum over these Dehn twists at every fixed-area surface, quantising every angular momentum. 

Consider a third example now, involving $D^{2} \cup D^{2}$ case. This basis element consists of two copies of $\mathds{H}^{2}$, and its norm is $1$.
Here we can take the ket to be $S^{1} \times \mathds{R}$ and the bra to be $D^{2} \cup D^{2}$.
The inner product between them is given by a saddle that is topologically
\begin{equation}
  \pmel{
		\begin{tikzpicture}[baseline]
		  \node[cyan,circle,inner sep=.5mm,fill] (L) at ( 0,0) {};
		  \node[cyan,circle,inner sep=.5mm,fill] (R) at (.5,0) {};
		\end{tikzpicture}
	}{\eta}{
		
\begin{tikzpicture}[baseline=-.2cm]
	\draw[cyan] (0,0) to[out=-90,in=-90] node[pos=.5,below] {\tiny $\ell$} (1,0);
\end{tikzpicture}

	}
	=
	\exp \left\{ - I \left[
		\input{figs/jafferis-ip.tex} 
	\right] \right\}.
  \label{eqn:jafferis-case}
\end{equation}
However, notice that the blue line is contractible in the bulk.
This we will now explain is a contradiction, because in hyperbolic space there are no contractible geodesics.

One way to see this is to note that the length of a geodesic in a homotopy class is given by a Wilson loop in the $PSL(2,\mathds{R}) \times PSL(2,\mathds{R})$ Chern-Simons description of 3d GR \cite{Ammon:2013hba,Jackson:2014nla} which is a homotopy invariant.
Since the homotopy class is contractible, we can shrink the Wilson loop to zero size and get that the holonomy is given by the identity group element, which corresponds to zero length.
We thus conclude that there is no saddle for these boundary conditions.

We can summarise all the above cases neatly as follows. Noticing that $D^{2} \cup D^{2}$ is the same\footnote{
To see this, note that the $K_{ij} = 0$ slice metric of global $\text{AdS}_{3}$ as well as that of one exterior of a non-rotating BTZ can be put in the form $\dd{r}^{2}/f(r) + r^{2} \dd{\phi}^{2}$, where $f(r) = r^{2} + 1$ for global and $r^{2} - r_{h}^{2}$ for the black hole.
It is clear that taking $r_{h} = i$ gives us the global metric.
}
as $S^{1} \times \mathds{R}$ with $\ell = 2\pi i$, we can write
\begin{equation}
  \pmel{
\begin{tikzpicture}[baseline=-.2cm]
	\draw[cyan] (0,0) to[out=-90,in=-90] node[pos=.5,below] {\tiny $\ell'$} (1,0);
\end{tikzpicture}
}{\eta} {
\begin{tikzpicture}[baseline=-.2cm]
	\draw[cyan] (0,0) to[out=-90,in=-90] node[pos=.5,below] {\tiny $\ell$} (1,0);
\end{tikzpicture}
} = \delta^{(2)} (\ell - \ell') \delta (r_{+} r_{-} \in 2 G_{N} \mathds{Z})
  \label{eqn:2-bd-ip-gen}
\end{equation}
where $\ell, \ell'$ can now also take the value $2\pi i$ and the $\delta$-function is the natural one for the set of allowed values of $\ell$ (i.e. it is a Kronecker delta when $\ell = 2\pi i$).

\paragraph{Three Boundaries}

Let us now turn to the three-boundary case.
The types of slices are (a) $D^{2} \cup D^{2} \cup D^{2}$, (b) $D^{2} \cup S^{1} \times \mathds{R} = \Sigma_1 \cup \Sigma_2$ and (c) a three-boundary wormhole $\Sigma_3.$ 
The simplest inner product, $\pmel{\Sigma_{3}}{\eta} {\Sigma_{3}}$, is dealt with in the main text.

As a new example, consider the inner product involving $\Sigma_1 \cup\Sigma_2$. 
\begin{equation}
  \pmel{
\begin{tikzpicture}[baseline=-.2cm]
	\node[cyan,fill,circle,inner sep=.5mm] (1) at (0,0) {};
	\draw[cyan] (.5,0) to[out=-90,in=-90] node[below] {\tiny $\ell'$} (1,0);
\end{tikzpicture}
}{\eta} {
\begin{tikzpicture}[baseline=-.2cm]
	\draw[cyan] (.5,0) to[out=-90,in=-90] node[below] {\tiny $\ell$} (1,0);
	\node[cyan,fill,circle,inner sep=.5mm] (1) at (1.5,0) {};
\end{tikzpicture}
} = \delta(\ell-\ell') 
\begin{tikzpicture}[baseline]
	\node[cyan,fill,circle,inner sep=.5mm] (1) at (0,0) {};
	\node[cyan,fill,circle,inner sep=.5mm] (1) at (1,0) {};
	\draw[cyan] (0,0) to[out=90,in=90] (.5,0) to[out=-90,in=-90] (1,0);
\end{tikzpicture}
 = 0.
  \label{eqn:2-bd-ip-snake}
\end{equation}
Using the argument given for \eqref{eqn:jafferis-case}, the geodesics must be contractible in the bulk and therefore have zero length.

The next interesting one is the inner product of $\Sigma_3$ with $\Sigma_1 \cup \Sigma_2$: 
\begin{align}
  \pmel{
\begin{tikzpicture}[baseline=-.2cm]
	\node[cyan,fill,circle,inner sep=.5mm] (1) at (0,0) {};
	\draw[cyan] (.5,0) to[out=-90,in=-90] node[below] {\tiny $\ell'$} (1,0);
\end{tikzpicture}
}{\eta} {
\begin{tikzpicture}[baseline=.5cm]
	\draw[cyan] (0,0) to[out=120,in=-90] node[pos=.9,left]  {\small $1$} (-.5,1);
	\draw[cyan] (0,0) to[out= 90,in=-90] node[pos=.9,left]  {\small $2$} (  0,1);
	\draw[cyan] (0,0) to[out= 60,in=-90] node[pos=.9,right] {\small $3$} ( .5,1);
\end{tikzpicture}
} &=  \delta(\ell'-\ell_{2}) \delta(\ell'- \ell_{3}) 
\begin{tikzpicture}[baseline=-.5cm]
	\node[cyan,fill,circle,inner sep=.5mm] (1) at (0,0) {};
	\draw[cyan] (0,0) to[out=-90,in=120] (.5,-1);
	\draw[cyan] (.5,-1) -- (.5,0) to[out=90,in=90] (1,0) to[out=-90,in=60] (.5,-1);
\end{tikzpicture}
 \nonumber\\
  &= \delta(\ell' - \ell_{1}) \delta (\ell' - \ell_{2}) \quad  \input{figs/lasso-ip-3d.tex} \nonumber\\
  &= 0.
  \label{eqn:3-bd-ip-lasso}
\end{align}
The diagrammatic steps above highlight that yet again, $\gamma_{1}$ is contractible in the bulk and so cannot be a non-zero length geodesic.

An interesting observation of the analysis till now is that inner products involving elements of different topologies have been zero. 
This is reflected in the formulas we have, see  
\eqref{eqn:2-bd-ip-gen}, where we have the proviso that the $\delta$-function is Kronecker delta for $\ell = 2\pi i$.
In four boundary case, we will see non-zero overlap between different topologies. 

\paragraph{Four Boundaries}

In the case of four boundaries, we have non-trivial inner products already with a single topology, which is the four-boundary wormhole $\Sigma_4$.
We can consider the following basis elements,
\begin{equation}
  \pket{\input{figs/fa-4-bd-s-no-l.tex}}, \quad \pket{\input{figs/fa-4-bd-t.tex}}, \quad \pket{\input{figs/fa-4-bd-u.tex}},
  \label{eqn:4-bd-basis-elts}
\end{equation}
which we label as $s,t,u$ channels.
We denote these using $\Sigma_{4 s}, \Sigma_{4 t}, \Sigma_{4 u}$ for obvious reasons.
There are actually two $u$-channel networks, the one shown here and the one where the $2$ line goes below the $3$ line.

Inner products involving the first two have been described in the main text, so we focus on the $s-u$ inner product, when all angular momenta are zero.
We will find
\begin{equation}
  \pmel{\input{figs/fa-4-bd-sp.tex}}{\eta} {\input{figs/fa-4-bd-u.tex}} = \prod_{i=1}^{4} \delta (\ell_{i} - \ell_{i}') e^{\frac{\ell_{s} + \ell_{u}}{8 G_{N}}}
	\begin{Bmatrix}
	1 & 2 & s \\
	4 & 3 & u \\
	\end{Bmatrix}^{2}.
  \label{eqn:fa-su-xing}
\end{equation}
This can be derived in our line diagrams as follows:
\begin{align}
  e^{- \frac{\ell_{s} + \ell_{u}}{8 G_{N}}} \pmel{\input{figs/fa-4-bd-sp.tex}}{\eta} {\input{figs/fa-4-bd-u.tex}} &=  \prod_{i=1}^{4} \delta (\ell_{i} - \ell_{i}') \input{figs/su-ip-1.tex} \nonumber\\
  &\xrightarrow{\text{braid around $s,u$}}  \prod_{i=1}^{4} \delta (\ell_{i} - \ell_{i}') \input{figs/su-ip-2.tex} \nonumber\\
  &\xrightarrow{\text{rearrange}}  \prod_{i=1}^{4} \delta (\ell_{i} - \ell_{i}') \input{figs/su-ip-3.tex}
  \label{eqn:fa-su-xing-line}
\end{align}
In going to the third line, we have only rearranged the diagram to make it look like an s-t inner product discussed around \eqref{eqn:fa-st-xing-main}.
The braiding phase for non-zero angular momentum can easily be derived by following the braidings above --- we find $(-1)^{J_{1} + J_{4} + J_{s} + J_{u}}$ since the $J_{3}$s cancel out in the two braidings above.

Now, we will see that the second contribution in \eqref{eqn:4-bd-ip-s2s1-2} also gives rise to our first non-zero example of topology change.
It appears in the inner product between $\Sigma_2 \cup \Sigma_2$ and $\Sigma_{4}$:
\begin{equation}
  \pmel{
\begin{tikzpicture}[baseline=-.5cm]
	\draw[cyan] (  0,0) to[out=-90,in=-90] node[pos=.5,above] {\small $1'$} (  1,0);
	\draw[cyan] (-.5,0) to[out=-90,in=-90] node[pos=.5,below] {\small $2'$} (1.5,0);
\end{tikzpicture}
}{\eta} {\input{figs/fa-4-bd-s-no-l.tex}} = (\delta\,\text{fns})\  
\begin{tikzpicture}[baseline=.5cm]
	\draw[cyan] (0,0) -- node[below] {\small $s$} (1,0);
	\draw[cyan] (0,0) arc(180:0:.5 and .5) node[above,pos=.5] {\small $1$};
	\draw[cyan] (0,0) arc(180:0:.5 and  1) node[above,pos=.5] {\small $3$};
\end{tikzpicture}
\ .
  \label{eqn:topology-change-ip}
\end{equation}
The right side here, using a careful drawing which  (see figure \ref{fig:2-2-4} for example), can be seen to be exactly a degenerate double torus, giving us
\begin{equation}
  \pmel{
\begin{tikzpicture}[baseline=-.5cm]
	\draw[cyan] (  0,0) to[out=-90,in=-90] node[pos=.5,above] {\small $1'$} (  1,0);
	\draw[cyan] (-.5,0) to[out=-90,in=-90] node[pos=.5,below] {\small $2'$} (1.5,0);
\end{tikzpicture}
}{\eta} {\input{figs/fa-4-bd-s-no-l.tex}} = (\delta\,\text{fns}) e^{\frac{\ell_{s} - \ell_{1} - \ell_{3}}{8 G_{N}}}.
  \label{eqn:topology-change-ip-2}
\end{equation}
Notice that the bra element above can be seen as t-channel network with $\ell_{t} = 2\pi i$.
The $\ell_{t} = 2 \pi i$ limit of the $6j$-symbol in \eqref{eqn:fa-st-xing-main} has exactly this limit as we take $\ell_{t} \to 2\pi i$.

\paragraph{Higher Genus}

The simplest higher genus surface is the so-called torus wormhole or toplogical geon, which is a genus-1 surface with one boundary.
We can consider its inner product with a disk: 
\begin{equation}
  \pmel{\tikz{\node[cyan,fill,circle,inner sep=.5mm] (a) at (0,0) {};}}{\eta} {
\begin{tikzpicture}[baseline=-.75cm]
	\draw[cyan] (0,0) -- node[right] {\small $1$} (0,-.5);
	\draw[cyan] (0,-.5) arc(90:-270:.5);
	\node[cyan] (2) at (.6,-1) {\small $2$};
\end{tikzpicture}
} = 
\begin{tikzpicture}[baseline=-.5cm]
	\node[cyan,fill,circle,inner sep=.5mm] (1) at (0,0) {};
	\draw[cyan] (0,0) to[out=-90,in=120] (.5,-1);
	\draw[cyan] (.5,-1) -- (.5,0) to[out=90,in=90] (1,0) to[out=-90,in=60] (.5,-1);
\end{tikzpicture}
 = 0 \quad \text{if } \ell_{1} > 0.
  \label{eqn:t2-d2-ip}
\end{equation}
following similar arguments as above.

A more non-trivial case is the inner product between two torus wormholes.
For $\ell_{1} > 0$, $\Sigma$ has the topology of a torus with a hole; but taking $\ell_{1} = 2\pi i$ corresponds to degenerating $\Sigma$ into $D^{2} \cup T^{2}$.
By definition, the $T^{2}$ piece satisfies our gauge conditions, so that it contains a geodesic $\gamma$ and $T^{2} \setminus \gamma$ is a hyperbolic cylinder whose ends are closed geodesics.
But a hyperbolic cylinder has metric $\dd{\rho}^{2} + r_{h}^{2} \cosh^{2} \rho \dd{\phi}^{2}$, which contains only one closed geodesic (at $\rho = 0$).
So, we conclude that both $\Sigma,\Sigma'$ degenerate into $S^{1} \cup D^{2}$, with the $D^{2}$s being identified in the inner product.
The two $S^{1}$s cannot be identified, since that would correspond to a manifold that is locally one-dimensional and therefore not in the class of metrics being extremised over in the GPI.
We thus find that the internal angle at each $S^{1}$ is $0$ and so the inner product is $\exp{(\ell_{2} + \ell'_{2})/8 G_{N}}$.

For $\ell_{1} > 0$, $\Sigma$ is now a connected surface of Euler characteristic $\chi = -1$, and so there are admissible hyperbolic metrics that are non-degenerate.
The bulk solution is such that $\gamma_{1}$ is homotopic to some $PSL(2,\mathds{Z})$ image of $\gamma'_{1}$.
Using the decomposition shown in \cite{Hartman:2025cyj,Hartman:2025ula}, the region in between looks like a generalised tetrahedron and we find
\begin{equation}
	
\begin{tikzpicture}[baseline]
	\draw[cyan] (0,-.5) arc(90:450:.5) node[pos=.75,right] {\small $2$} -- node[right] {\small 1} (0,.5) arc(-90:270:.5) node[pos=.25,right] {\small $2'$};
\end{tikzpicture}
 = \delta_{\ell_{1}, 2\pi i} e^{\frac{\ell_{2} + \ell_{2}'}{8 G_{N}}} + \delta(\ell_{2} - \ell_{2}') + \widehat{\mathds{S}}_{22'} [1] + \dots
  \label{eqn:t2-t2-ip}
\end{equation}
Here again the function $\widehat{\mathds{S}}$ is a known function and the $\dots$ indicates a sum over the modular group.

\paragraph{Some Virasoro TQFT Identities}

We have seen some examples and also many building blocks for calculating complicated inner products in section \ref{ssec:inner-products} and in this section.
There are two identities in Virasoro TQFT that have simple descriptions in the classical limit, and we show these since they might be interesting.
They are the Wilson bubble and the fictitious identity line \cite{Collier:2023fwi}.

\begin{figure}[ht!]
  \centering
  \includegraphics[width=0.96\linewidth]{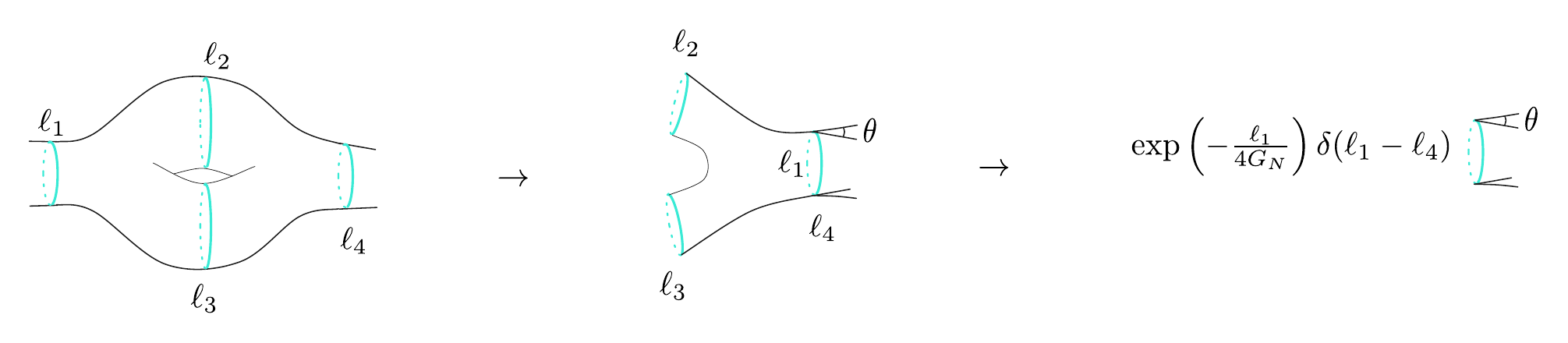}
  \caption{The figure on left represents part of a boundary condition inside a closed ball $D^3$ in the `inside out' picture. Here $\ell_1, \ell_4$ are homotopic to each other. On the right we show what this becomes, with $\ell_1, \ell_4$ identified.  }
  \label{fig:wilson-bubble}
\end{figure}

For the Wilson bubble identity, we will find the inside out depiction useful.
Figure \ref{fig:wilson-bubble} shows a boundary condition on the left side, drawn as an inside-out diagram in the style of \cite{Belin:2026pko}, inside $D^3$.
The full boundary condition may be bigger than shown here, and this is meant to be just a part of the whole. 
As the geodesics $1$ and $4$ are homotopic to each other through $D^3$, they get identified on-shell, and this makes this component in the usual picture look like the middle figure.
We have a degenerate double torus arising which gives contributions only from the corner of $\ell_1, \ell_4$ geodesics, where the total angle is $2\pi + \theta$; see the figure \ref{fig:I-half-pic} for the derivation.
Here $\theta$ is the internal angle at the $1$ geodesic after identification and removing the degenerate double torus.
Therefore the degenerate double torus gives $\exp \left( - \frac{\ell_1}{4G_N}\right) \delta(\ell_1 - \ell_4)$. 

The fictitious identity line has the following classical saddle.
Again we use the inside out depiction.
\begin{equation}
	\begin{tikzpicture}[baseline]
	  \draw[cyan] (0, .25) -- node[above] {\small $1$} (1, .25);
	  \draw[cyan] (0,-.25) -- node[below] {\small $2$} (1,-.25);
	\end{tikzpicture}
	\quad \supset \quad
    \includegraphics[height=20mm,valign=c]{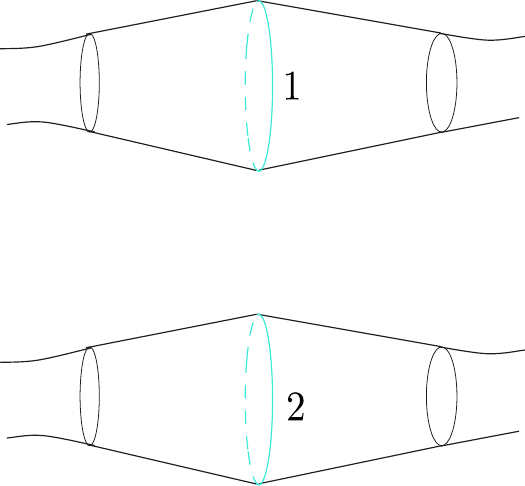} \quad = \quad \rho_{\mathrm{p}} (t) \quad  \includegraphics[height=20mm,valign=c]{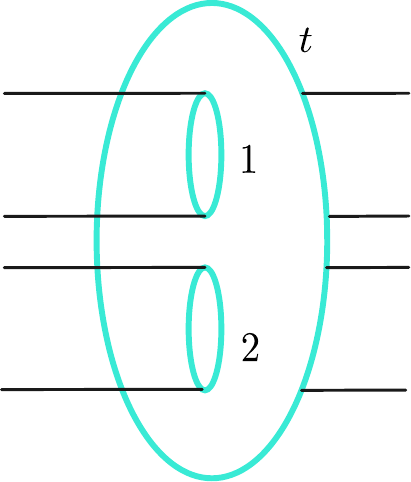} \quad \subset \quad \rho_{\mathrm{p}} (t) 
		\begin{tikzpicture}[baseline]
		  \draw[cyan] (0,0) -- node[pos=1,above] {\small $1$} ++ (150:.5);
		  \draw[cyan] (1,0) -- node[pos=1,above] {\small $1$} ++ ( 30:.5);
		  \draw[cyan] (0,0) -- node[pos=1,below] {\small $2$} ++ (210:.5);
		  \draw[cyan] (1,0) -- node[pos=1,below] {\small $2$} ++ (-30:.5);
			\draw[cyan] (0,0) -- node[above] {\small $t$} (1,0);
		\end{tikzpicture}
  \label{eqn:fake-identity}
\end{equation}
First we take two geodesics that can be surrounded by a $D^{3}$, as shown in the LHS.
There is a geodesic surrounding $1$ and $2$, which we call $t$.
We are free to add a degenerate double torus bordered by the $1,2,t$ geodesics without changing the bulk saddle at all.
The only change is that when we have added it, we get a corner term from the $t$ geodesic.
Since the internal angle is $2\pi$, the corner term is $\exp(- \ell_{t}/4 G_{N})$.
So we have to explicitly multiply by $e^{\ell_{t}/4 G_{N}}$ to get back the original solution, as shown in the middle expression.
This is a contribution to the partial boundary condition shown on the right.

\subsection{Examples with ETW Branes} \label{ssec:egs-open}

We will now discuss examples of elements involving end of the world branes.
These will include special cases of single boundary with end of the world branes, a case involving both open and closed geodesics, and multiple open geodesic cases.

\begin{figure}[h!]
  \centering
  \includegraphics[width=0.9\linewidth]{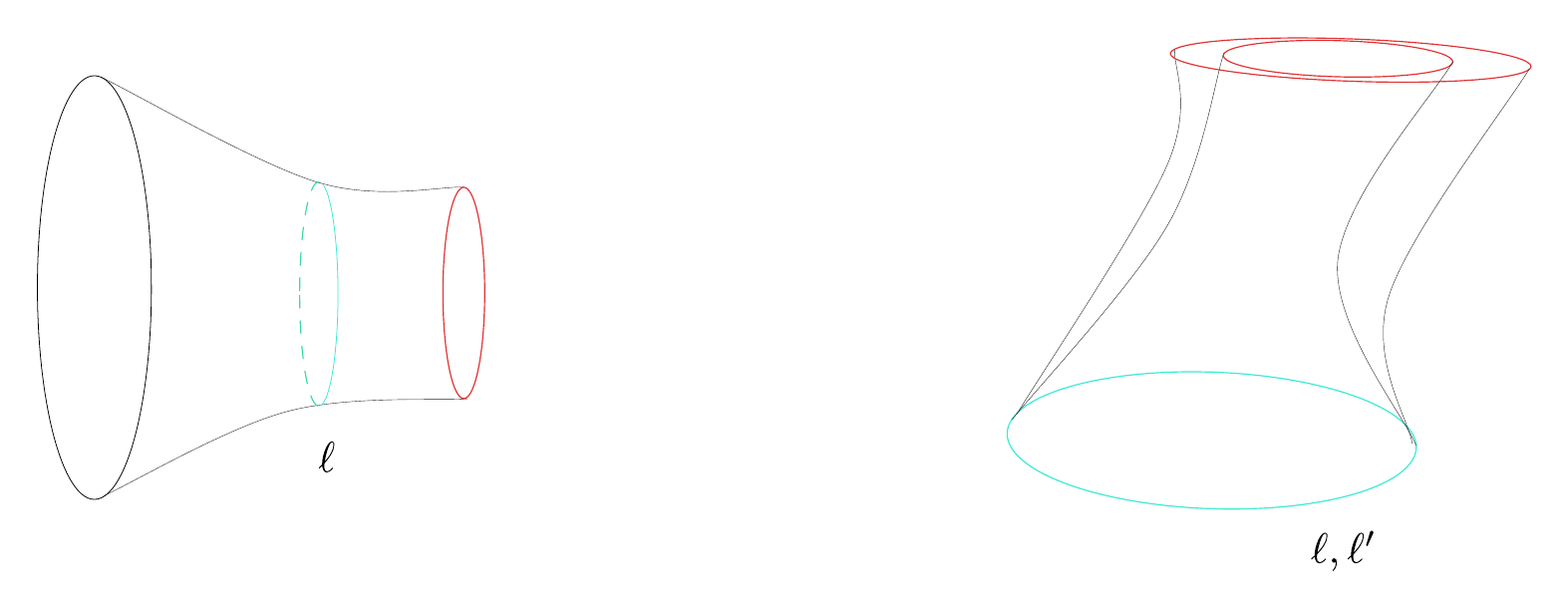}
  \caption{$\Sigma_{1b}:$ An alternative single boundary element compared to $\Sigma_1$, which possesses a closed geodesic. On right, we have the inner product picture with the asymptotic ``trumpets" removed, and the ETW have been drawn separately to illustrate both the pieces. Since the tension is identical, the two branes will forced to overlap. }
  \label{fig:alternative-1-bdy}
\end{figure}

\paragraph{Single CFT boundary with ETW brane}

We earlier mentioned $\Sigma_1 = \tikz{\node[cyan,fill,circle,inner sep=.5mm] (D) at (0,0) {};}$ as the simplest, single, full CFT boundary case. If we allow for an ETW brane, we can add a new case, shown in figure \ref{fig:alternative-1-bdy}, which looks like $\Sigma_2$ from only one end. 

Unlike $\Sigma_1$, this case has a closed geodesic. We will call this case $\Sigma_{1b}$, and here look at inner product of two $\Sigma_{1b}$ elements, again illustrated in the right side of \ref{fig:alternative-1-bdy}. Because of the homotopy argument, we will be forced to identify the two geodesics, and since tensions are equal, the volume of the integrated region will be zero, with zero angles at the corners, giving us 
\begin{equation}
	(\Sigma_{1b}; \ell | \Sigma_{1b}; \ell')	 = \delta(\ell - \ell')
.
\end{equation}

\paragraph{Element with both open and closed geodesics}

\begin{figure}[h!]
  \centering
  \includegraphics[width=0.9\linewidth]{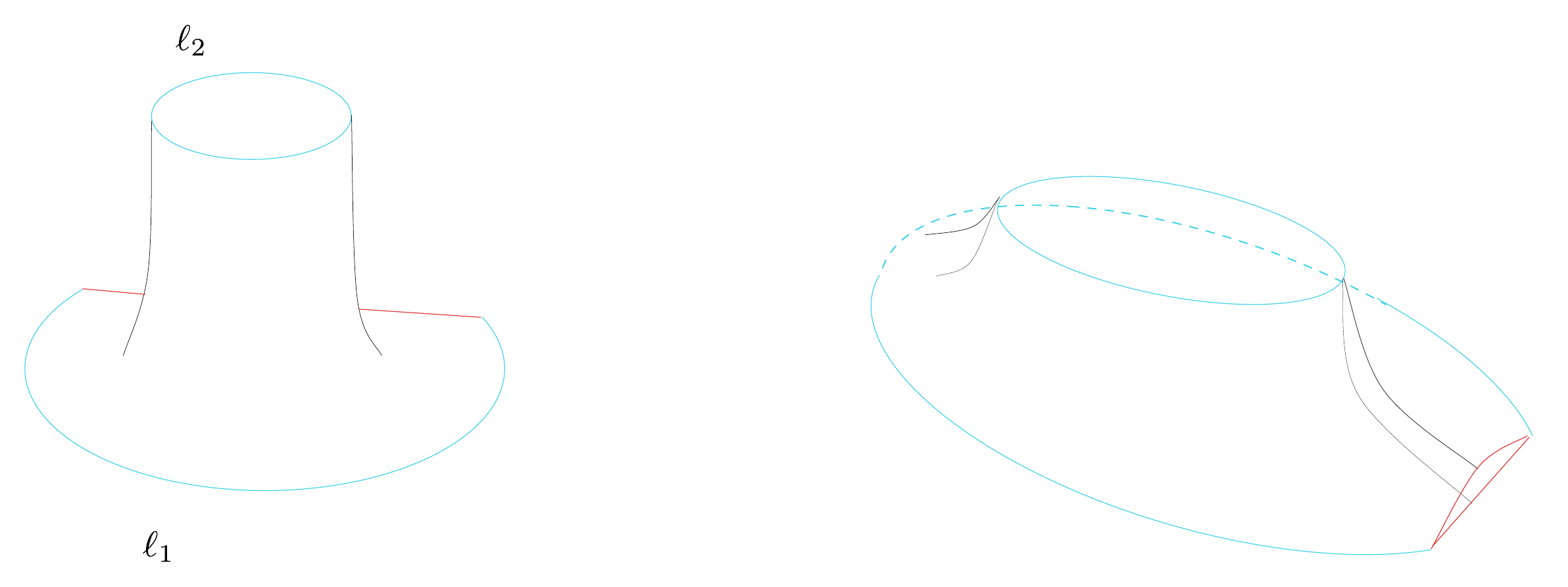}
  \caption{Inner product between two $\Sigma^{1}_1$ elements; candidate shown on left. In the figure on right we have identified the open and closed geodesics geodesics, however the ETW branes are shown separately for clarity. In the solution they get identified because of equal tension. }
  \label{fig:Sigma-1-1}
\end{figure}

The other special example is $\Sigma^{1}_1$, which first appeared as third lego in figure \ref{fig:S-legos}, also drawn again in figure \ref{fig:Sigma-1-1}. Here we will consider inner product of two such elements. 

Due to homotopy, both the open and closed geodesic get identified to each other, and because of the tension being the same, the ETW branes get identified, giving us again a zero volume, zero angles case, with answer
\begin{equation}
	(\Sigma_{1}^{1} ; \ell_1, \ell_2 | \Sigma_{1}^{1}; \ell_1', \ell_2') = \delta(\ell_1 - \ell_1') \delta(\ell_2 - \ell_2')
.
\end{equation}

\begin{figure}[h!]
  \centering
  \includegraphics[width=\linewidth]{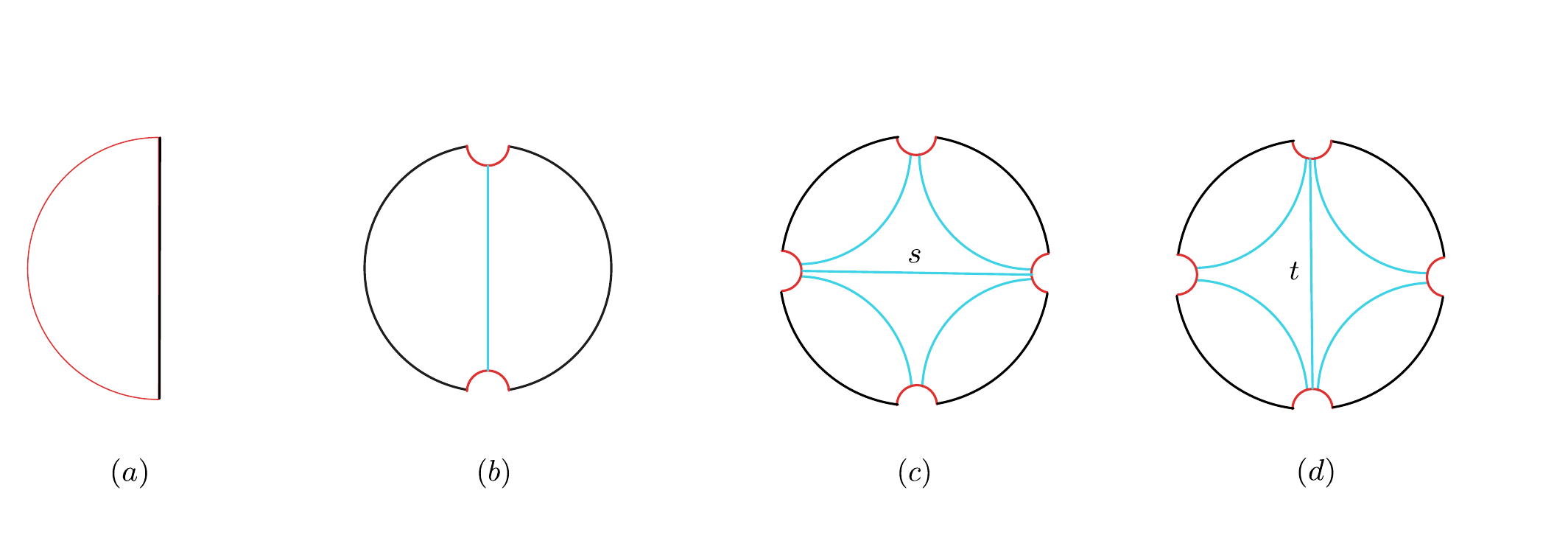}
  \caption{Geometries; the red curves represent end of the world branes, the blue ones represent geodesics.}
  \label{fig:geometries}
\end{figure}

Next, we will focus on just BCFT cases. We will use the following basis elements for illustrations:

\begin{align}
\pket{\textcolor{cyan}{\circ}}, \quad
\pket{
    \begin{tikzpicture}[baseline]
      \draw[double,cyan] (0,0) -- node[above] {\small $\ell$} ++(  0:1);
    \end{tikzpicture}
}, \quad
\pket{
    \begin{tikzpicture}[baseline]
      \draw[double,cyan] (0,0) -- node[below] {\tiny $\ell_{s}$} ++(  0:1);
      \draw[double,cyan] (0,0) -- node[pos=1,above] {\tiny $\ell_{1}$} ++(120:.5);
      \draw[double,cyan] (0,0) -- node[pos=1,above] {\tiny $\ell_{2}$} ++( 60:.5);
      \draw[double,cyan] (1,0) -- node[pos=1,above] {\tiny $\ell_{3}$} ++(120:.5);
      \draw[double,cyan] (1,0) -- node[pos=1,above] {\tiny $\ell_{4}$} ++( 60:.5);
    \end{tikzpicture}
}, \quad
\pket{
\begin{tikzpicture}[baseline=.125cm]
	\draw[double, cyan] (0,-.5) -- node[left,pos=.75] {\tiny $\ell_t$} (0,.5);
	\draw[double, cyan] (0,-.5) -- node[pos=1,above] {\tiny $\ell_1$} (-.75,1);
	\draw[double, cyan] (0, .5) -- node[pos=1,above] {\tiny $\ell_2$} (-.25,1);
	\draw[double, cyan] (0, .5) -- node[pos=1,above] {\tiny $\ell_3$} ( .25,1);
	\draw[double, cyan] (0,-.5) -- node[pos=1,above] {\tiny $\ell_4$} ( .75,1);
\end{tikzpicture}}
\end{align}
which represent respectively the geometries in figure \ref{fig:geometries}. In our notation these are $\Sigma^{1},\Sigma^{2}, \Sigma^{4 s}, \Sigma^{4t}$ respectively.

\paragraph{Two BCFTs}

Let us start by considering the simplest possible inner product, in a system of 2 BCFTs of $\Sigma^{1} \cup \Sigma^{1}$ with $\Sigma^{2}$.
\begin{equation}
  \pmel{\textcolor{cyan}{\circ \circ}}{\eta} {
    \begin{tikzpicture}[baseline]
      \draw[double,cyan] (0,0) -- node[above] {\small $\ell$} ++(  0:1);  
    \end{tikzpicture}
} 
\end{equation}
Following the rule \eqref{eqn:physical-ip}, this is given by the path integral
\begin{equation}\label{eq:2-bdy-bcft}
	\int^{*} Dg \exp{ - I_{\mathrm{full}} \left[ 
			\includegraphics[width=.2\textwidth,valign=c]{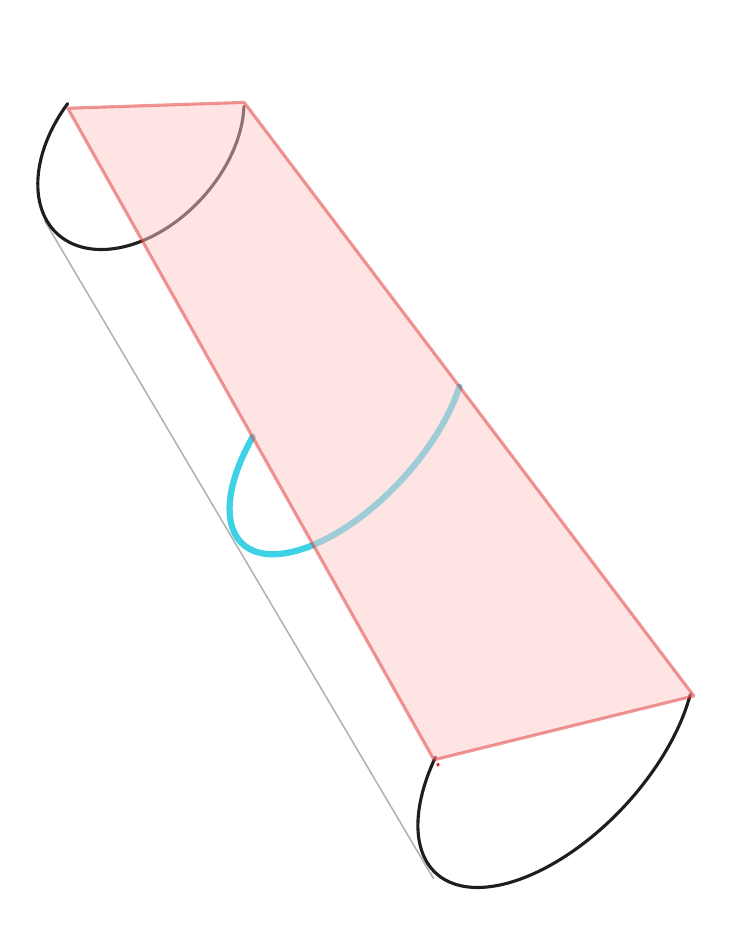}
	\right]}
\end{equation}
where the red rectangle is a 2d ETW. The figure is obtained by using two copies of (a) and one (b) from figure \ref{fig:geometries}. Note that the geometry is what you will get by cutting \eqref{eqn:jafferis-case} geometry down the length of hyperbolic cylinder, after inserting the 2d ETW brane on the open part. This pattern will repeat in the examples below.

For reasons similar to \eqref{eqn:jafferis-case}, this inner product is zero. As we have discussed above, a non-zero answer requires boundary conditions to be compatible with a hyperbolic spacetime, which are known to have non-contractible, unique geodesics for each homotopy class. Demanding interior to be filled, cyan geodesic can be homotopically deformed and shrunk in size. Its homotopy class is unchanged even if the end points are moved on the ETW brane.\footnote{This is the natural way to define the homotopy class given that this is \eqref{eqn:jafferis-case} geometry cut in half.} Shrinking the geodesic is then possible to zero length, forcing answer to be 0. 

In a very similar way, \eqref{eqn:2-bd-ip-snake}, \eqref{eqn:3-bd-ip-lasso} analogues will be 0.

\paragraph{Four BCFTs: s-s' overlap}

\begin{figure}[h!]
  \centering
  \includegraphics[width=\linewidth]{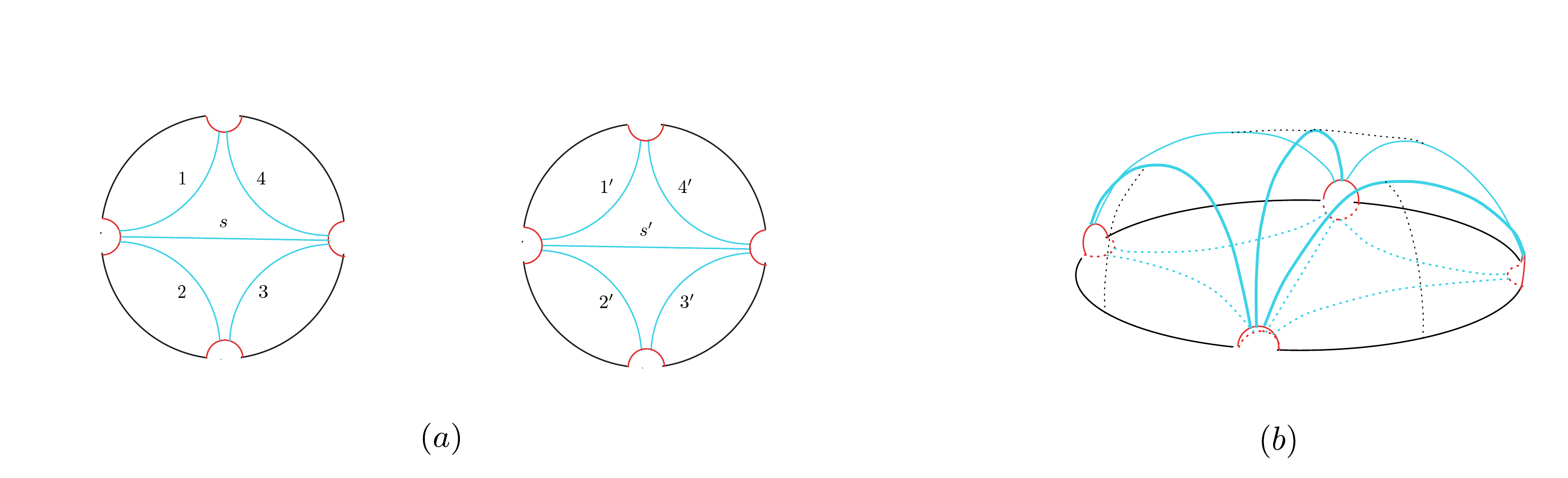}
  \caption{Graphical illustration to understand the inner product computation. (b) shows the boundary of the region that must be filled by the saddle.  }
\label{fig:s-s-ip}
\end{figure}

Let us concentrate on the more interesting cases of 4 BCFTs; the analogues of \eqref{eqn:4-bd-ip-main} and \eqref{eqn:fa-st-xing-main}.
Like in \eqref{eqn:4-bd-ip-main}, we will discuss two saddles for the path integral, and like in \eqref{eqn:fa-st-xing-main}, we will find a contribution from a generalised tetrahedon.
Let us start with 
\begin{equation}\label{eqn:s-ch-saddles-bcft}
  \pmel{
	\input{figs/fa-4-bd-I-sp.tex}
}{\eta} {
	\input{figs/fa-4-bd-I-s.tex}
} 
\end{equation}
Figure \ref{fig:s-s-ip}, (b), shows the geometry associated to the saddle in which $s, s'$ are homotopic, which contributes
\begin{equation}
\text{First saddle} \to  \delta(s-s') \cdot \prod_{i}\delta(\ell_i - \ell_i')
\end{equation}

\begin{figure}[h!]
  \centering
  \includegraphics[width=0.7\linewidth]{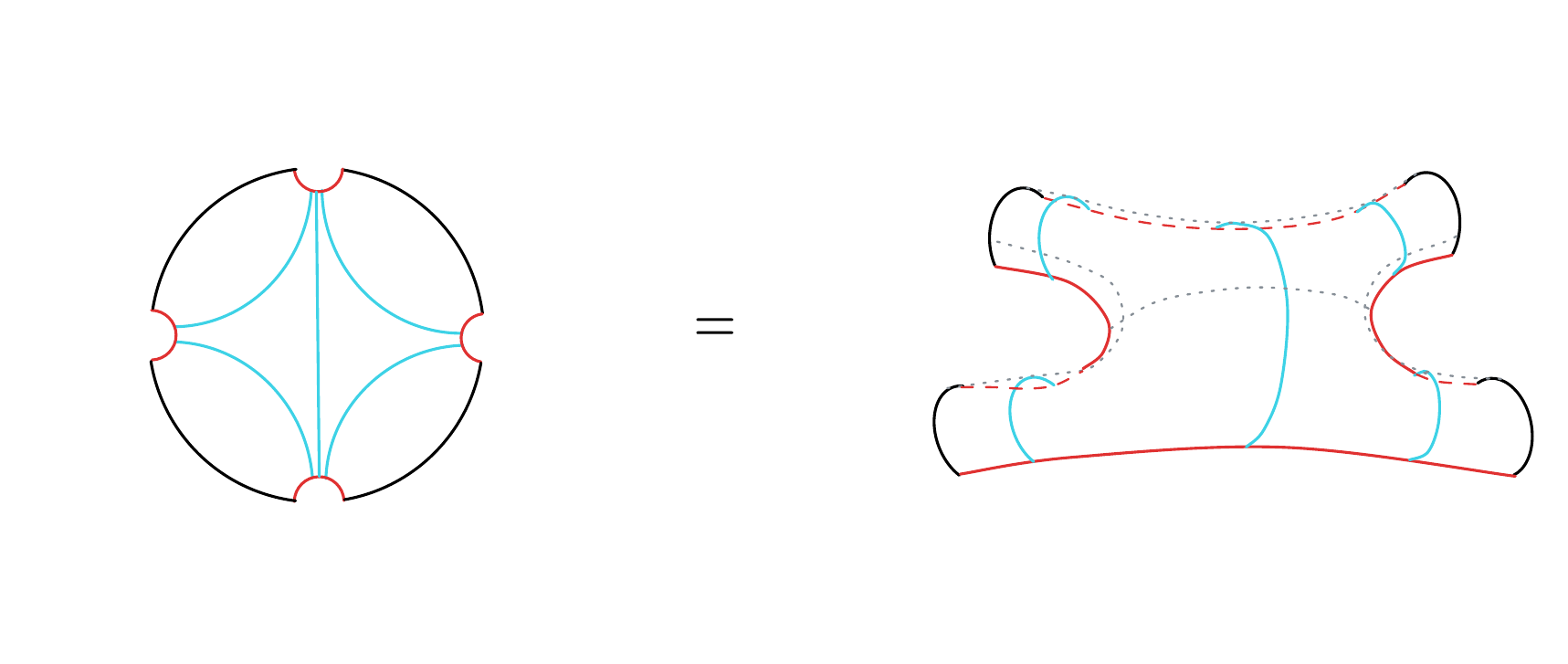}
  \caption{An alternate way to see the 4 BCFT slice}
\label{fig:redraw}
\end{figure}

\begin{figure}[h!]
  \centering
  \includegraphics[width=0.8\linewidth]{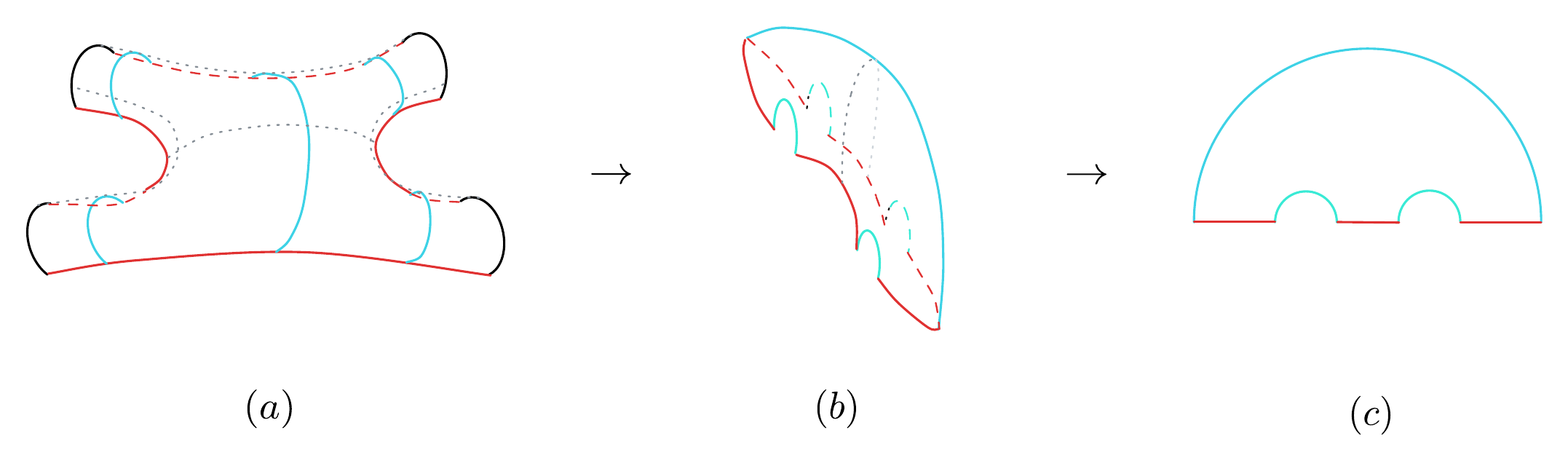}
  \caption{Visualizing transformation of the 4 BCFT slice to understand the  second saddle for \eqref{eqn:s-ch-saddles-bcft}. In the (b), (c) figures we've removed the asymptotic trumpets.}
  \label{fig:second-saddle-bcft}
\end{figure}

\begin{figure}[h!]
  \centering
  \includegraphics[width=0.7\linewidth]{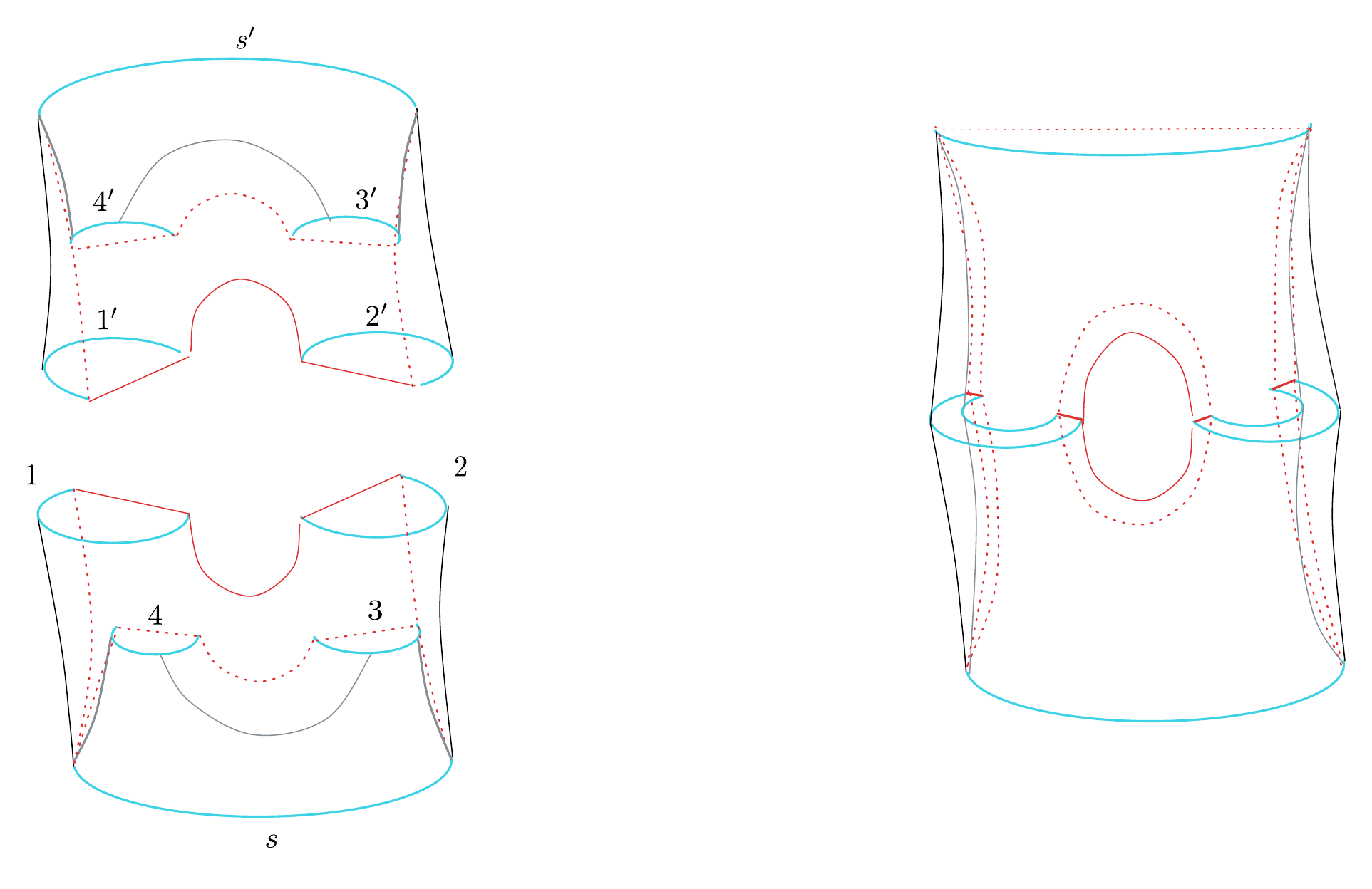}
  \caption{Inner product contribution from the second saddle can be understood from the right figure, where the volume of the gap between ket and bras tends to zero, however the angles at each of the corners is non-zero, giving the contribution in \eqref{eq:second-saddle-bcft}, like in figure ..}
\label{fig:ip-s-s}
\end{figure}

To see the second saddle, it is useful to make the observation in figure \ref{fig:redraw}. Further deformation as shown in figure \ref{fig:second-saddle-bcft} illustrates how this becomes half of a two-hole disk. However, visualizing the boundary conditions for computing the inner product requires us to consider both the ket and the bra; this is shown in figure \ref{fig:ip-s-s}. Note the ETW brane regions shrink to zero size when the volume shrinks to zero. 
The contribution from this saddle 
\begin{equation}\label{eq:second-saddle-bcft}
\prod_{i} \delta(\ell_i - \ell_i') \cdot \exp\left(\frac{1}{8 G_N} (\ell_s + \ell_s' - 2 \ell_1 - 2 \ell_3)\right)\cdot \delta(\ell_1 - \ell_4) \delta(\ell_2- \ell_3)
\end{equation}
is same as before in form, only now $\ell_i$ represent lengths of open geodesics instead of closed geodesics.

\paragraph{Four BCFTs: s-t overlap}

\begin{figure}[h!]
  \centering
  \includegraphics[width=\linewidth]{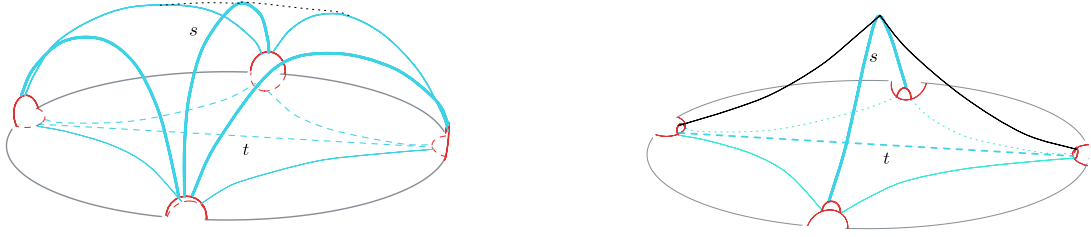}
  \caption{Geometries for inner product for $s-t$ case}
  \label{fig:s-t-geo}
\end{figure}

We can now look at the inner product for $s-t$ case, 
\begin{equation}\label{eqn:s-t-saddles-bcft}
  \pmel{
\begin{tikzpicture}[baseline=.125cm]
	\draw[double, cyan] (0,-.5) -- node[left,pos=.75] {\tiny $\ell_t$} (0,.5);
	\draw[double, cyan] (0,-.5) -- node[pos=1,above] {\tiny $\ell_1$} (-.75,1);
	\draw[double, cyan] (0, .5) -- node[pos=1,above] {\tiny $\ell_2$} (-.25,1);
	\draw[double, cyan] (0, .5) -- node[pos=1,above] {\tiny $\ell_3$} ( .25,1);
	\draw[double, cyan] (0,-.5) -- node[pos=1,above] {\tiny $\ell_4$} ( .75,1);
\end{tikzpicture}
}{\eta} {
    \begin{tikzpicture}[baseline]
      \draw[double,cyan] (0,0) -- node[below] {\tiny $\ell_{s}$} ++(  0:1);
      \draw[double,cyan] (0,0) -- node[pos=1,above] {\tiny $\ell_{1}$} ++(120:.5);
      \draw[double,cyan] (0,0) -- node[pos=1,above] {\tiny $\ell_{2}$} ++( 60:.5);
      \draw[double,cyan] (1,0) -- node[pos=1,above] {\tiny $\ell_{3}$} ++(120:.5);
      \draw[double,cyan] (1,0) -- node[pos=1,above] {\tiny $\ell_{4}$} ++( 60:.5);
    \end{tikzpicture}
} 
\end{equation}
for which the geometries are represented in figure \ref{fig:s-t-geo}. After accounting for the equivalence under homotopy for geodesics $i$ and $i'$, we're left with two non-homotopic geodesics prominently visible in right figure, giving us a hyperbolic tetrahedron. That makes the contribution 
\begin{equation}
  \pmel{
\begin{tikzpicture}[baseline=.125cm]
	\draw[double, cyan] (0,-.5) -- node[left,pos=.75] {\tiny $\ell_t$} (0,.5);
	\draw[double, cyan] (0,-.5) -- node[pos=1,above] {\tiny $\ell_1$} (-.75,1);
	\draw[double, cyan] (0, .5) -- node[pos=1,above] {\tiny $\ell_2$} (-.25,1);
	\draw[double, cyan] (0, .5) -- node[pos=1,above] {\tiny $\ell_3$} ( .25,1);
	\draw[double, cyan] (0,-.5) -- node[pos=1,above] {\tiny $\ell_4$} ( .75,1);
\end{tikzpicture}
}{\eta} {
    \begin{tikzpicture}[baseline]
      \draw[double,cyan] (0,0) -- node[below] {\tiny $\ell_{s}$} ++(  0:1);
      \draw[double,cyan] (0,0) -- node[pos=1,above] {\tiny $\ell_{1}$} ++(120:.5);
      \draw[double,cyan] (0,0) -- node[pos=1,above] {\tiny $\ell_{2}$} ++( 60:.5);
      \draw[double,cyan] (1,0) -- node[pos=1,above] {\tiny $\ell_{3}$} ++(120:.5);
      \draw[double,cyan] (1,0) -- node[pos=1,above] {\tiny $\ell_{4}$} ++( 60:.5);
    \end{tikzpicture}
}
 = \prod_{i=1}^{4} \delta (\ell_{i} - \ell_{i}') 
	\begin{Bmatrix}
	1 & 2 & s \\
	4 & 3 & u \\
	\end{Bmatrix}.
  \label{eqn:6j-bcft-s-t}
\end{equation}

\paragraph{Topology change example}

\begin{figure}
  \centering
  \includegraphics[width=\linewidth]{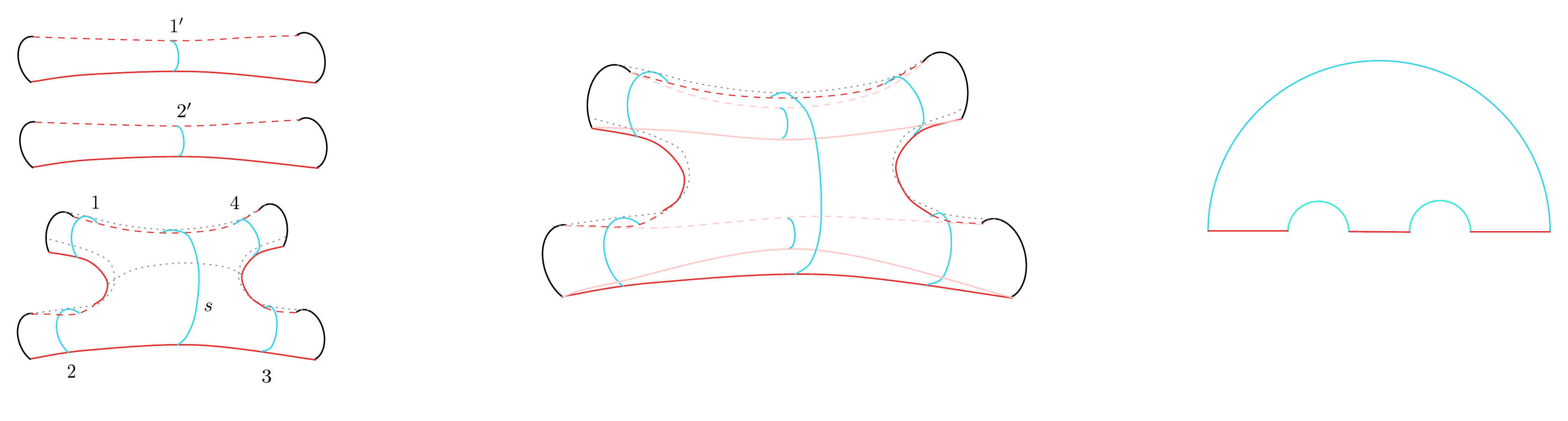}
  \caption{The analogue of \eqref{eqn:topology-change-ip-2}. In the right figure we have removed the asymptotic trumpet regions.}
  \label{fig:2-2-4}
\end{figure}

Another case we will study is $(\Sigma^{2} \cup \Sigma^2 | \Sigma^4)$, the analogue of \eqref{eqn:4-bd-ip-main}. The associated geometries are shown in figure \ref{fig:2-2-4}. More specifically we have $\Sigma^{2}(\ell_1') \cup \Sigma^{2}(\ell_2')$ as one element, and $\Sigma^{4s}(\ell_1, \ldots \ell_4; \ell_s)$ as the other. 

In this example, the middle figure shows the bulk region to be filled with a hyperbolic space, in order to obtain the non-zero inner product.
Here, in a manner similar to \eqref{eq:2-bdy-bcft}, it is seen that geodesics $1, 4, 1'$ are in one homotopy class, and $2, 3, 2'$ in another.
To recall, this is because we can move the geodesics ends on the ETW brane for free without changing its homotopy class.
Thus we must identify these geodesics, ensuring proportionality to $\delta(\ell_1 - \ell_4) \delta(\ell_2 -\delta_3) \cdot \delta(\ell_1 - \ell_1') \delta(\ell_2- \ell_2')$. 

The right figure shows the final shape of the saddle, which is a flat surface, with two sides. As $s$ is an internal geodesic, the associated corner contributes $\exp(\ell_s/8G_N)$ to the product, but since $1, 2$ are external geodesics, angle between two normals, $\pi$, must be subtracted by $2\pi$ to get $-\pi$, giving the total to be
\begin{equation}
 \delta(\ell_1 - \ell_4) \delta(\ell_2 -\delta_3) \cdot \delta(\ell_1 - \ell_1') \delta(\ell_2- \ell_2')\exp\left(\frac{1}{8 G_N}(\ell_s - \ell_1 - \ell_2)\right)
.
\end{equation}
This is again the same form as \eqref{eqn:topology-change-ip-2}, with $\ell_i$ representing the open geodesics now. 

\section{Introduction to Calculating OPE Statistics in Holographic CFT} \label{app:ope-statistics}

In this appendix we cover some basics of conformal blocks and bootstrap that are relevant for this work. This will primarily involve discussion of partition functions of CFTs with no primary insertions.
In the case of a full CFT, these partition functions give rise to Riemann surfaces without punctures, whereas in BCFT, additional $\sigma$ boundaries arise; these boundaries will always be red in our diagrams.

As is well-known, a conformal block is the part of the partition function completely fixed by symmetry.
To obtain a conformal block, there are two steps involved: (a) inserting a projector onto the Virasoro module of a fixed primary along each ``cycle" and (b) stripping off all OPE coefficients. To clarify this statement and explain what we mean by a cycle, we will recall the process of pair of pants decomposition of a general manifold. 

Given a manifold $\Sigma$, we define a cycle as either a non-contractible $S^{1}$ or an interval with both ends on $\sigma$ boundaries.
Each cycle hosts a Hilbert space of the form \eqref{eqn:H-S1} or \eqref{eqn:H-I} depending on the topology.
A generalised pair-of-pants decomposition is given by marking enough non-intersecting cycles such that cutting along them results in a decomposition of the surface into the three building blocks shown in figure \ref{fig:S-legos}.
A pair-of-pants decomposition (we will drop the `generalised') leads to a Hilbert space interpretation of the partition function.
Therefore, it allows considering projectors involved in step (a) above. 
A surface generically admits an infinite number of pair-of-pants decompositions and therefore an infinite number of Hilbert space interpretations; we call each such decomposition a channel.

\paragraph{Bootstrap without the blocks}

Traditionally, the conformal bootstrap uses the fact that the partition function is independent of the channel to constrain the spectrum and OPE coefficients of the CFT.
In the literature on holographic CFT, however, it is used in a subtly different way: to extract \emph{statistics} of the spectrum and OPE coefficients.
The main technique for this can be called `bootstrap without the blocks' \cite{Collier:2019weq}, and here we review the main idea.

To start, note that the partition function on a surface $\Sigma$ in a given channel $\mathcal{C}$ is
\begin{equation}
  Z (\Sigma) = Z_{\mathrm{W}} \,  \sum_{P_{\mathcal{C}}} \prod_{ijk \in \mathcal{C}} C_{ijk} \,  \mathcal{F}_{\mathcal{C}} (P_{\mathcal{C}}, \Sigma),
  \label{eqn:Z-channel}
\end{equation}
where $Z_{\mathrm{W}}$ is the Weyl anomaly contribution (independent of the channel or the primary weights) and $\mathcal{F}$ is the conformal block (which sums up the contribution of the descendants to the partition function). 
The conformal block depends only on the central charge, the weights of the primaries and the moduli of $\Sigma$ and is completely fixed by Virasoro symmetry once these data are given.
With a slight abuse of terminology, we will say that the conformal block is `theory-independent;' by this, we mean that it depends only on the central charge of the theory and not on its spectrum or OPE coefficients.
In \eqref{eqn:Z-channel}, the theory-dependence appears (a) explicitly in the OPE coefficients and (b) implicitly in the sum over primaries.
Let us define an OPE density $\rho^{\mathcal{C}}$ in the channel $\mathcal{C}$ to satisfy
\begin{equation}
  Z (\Sigma) = Z_{\mathrm{W}} \int \dd[\abs{\mathcal{C}}]{P} \rho^{\mathcal{C}} (P_{\mathcal{C}}) \mathcal{F}_{\mathcal{C}} (P_{\mathcal{C}}, \Sigma),
  \label{eqn:Z-channel-ope-dens}
\end{equation} 
where $\abs{\mathcal{C}}$ is the number of distinct cycles in the channel.
The OPE density contains enough $\delta$-functions to reproduce \eqref{eqn:Z-channel}.
Now, the theory dependence has been \emph{completely} absorbed into the OPE density.
Given two different channels, we obtain the bootstrap equation
\begin{equation}
  \int \dd[\abs{\mathcal{C}}]{P} \rho^{\mathcal{C}} \mathcal{F}_{\mathcal{C}} = \int \dd[\abs{\mathcal{C}'}]{P} \rho^{\mathcal{C}'} \mathcal{F}_{\mathcal{C}'}.
  \label{eqn:bootstrap-eqn-1}
\end{equation}

This trick of absorbing the theory-dependence into the OPE density is useful because of two facts.
The first is that any two conformal blocks can be transformed into each other by a \emph{known} integral kernel,
\begin{equation}
  \mathcal{F}_{\mathcal{C}} = \int \dd[\mathcal{C}']{P} \mathds{K}_{\mathcal{C}}^{\mathcal{C}'} \mathcal{F}_{\mathcal{C}'}.
  \label{eqn:conf-block-trans}
\end{equation}
Here, the crossing kernel $\mathds{K}$ is a known function for every pair of channels; it can be built out of a few elementary `moves' as we will review in more detail below \cite{Moore:1988qv,Lewellen:1991tb,Eberhardt:2023mrq,Behrend:1999bn}.
The second fact that makes it useful is that conformal blocks in a fixed channel with different momenta form a complete linearly independent basis for the set of functions with the correct transformation properties \cite{Eberhardt:2023mrq}.\footnote{
	This can be demonstrated either by equipping the space with an inner product and showing that different conformal blocks are orthogonal as in \cite{Eberhardt:2023mrq}, or by going to a boundary of moduli space where the different blocks are manifestly linearly independent.
}
Plugging \eqref{eqn:conf-block-trans} into \eqref{eqn:bootstrap-eqn-1} and changing the order of integration, we then find
\begin{equation}
	 \left[ \rho^{\mathcal{C}'} -  \mathds{K}^{\mathcal{C}'}_{\mathcal{C}} \cdot \rho^{\mathcal{C}} \right] \cdot  \mathcal{F}_{\mathcal{C}'} = 0 \quad \implies \quad \rho^{\mathcal{C}'} = \mathds{K}^{\mathcal{C}'}_{\mathcal{C}} \cdot  \rho^{\mathcal{C}},
  \label{eqn:bootstrap-eqn-2}
\end{equation}
where $\cdot $ represents an integral over the `contracted' channel.
Thus, we can directly relate the OPE coefficient densities.

The equation \eqref{eqn:bootstrap-eqn-2} becomes particularly useful when $\mathcal{F}_{\mathcal{C}}$ is dominated by some known primaries; usually when all of the primaries propagating through the cycles are the identity.
In a generic CFT, this happens when all the cycles become really `thin.'\footnote{
	Take a cycle and draw the biggest cylinder (in the $S^{1}$ case) or rectangle (in the $I$ case) around it.
	The height of this cylinder/rectangle should be much larger than its radius/width.
}
In the holographic case, we assume that any cycle that is even a little bit `thin' may be dominated by the identity.
In that case, since $\rho^{\mathcal{C}} \cdot \mathcal{F}_{\mathcal{C}} \approx \rho^{\mathcal{C}} (\mathds{1}_{\mathcal{C}}) \mathcal{F}_{\mathcal{C}} (\mathds{1}_{\mathcal{C}}) = \rho^{\mathcal{C}} (\mathds{1}_{\mathcal{C}}) \mathds{K}_{\mathcal{C}}^{\mathcal{C}'} (P_{\mathcal{C}'}, \mathds{1}_{\mathcal{C}}) \cdot \mathcal{F}_{\mathcal{C}'} (P_{\mathcal{C}'})$ and $\rho^{\mathcal{C}} (\mathds{1}_{\mathcal{C}}) = 1$, \eqref{eqn:bootstrap-eqn-2} simplifies to
\begin{equation}
  \rho^{\mathcal{C}'} (P_{\mathcal{C}'}) = \mathds{K}^{\mathcal{C}'}_{\mathcal{C}} (P_{\mathcal{C}'},\mathds{1}_{\mathcal{C}}).
  \label{eqn:bootstrap-eqn-3}
\end{equation}
Thus we only need to calculate the crossing kernel that transforms conformal blocks from the \emph{target} channel $\mathcal{C}'$ to the \emph{source} channel $\mathcal{C}$.
This inversion is because of \eqref{eqn:Z-channel-ope-dens}: the OPE densities formally live in a dual space of the blocks.

Note that, though the final form \eqref{eqn:bootstrap-eqn-3} doesn't in any way depend on the moduli of $\Sigma$, its derivation crucially uses the existence of some $\Sigma$ such that the channel $\mathcal{C}$ is dominated by the vacuum block and the channel $\mathcal{C}'$ is dominated by $P_{\mathcal{C}'}$.
This is of course a hard problem, so the procedure is simplified as follows.
We consider all potential identity channels $\mathcal{C}_{i}$ and then write
\begin{equation}
  \rho^{\mathcal{C}'} (P_{\mathcal{C}'}) = \sum_{\mathcal{C}_{i}} K_{\mathcal{C}_{i}}^{\mathcal{C}'} (P_{\mathcal{C}'}, \mathds{1}_{C_{i}}).
  \label{eqn:bootstrap-logic-final}
\end{equation}
The notion of `potential identity channel' will be clarified below.
It is important to note that this is not necessarily the correct answer.
At genus $2$, for example, \cite{Belin:2017nze,Dong:2018esp} found that a sufficiently light scalar in the spectrum can be a barrier to the dominance of the identity in \emph{any} channel.

\paragraph{Diagrammatic Notation for Conformal Blocks}

We adopt the following notation for the conformal block.
Let each $S^{1}$ cycle be represented as a line, like in our line diagrams for fixed area elements.
Each cycle with the topology of an interval will be represented by two parallel red lines, similar to the double line representation used in section \ref{sec:H-CFT}.
The three building blocks of figure \ref{fig:S-legos} will be represented as
\begin{equation}
  
\begin{tikzpicture}[baseline]
	\draw (0,0) to (  0:1cm);
	\draw (0,0) to (120:1cm);
	\draw (0,0) to (240:1cm);
\end{tikzpicture}
\ , \quad 
\begin{tikzpicture}[baseline]
	\draw[chillred,double=white] (0,0) -- ++(  0:1cm);
	\draw[chillred,double=white] (0,0) -- ++(120:1cm);
	\draw[chillred,double=white] (0,0) -- ++(240:1cm);
\end{tikzpicture}
\ , \quad 
\begin{tikzpicture}[baseline]
	\draw (0,0) -- (90:1cm);
	\draw[chillred,double=white] (0,0) -- (-90:1);
\end{tikzpicture}

  \label{eqn:pop-reps}.
\end{equation}
As an example, a block on the following surface with three $\sigma$ boundaries is represented as
\begin{equation}
  
\begin{tikzpicture}[baseline,scale=.3]
	\draw[chillred] ( 0,0) circle ( 2cm);
	\draw[chillred] ( 1,0) circle (.5cm);
	\draw[chillred] (-1,0) circle (.5cm);

	\draw[thin,gray] ( -.5,0) -- (.5,0);
	\draw[thin,gray] ( 1.5,0) -- ( 2,0);
	\draw[thin,gray] (-1.5,0) -- (-2,0);
\end{tikzpicture}

	\quad \longrightarrow \quad 
	\input{figs/g2-eg-3.tex}
	\ ,
  \label{eqn:conf-block-cons}
\end{equation}
where grey lines denote cycles. Further, as part of our notation, if we draw the full left figure, it will represent the partition function on the given surface, with the grey lines representing the insertion of a projector onto a fixed primary in the corresponding Hilbert space.
The diagram on the right will denote the conformal block.
The conformal block and the projected partition function differ by factors of OPE coefficients,
\begin{equation}
  
\begin{tikzpicture}[baseline,scale=.3]
	\draw[chillred] ( 0,0) circle ( 2cm);
	\draw[chillred] ( 1,0) circle (.5cm);
	\draw[chillred] (-1,0) circle (.5cm);

	\draw[thin,gray] ( -.5,0) -- (.5,0);
	\draw[thin,gray] ( 1.5,0) -- ( 2,0);
	\draw[thin,gray] (-1.5,0) -- (-2,0);
\end{tikzpicture}

	= C^{123} C^{213}
	\input{figs/g2-eg-3.tex}
	\equiv
	\input{figs/g2-eg-3-dot.tex}\ .
  \label{eqn:conf-block-cons-exact}
\end{equation}
In this equation, we have introduced yet another notation involving dots on the right side. These dots will represent the OPE coefficients.

An important part of the notation is the ordering of the lines at a vertex.
We always associate to a vertex an OPE coefficient whose labels agree with the three lines arranged in a \emph{clockwise} order; a rule used in section \ref{sec:H-CFT}.
For example, in the block shown in \eqref{eqn:conf-block-cons}, the bottom vertex has a $C^{123}$ and the top one a $C^{213}$.
The OPE coefficients are invariant under cyclic permutations and get complex conjugated under anti-cyclic ones.

\paragraph{Elementary Crossing Transformations}

For BCFT conformal blocks, we follow the conventions of \cite{Kusuki:2021gpt}.
\cite{Kusuki:2021gpt} normalises boundary primaries so that
\begin{equation}
  \ev{O^{i} (x) O^{j} (y)} = g \frac{\delta^{ij}}{|x-y|^{2 \Delta_{i}}}.
  \label{eqn:kusuki-nomralisation}
\end{equation}
The $g$ in this definition ensures that $C^{\mathds{1}}_{\mathds{1}} = C^{\mathds{1} ii} = 1$.
This resembles the case without boundary, where $C_{\mathds{1} ii} = 1$.
Our $C$s are OPE coefficients and not the coefficients that appear in correlation functions; for example,
\begin{equation}
  O_{i} O_{j} \sim C^{ijk} O_{k} \quad \implies \quad \ev{O_{i} O_{j} O_{k}} \sim g C^{ijk}
  \label{eqn:C-meaning}
\end{equation}

Using this notation, we can list out the elementary crossing transformations that can be used to relate any two channels.
There are four of these in the theory without boundaries, the four-point crossing, the modular S-transformation, the braiding and the Dehn twist.
They are given by
\begin{align}
  \input{figs/s-channel-open-conf.tex}
	&=\, \int \dd{\mu(P_{t})} \VF{s}{t}{2}{3}{1}{4}
	\input{figs/t-channel-open-conf.tex} \,,
  \label{eqn:F-defn} \\
	
\begin{tikzpicture}[baseline]
	\draw (0,0) -- node[above] {\small $1$} (.5,0) arc(180:540:.5) node[pos=.75,above] {\small $2$};
\end{tikzpicture}

	&= \int \dd{\mu (P_{2'})} \mathds{S}_{22'} [1] \ 
	
\begin{tikzpicture}[baseline]
	\draw (0,0) -- node[above] {\small $1$} (.5,0) arc(180:540:.5) node[pos=.75,above] {\small $2$};
	\node (S) at (1,0) {\small $\mathrm{S}$};
\end{tikzpicture}
\,, 
	\label{eqn:S-defn} \\
	
\begin{tikzpicture}[baseline]
	\draw (0,-.75) -- node[right] {\small $3$} (0,0);
	\draw (0,   0) -- node[left ] {\small $1$} (120:.75);
	\draw (0,   0) -- node[right] {\small $2$} ( 60:.75);
\end{tikzpicture}

	&= \mathds{B}_{3}^{12}\ 
	
\begin{tikzpicture}[baseline]
	\draw (0,-.5) -- node[right] {\small $3$} (0,0);
	\draw (0,0) .. controls (-.5,0) and (-.5,.25) .. node[pos=.9,right] {\small $2$} (60:.75);
	\draw[preaction={draw, white, line width=2pt}] (0,0) -- node[pos=.9,left] {\small $1$} (120:.75);
\end{tikzpicture}
 \,
	= \mathds{B}^{3}_{12}
	
\begin{tikzpicture}[baseline]
	\draw (0,-.5) -- node[right] {\small $3$} (0,0);
	\draw (0,0) .. controls ( .5,0) and ( .5,-.25) .. node[pos=.9,left] {\small $1$} (120:.75);
	\draw[preaction={draw, white, line width=2pt}] (0,0) -- node[pos=.9,right] {\small $2$} ( 60:.75);
\end{tikzpicture}

	\label{eqn:B-defn} \\
	\tikz{\draw (0,0) -- node[above] {\small $1$} (.75,0);} &= \mathds{T}_{1}\ 
	
\begin{tikzpicture}[baseline]
	\draw (0,0) -- (.5,0);
	\draw (.5,-.4) to[out=180,in=180] node[pos=.9,above] {\small $1$} (.75,0);
	\draw[preaction={draw, white, line width=2pt}] (0,0) -- (.5,0) arc(90:-90:.2);
\end{tikzpicture}
.
	\label{eqn:T-defn}
\end{align}
where $\mathds{F}, \mathds{S}, \mathds{B}$ are known as the Virasoro crossing kernel, modular crossing kernel and braiding phase respectively \cite{Ponsot:1999uf,Eberhardt:2023mrq}.
In equations like these, the numbers should be understood as labels of Liouville momenta, e.g. $1 \to P_{1}$.
The measure $\dd{\mu(P)}$ is defined by
\begin{equation}
  \int \dd{\mu(P)} f(P) = f(i/2b) + \int_{\mathds{R}^{+}} \dd{P} f(P).
  \label{eqn:P-meas}
\end{equation}
These moves can be applied to any subdiagram in a conformal block.
We will need the following values of these kernels:
\begin{align}
	\mathds{B}^{3}_{12} = \left( \mathds{B}^{12}_{3} \right)^{-1} &= e^{\pi i (h_{1} + h_{2} - h_{3})} \nonumber\\
	T_{1} &= e^{2\pi i h_{1}} \nonumber\\
	\VF{\mathds{1}}{t} {2} {1} {1} {1} &= \rho_{\mathrm{p}} (t) \mathfrak{C}_{P12} 
	\label{eqn:F-C} \\
	\mathds{S}_{\mathds{1} 1} [\mathds{1}] &= \rho_{\mathrm{p}} (P_{2}) \\
	\VF{s} {t} {2} {1} {3} {4} &= \rho_{\mathrm{p}} (s) \sqrt{ \frac{\mathfrak{C}_{12s} \mathfrak{C}_{s34}}{\mathfrak{C}_{41t} \mathfrak{C}_{t23}}} 
	\begin{Bmatrix}
		2 & 3 & s \\
		1 & 4 & t \\
	\end{Bmatrix}
	.
	\label{eqn:6j-F-reln}
\end{align}
The last factor is known as the Virasoro $6j$-symbol, already seen in the bulk inner product.

The new elementary transformations in the case with boundary are listed in \cite{Lewellen:1991tb}, and the corresponding crossing kernels (which can be built out of $\mathds{F},\mathds{S},\mathds{B}$) in \cite{Behrend:1999bn}.
The two simplest ones are
\begin{align}
  \input{figs/bcft-4-pt-s.tex}
	&= \int \dd{\mu(P_{t})} \VF{s}{t}{2}{3}{1}{4}
	\input{figs/bcft-4-pt-t.tex}
  \label{eqn:bcft-xing-F} \\
	
\begin{tikzpicture}[baseline]
	\draw[chillred,double=white] (0,0) -- node[above] {\small $1$} (1,0);
	\draw (1,0) -- node[above] {\small $2$} (2,0);
	\draw[chillred,double=white,dotted,thick] (2,0) -- (2.4,0);
\end{tikzpicture}

	\equiv
	
\begin{tikzpicture}[baseline]
	\draw[chillred,double=white] (0,0) -- node[above] {\small $1$} (1,0);
	\draw (1,0) -- node[above] {\small $2$} (2,0);
	\draw[chillred,double=white] (2,0) -- node[pos=1,right] {\small $\mathds{1}$} (2.4,0);
\end{tikzpicture}

	&= \int \dd{\mu(P_{2'})} \mathds{S}_{2 2'} [P_{1}]\ 
	
\begin{tikzpicture}[baseline]
	\draw[chillred,double=white] (0,0) -- node[above] {\small $1$} (1,0) arc(180:-180:.5cm);
	\node[chillred] (2p) at (2.2,0) {\small $2'$};
\end{tikzpicture}

	\ .
	\label{eqn:bcft-xing-s}
\end{align}
The second equation is about the cylinder one-point function; the reader may find the depiction in \cite{Numasawa:2022cni} useful.
We have also taken the opportunity to introduce an extra piece of our diagrammatic notation: a dotted line in a conformal block means the corresponding primary is the identity.

While the Moore-Seiberg-Lewellen moves are all the crossing transformations for conformal blocks, there are more moves one can make when calculating OPE statistics.
Suppose we are calculating the microcanonical average of $C^{123} C^{213}$, which appears in the partition function of a sphere with three holes as in \eqref{eqn:conf-block-cons}.
The same combination of OPE coefficients appears in the partition function of a sphere with four holes, if one of the open cycles is fixed to be the identity operator,
\begin{equation}
  \input{figs/g3-open-id-dotted.tex}
	\ = C^{123} C^{213}
	
\begin{tikzpicture}[baseline]
	\draw[chillred,double=white] (0,-.5) -- node[left] {\small $2$} (0,.5) arc(90:540:.5cm);
	\draw[chillred,double=white,dotted,thick] (0,0) -- (.5,0);
	\node[chillred] (1) at (-.7,0) {\small $1$};
	\node[chillred] (3) at ( .7,0) {\small $3$};
\end{tikzpicture}
\ 
	.
  \label{eqn:add-id}
\end{equation}
So, any identity-dominated channels for this partition function also contribute to the OPE density.

\subsection{OPE statistics calculations}

As an example, let us calculate $\mathds{E} \left[ C^{123} C^{213} \right]$ when all three primaries are heavy, using these.
We will use the partition function on the sphere with three holes, shown in \eqref{eqn:conf-block-cons}.
The OPE coefficients appearing in the channel shown there are precisely the ones we want to calculate.
To ensure that all three primaries are heavy, we need that all three strips depicted in the conformal block in \eqref{eqn:conf-block-cons} are `thin.'
As a result, the channel in which all three are appropriately dualised are dominated by the identity.
This channel is
\begin{equation}
  
\begin{tikzpicture}[baseline]
	\draw[thick,chillred] (  0,0) circle (   1cm);
	\draw[thick,chillred] ( .5,0) circle (.125cm);
	\draw[thick,chillred] (-.5,0) circle (.125cm);

	\draw[gray] (0,1) -- (0,-1);
	\draw[gray] ( .5,0) circle (.2cm);
	\draw[gray] (-.5,0) circle (.2cm);
\end{tikzpicture}

	\quad \longrightarrow \quad 
	
\begin{tikzpicture}[baseline]
	\draw[chillred,double=white] (-.2,0) -- node[pos=0,left] {\small $\mathds{1}$} (0,0);
	\draw (0,0) -- (.5,0);
	\draw[chillred,double=white] (.5,0) -- (1.5,0);
	\draw (1.5,0) -- (2,0);
	\draw[chillred,double=white] (2,0) -- node[pos=1,right] {\small $\mathds{1}$} (2.2,0);
\end{tikzpicture}

  \label{eqn:g2-id-ch}
\end{equation}
We now have to transform this channel into the one depicted in \eqref{eqn:conf-block-cons}:
\begin{align}
  g^{4} \input{figs/g2-eg-1.tex}
	&= g^{4} \int \dd{\mu(P_{1})} \dd{\mu(P_{3})} S_{\mathds{1} 1} [\mathds{1}] S_{\mathds{1} 3} [\mathds{1}] \ 
	\input{figs/g2-eg-2.tex} \nonumber\\
	&= g^{4} \int \dd[3]{\mu(P)} S_{\mathds{1} 1} [\mathds{1}] S_{\mathds{1} 3} [\mathds{1}] \VF{\mathds{1}}{2}{1}{3}{1}{3}
	\input{figs/g2-eg-3.tex}
	\nonumber\\
	\implies \quad \prod_{i=1}^{3} \rho_{\mathrm{p}} (g,P_{i}) \mathds{E} \left[ C^{123} C^{213} \right] &= g^{4} S_{\mathds{1} 1} S_{\mathds{1} 3} \VF{\mathds{1}}{2}{1}{3}{1}{3}
  \label{eqn:g2-bcft-eg}
\end{align}
The $g^{4}$ is technically a part of the OPE density in the left channel; we are skipping one step by considering the product.
The power is decided as follows: there is one factor of $g$ for every \tikz{
	\draw[chillred,double=white] (-.2,0) -- node[pos=0, left] {\small $\mathds{1}$} ( 0,0);
	\draw                        (  0,0) -- node[pos=1,right] {\small $\mathds{1}$} (.2,0);
}
junction.
Now, we use the relations
\begin{align}
  \rho_{\mathrm{p}} (g,P_{i}) &= g^{2} S_{\mathds{1} i}, \nonumber\\
  \VF{\mathds{1}}{2}{1}{3}{1}{3} &= S_{\mathds{1}2} C_{0} (1,2,3)
  \label{eqn:xing-kernel-id-relations}
\end{align}
to find that \cite{Kusuki:2021gpt,Numasawa:2022cni}
\begin{equation}
	\mathds{E} \left[ C^{123} C^{213} \right] = g^{-2} C_{0} (1,2,3).
  \label{eqn:g2-bcft-ans}
\end{equation}

\paragraph{The Emergence of the Bulk Inner Product}

One reason that the OPE block representation of states is so useful is that the OPE block is related to the geometry of a maximal slice in the bulk \cite{Chandra:2023dgq}.
In particular, the blue lines in the OPE blocks correspond to extremal surfaces on this slice that are not homotopic to any boundary region; and the conformal weight is related to the area of the extremal surface \cite{Chandra:2023dgq}.\footnote{
	In the presence of matter, there might be homotopy classes where there are no extremal surfaces.
	In this case, the blue line still corresponds to the homotopy class and the conformal weight is related to the holographic covariant entropy bound \cite{Chandra:2023dgq,Soni:2024aop}.
}
`Pulling back' through this correspondence the Wheeler-DeWitt inner product natural to bulk slices, the inner products between OPE blocks should have the property that the internal lines also agree between the bras and the kets; for example, \eqref{eqn:norm-conf-block} should have a $\delta (P_{s} - P_{s'})$ hiding in it.\footnote{
	It should be noted that this sort of inner product is also found in Virasoro TQFT \cite{Collier:2023fwi}.
}
We will find that this orthogonality emerges via the averaged OPE coefficients.

To see this, let us try to learn the OPE density in \eqref{eqn:norm-conf-block} in the limit where all the weights that appear are heavy.
For simplicity, we work in the BCFT case; the story is completely parallel.
Going to a sphere with four holes such that the cycles corresponding to $1,2,3,4,s,s'$ are `thin,' we can dualise all these cycles to find a channel where all the cycles are `wide.'
We write
\begin{align}
  g^{6} \input{figs/ssp-eg-1.tex}
	&= g^{6} \int \dd{P_{1}} \dd{P_{4}} \dd{P_{s}} S_{\mathds{1} 1} S_{\mathds{1} 4} S_{\mathds{1} s}
	\input{figs/ssp-eg-2.tex}
	\nonumber\\
	&= g^{6} \int \dd[4]{P_{1,4,s,\tilde{s}}} S_{\mathds{1} 1} S_{\mathds{1} 4} S_{\mathds{1} s} \VF{\mathds{1}}{ \tilde{s}}{s}{\mathds{1}}{s}{\mathds{1}}
	\input{figs/ssp-eg-3.tex}
	\nonumber\\
	&= g^{6} \int \dd[6]{P_{1,2,3,4,s,\tilde{s}}} S_{\mathds{1} 1} S_{\mathds{1} 4} S_{\mathds{1} s} \VF{\mathds{1}}{ \tilde{s}}{s}{\mathds{1}}{s}{\mathds{1}} \VF{\mathds{1}}{2}{1}{\tilde{s}}{1}{s} \VF{\mathds{1}}{3}{\tilde{s}}{4}{s}{4} \nonumber\\
	& \qquad \qquad \qquad \qquad \qquad 
	\times \input{figs/ssp-eg-4.tex}
	\ .
  \label{eqn:bulk-ip-emergence}
\end{align}
The integrand in the RHS should be equal to $\prod_{I = 1,2,3,4,s,\tilde{s}} \rho_{\mathrm{p}}(P_{I}) \mathds{E} \left[ C^{12s} C^{21 \tilde{s}} C^{ \tilde{s} 43} C^{s34} \right]$.
Plugging in $\rho_{\mathrm{p}}$ from \eqref{eqn:xing-kernel-id-relations}, we find
\begin{equation}
	\mathds{E} \left[ C^{12s} C^{21 \tilde{s}} C^{ \tilde{s}43} C^{s34} \right] = g^{-6} \frac{\VF{\mathds{1}}{ \tilde{s}}{s}{\mathds{1}}{s}{\mathds{1}} \VF{\mathds{1}}{2}{1}{\tilde{s}}{1}{s} \VF{\mathds{1}}{3}{\tilde{s}}{4}{s}{4}}{S_{\mathds{1} \tilde{s}} S_{\mathds{1}2} S_{\mathds{1}3}}
  \label{eqn:bulk-ip-emergence-2}
\end{equation}
The crucial fact is that \cite{Belin:2021ryy}
\begin{equation}
  \VF{\mathds{1}}{ \tilde{s}}{s}{\mathds{1}}{s}{\mathds{1}} = S_{\mathds{1} \tilde{s}} \delta (P_{s} - P_{ \tilde{s}}),
  \label{eqn:F-delta}
\end{equation}
which gives
\begin{equation}
	\mathds{E} \left[ C^{12s} C^{21 \tilde{s}} C^{ \tilde{s}43} C^{s34} \right] = g^{-6} C_{0} (1,2,s) C_{0} (3,4,s) \delta (P_{s} - P_{ \tilde{s}}).
  \label{eqn:bulk-ip-emergence-3}
\end{equation}

\bibliographystyle{JHEPmod}
\bibliography{refs.bib}
\end{document}